\documentclass[letterpaper,10pt,journal,twoside]{ieeeconf}
\IEEEoverridecommandlockouts
\usepackage{xcolor}
\usepackage{amsmath,amssymb,mathtools}

\usepackage{amsthm}
\newtheorem{theorem}{Theorem}
\newtheorem{corollary}{Corollary}
\newtheorem{proposition}{Proposition}
\newtheorem{lemma}{Lemma}
\newtheorem{assumption}{Assumption}
\newtheorem{example}{Example}
\theoremstyle{remark}
\newtheorem{remark}{Remark}
\newtheorem{problem}{Problem}
\usepackage{url}
\usepackage{graphicx}
\usepackage[hidelinks,hypertexnames=false]{hyperref}
\usepackage[capitalize,noabbrev,nameinlink]{cleveref}
\crefname{theorem}{Theorem}{Theorems}
\crefname{corollary}{Corollary}{Corollaries}
\crefname{proposition}{Proposition}{Propositions}
\crefname{lemma}{Lemma}{Lemmas}
\crefname{assumption}{Assumption}{Assumptions}
\crefname{remark}{Remark}{Remarks}
\usepackage{cite}
\usepackage{booktabs}
\usepackage{array}
\usepackage{makecell}
\usepackage{tikz}
\usetikzlibrary{tikzmark,calc}
\usepackage{etoolbox}

\DeclareMathOperator*{\argmin}{arg\,min}
\newcommand{\Prob}{\mathbb{P}}
\newcommand{\E}{\mathbb{E}}
\newcommand{\R}{\mathbb{R}}
\newcommand{\Fc}{\mathcal{F}}
\newcommand{\Xc}{\mathcal{X}}
\newcommand{\Uc}{\mathcal{U}}
\newcommand{\Wc}{\mathcal{W}}
\newcommand{\Cc}{\mathcal{C}}
\definecolor{vl}{HTML}{009E73}

\makeatletter
\patchcmd{\@maketitle}{\vskip1.0em}{\vskip0.5em}{}{}
\IEEEaftertitletext{\vspace{-2\baselineskip}}
\newenvironment{inlinetable}{\par\addvspace{3pt}\begingroup\def\@captype{table}\centering}
  {\par\endgroup\addvspace{3pt}}
\newenvironment{inlinefigure}{\par\addvspace{2pt}\noindent
  \begin{minipage}[b]{\textwidth}\def\@captype{figure}\centering}
  {\end{minipage}\par\addvspace{2pt}}
\newcommand{\guaranteeboxrow}[1]{\multicolumn{9}{@{}l}{}&\multicolumn{2}{@{}l}{\tikzmark{gb#1a}}&
  \multicolumn{2}{r@{}}{\tikzmark{gb#1c}}\\}
\newcommand{\guaranteeboxend}[1]{\noalign{\hbox to0pt{\tikzmark{gb#1b}\tikz[remember picture,overlay]{%
  \path(pic cs:gb#1a)coordinate(gbA)(pic cs:gb#1b)coordinate(gbB)([xshift=\tabcolsep]pic cs:gb#1c)coordinate(gbC);
  \path([yshift=\ht\strutbox]gbA)coordinate(gbTL)([yshift=-0.5pt]gbC|-gbB)coordinate(gbBR);
  \draw[line width=\lightrulewidth,rounded corners=1.5pt](gbTL)rectangle(gbBR);
  \node[anchor=base,inner sep=0pt,font=\footnotesize]at($(gbA)!0.5!(gbC)$)
    {Theorem~\ref*{prop:repeated-deployment}};}\hss}}}
\newif\if@raggedcolumn
\def\@texttop{\begingroup
  \count@\vbadness\vbadness\@M
  \setbox\z@\vbox to\@colht{\dimen@\dp\@outputbox\unvcopy\@outputbox\vskip-\dimen@}%
  \ifnum\badness>\count@\aftergroup\@raggedcolumntrue\else\aftergroup\@raggedcolumnfalse\fi
  \endgroup}
\def\@textbottom{\if@raggedcolumn\vskip\z@\@plus.0001fil\fi}
\makeatother

\title{\LARGE \bf
Anytime-Valid Certification of Robustness Margins\\
for Safe Control via E-Values%
}

\author{{Panagiotis S. Trakas$^{*}$, Omid Mirzaeedodangeh$^{*}$, and Lars Lindemann}
\thanks{The authors are with the Automatic Control Laboratory, ETH Z\"urich, Switzerland. E-mails: {\tt\small ptrakas@ethz.ch, omirzaeedoda@ethz.ch, llindemann@ethz.ch.}}
\thanks{$^{*}$Panagiotis S. Trakas and Omid Mirzaeedodangeh contributed equally.}
}

\begin{document}
\maketitle

\begin{abstract}
Robust control uses the error between a known model and the true system as a robustness margin. However, this error is usually unknown: conservative estimates restrict feasibility and performance, while underestimates can result in violations of the control property. We estimate this robustness margin from online data using e-values, a statistical tool for hypothesis testing, in the setting of discrete-time control barrier functions (CBFs). Here, we measure the robustness margin as the difference between the observed CBF and the model-based CBF prediction. We then design a robust CBF controller in three steps by: (1) selecting a set of candidate margins, (2) obtaining validity certificates for these candidates via sequential tests based on e-values, and (3) updating the CBF controller with the smallest certified candidate. The deployed tests accumulate evidence against the hypothesis that an exceedance risk, the probability of the robustness margin exceeding the candidate margin, remains at or above a pre-specified failure probability $\alpha$.  Our tests are anytime valid and thus enable the use of certified candidates in the CBF controller, guaranteeing that the probability that the exceedance risk of a certified margin exceeds $\alpha$ is at most a user-chosen confidence $\delta$. We validate our approach on navigation problems across different environments with visualizations in NVIDIA Isaac Sim.

\end{abstract}

\section{Introduction}
\label{sec:intro}

Learning-based controllers are increasingly deployed in a variety of applications, but often lack formal safety guarantees. This has motivated the use of safety filters via control barrier functions (CBFs) \cite{ames2017cbfqp}. However, their reliability depend on the availability of an accurate model of the system. In the presence of model mismatches, robust CBFs can still ensure safety by tightening the CBF safety constraint via a robustness margin, which  must be larger than the model error \cite{jankovic2018robust}.  Unfortunately, the model error is typically unknown and depends on the system's operating range. Robustness margins that are  chosen too small may  provide insufficient protection, whereas unnecessarily large choices lead to conservative or infeasible problems. This issue, and hence our contribution, is not limited to CBFs and applies to other tools, e.g., robust model predictive control and control Lyapunov functions.

\begin{figure}[t]
    \centering
    \includegraphics[width=0.99\columnwidth]{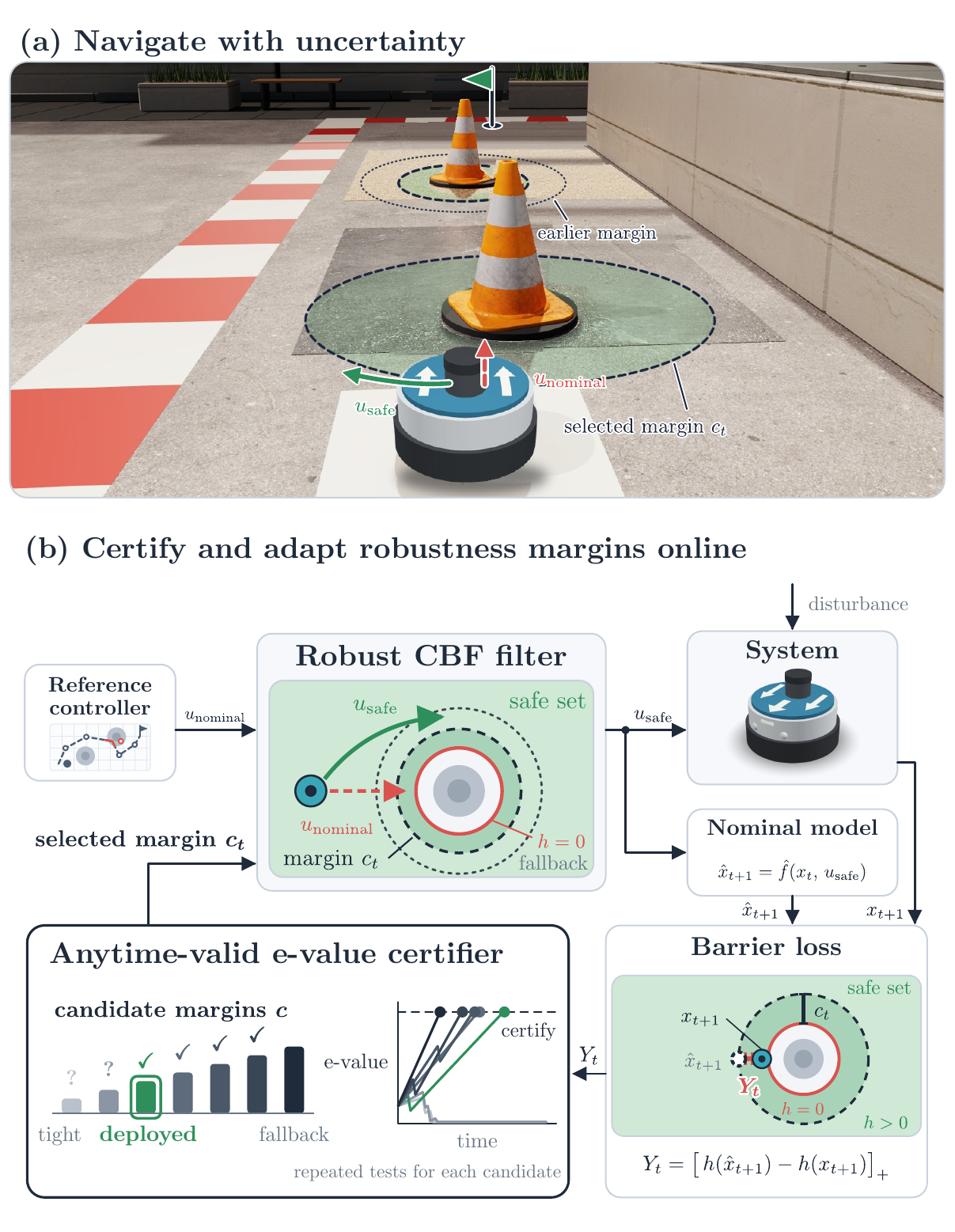}
    \caption[Overview of the proposed approach.]{Overview of the proposed approach.
(a) Robot navigation under model uncertainty, requiring a robustness margin $c_t$.
(b) Closed-loop barrier losses $Y_t$ provide evidence
for selecting valid robustness margins $c_t$.\protect\footnotemark}
    \label{fig:method-overview}
\end{figure}
\footnotetext{Videos from experiments at
\url{https://tinyurl.com/28bendsa} (all videos) and
\url{https://tinyurl.com/25qkw34h} (Maze robot).}

Data-driven estimates  of the model error can be obtained in practice, but are often heuristic and  approximate, resulting again in unsafe behavior. Therefore, recent work has studied how probabilistic safety guarantees can be obtained using Gaussian Processes \cite{lederer2019uniform,wabersich2023data} and conformal prediction \cite{hsu2025statistical}. While Gaussian Processes rely on a well-calibrated model assumption, conformal prediction struggles in dealing with shifts of the data generating distribution due to the closed-loop dependence between controller and data. This issue was addressed in \cite{mirzaeedodangeh2026robust,wei2025conformal,mirzaeedodangeh2026safe}, e.g., via robust conformal prediction, but only under additional regularity assumptions. Importantly, these works require the availability of an offline calibration dataset. We argue for departing from \emph{conformal prediction} in favor of \emph{e-values} \cite{ramdas2025hypothesis}, a statistical tool for hypothesis testing whose properties accommodate closed-loop distribution shifts without requiring an offline dataset.

Our proposed method is initialized with a conservative yet valid fallback robustness margin, see Fig.~\ref{fig:method-overview}. We then start collecting closed-loop data with the goal to find smaller robustness margins. Therefore, at each time step, we compute the \emph{barrier loss}, which is the amount by
which the observed CBF falls below its
model-based prediction. We then select a finite set of candidate margins. For each candidate, we test the hypothesis
that the exceedance risk, which is the probability of the barrier loss exceeding the candidate margin, remains at or above a user-chosen error confidence. As we need to check this hypothesis at each time step, we require anytime-valid tests for which we use sequential tests based on e-values, as opposed to often used p-values
\cite{ramdas2023savi,grunwald2024safe,vovk2021evalues}.

A certificate is now issued if sufficient evidence  against this hypothesis is obtained. Such a certificate tells us that the exceedance risk for the corresponding candidate was below the pre-specified failure probability at some point in the past. To transfer this certificate into a safe control action, we have to account for the fact that the exceedance risk may increase in the future, which is done by adding an exceedance buffer. The CBF controller then selects the smallest certified candidate.  Our main contributions are as follows:
\begin{enumerate}
    \item We develop anytime-valid certificates for robustness margins of CBF controllers using online data and sequential tests based on e-values. We also derive upper bounds for the expected certification time.

\item We propose a robust CBF controller that transfers certified
robustness margins into safe control actions. This transfer relies on a bound on the increase in exceedance risk, which introduces conservatism. Therefore, we permit parallel tests for efficient control design.

\item We compare our method against baseline algorithms on unicycle navigation problems in corridor and maze environments, visualized in NVIDIA Isaac Sim.
\end{enumerate}
\subsection{Related Work}
\label{sec:related_work}
Robust and stochastic safety filters account
for model uncertainty through error bounds or additional
distributional information.
Predictive safety filters use model-error bounds that
can be learned from data \cite{wabersich2021predictive},
while adaptive CBFs use state-estimation error
bounds \cite{das2026safe}.
Probabilistic CBFs use known moments or i.i.d. disturbance samples
\cite{mestres2026probabilistic}, while stochastic CBFs rely on a known disturbance distribution \cite{cosner2023robust}.An open challenge is to  quantify error bounds and to certify robustness
margins directly from online
observations.

Conformal prediction (CP) is increasingly used in
control to construct uncertainty bounds from observed
prediction errors without assuming a parametric
error distribution \cite{lindemann2025formal}.
Under i.i.d. assumptions, standard CP approaches provide
marginal coverage, so guarantees averaged over offline calibration
trajectories and a test trajectory. One of the main challenges in control, however, is  dealing with non i.i.d. data.

Online conformal prediction methods adapt uncertainty bounds
in tube-based MPC \cite{cherenson2026staggered} or switch thresholds in learned safety filters
\cite{huriot2026adaptive}. However, these provide  safety guarantees that are only valid on
average over time. For CBF controllers,
\cite{mirzaeedodangeh2026robust} provides an iterative procedure to deal with distribution shifts by updating robustness
margins using large sets of i.i.d. calibration
trajectories. Instead, we use anytime-valid tests
that require neither known disturbance
distributions, offline calibration data, nor i.i.d. observations.

Sequential hypothesis testing methods based on e-values have been
developed for change detection
\cite{shin2023edetectors} and
risk-controlling prediction sets \cite{xu2024active}. Conformal e-prediction provides anytime-valid coverage
for i.i.d. data arriving in batches
\cite{gauthier2025evalues}.
In robotics, martingale tests detect distribution
shifts with guarantees on false alerts
\cite{luo2024online}, while multiple hypothesis testing can help
identify the type of distribution shift
\cite{hindy2024diagnostic}. Sequential tests can also be used to compare robot policies
using i.i.d. evaluation trials \cite{snyder2026policy}.   Closest to our work are \cite{cosner2023robust,jagtap2020formal,santoyo2021barrier,fochesato2025synthesis,cosner2024bounding}, which formulate safety through enforcing a martingale condition. However, verifying the validity of this martingale condition requires precise knowledge of the disturbance distribution and must be performed offline.  To the best of our knowledge, no one has
used anytime-valid sequential certification of robustness margins from
online data to design safe feedback controllers.

\section{Problem Formulation}
\label{sec:problem}

\subsection{Stochastic System and Nominal Model}
\label{subsec:system}

Let \((\Omega,\Fc,\Prob)\) be a probability space, where \(\Omega\) is the
sample space, \(\Fc\) is the associated \(\sigma\)-algebra, and \(\Prob\) is
a probability measure. Let $\E$ denote expectation under $\Prob$. Relations between random variables are understood to hold almost surely.\footnote{An event holds
$\Prob$-almost surely if it holds with probability one under  $\Prob$.} We consider a discrete-time stochastic
system:
\begin{equation}
    x_{t+1}=f(x_t,u_t,w_t),
    \qquad t\in\mathbb N_0,
    \label{eq:true-stochastic-system}
\end{equation}
where \(x_t:\Omega\to\Xc\subseteq\mathbb R^n\),
\(u_t:\Omega\to\Uc\subseteq\mathbb R^m\), and
\(w_t:\Omega\to\Wc\subseteq\mathbb R^p\) denote the state,  control
input, and  disturbance, respectively. Moreover, $x_0:\Omega\to\Xc$ is the random initial state of \eqref{eq:true-stochastic-system}. The actual transition map
\(f:\Xc\times\Uc\times\Wc\to\Xc\) is unknown. Suppose, however, that a nominal one-step predictor
$\hat f:\Xc\times\Uc\to\Xc$ is available, e.g., obtained from first principles
or via system identification. Define the model error $e_t:=x_{t+1}-\hat f(x_t,u_t)$,
so that
\begin{equation}
    x_{t+1}=\hat f(x_t,u_t)+e_t.
    \label{eq:residual-decomposition}
\end{equation}
At each time \(t\), a causal, possibly randomized controller, such as an RL policy, generates a nominal input
\(u_t^{\mathrm{nom}}:\Omega\to \mathbb{R}^m\). The control input \(u_t\) is then
obtained by passing \(u_t^{\mathrm{nom}}\) through a CBF filter,  introduced below, which modifies the nominal input when necessary to enforce safety constraints.

The probability space supports the initial state, the disturbance process
$\{w_t\}_{t\geq0}$, and any randomization used to generate
$u_t^{\mathrm{nom}}$. Together with the system dynamics \eqref{eq:true-stochastic-system} and the
CBF controller, these random variables induce the state and input
processes $\{x_t\}_{t\geq0}$ and $\{u_t\}_{t\geq0}$, respectively. Let $\{\Fc_t\}_{t\geq0}$ denote the decision-time filtration, where
$\Fc_t$ contains all information available after $u_t$ is selected, but before
$x_{t+1}$ is observed. In particular, $\Fc_t$ contains the past states
and control inputs
$x_0,u_0,\ldots,x_{t-1},u_{t-1}$, the current state $x_t$, the nominal input $u_t^{\mathrm{nom}}$, and the
control input $u_t$. Thus, $x_t$, $u_t^{\mathrm{nom}}$, and $u_t$ are
$\Fc_t$-measurable, while $x_{t+1}$ and $e_t$ are
$\Fc_{t+1}$-measurable and are not available at  time $t$.

\subsection{Robust CBF Safety Filter}
\label{subsec:robust-cbf}

Let \(h:\Xc\to\mathbb R\) define the safe set $ \Cc:=\{x\in\Xc\mid h(x)\geq0\}.$
Our objective is to keep the state $x_t$ inside \(\Cc\) despite the mismatch
between the nominal  model $\hat{f}$ and the  system $f$. Therefore, we utilize robust CBFs for which we let \(c_t\geq0\) be a robustness margin. Consider now a control input \(u_t\)
satisfying the tightened nominal barrier condition
\begin{equation}
    h\bigl(\hat f(x_t,u_t)\bigr)-c_t
    \geq(1-\gamma)h(x_t),
    \qquad \gamma\in(0,1].
    \label{eq:tightened-cbf-condition}
\end{equation}
For a given margin $c\geq0$, define the feasible input set
\begin{equation*}
    \Uc(x_t,c)
    :=
    \left\{
        u\in\Uc:
        h\bigl(\hat f(x_t,u)\bigr)-c
        \geq(1-\gamma)h(x_t)
    \right\}.
\end{equation*}
The corresponding CBF safety filter modifies the nominal input
$u_t^{\mathrm{nom}}$ as little as possible:
\begin{equation}
    u_t(c)
    \in
    \argmin_{u\in\Uc(x_t,c)}
    \left\|u-u_t^{\mathrm{nom}}\right\|^2.
    \label{eq:cbf-safety-filter}
\end{equation}
If $\Uc(x_t,c)$ is nonempty, we assume that the minimum
is attained\footnote{If $\Uc$ is polyhedral
and the barrier constraint is affine in $u$, the filter is a
convex quadratic program with a unique minimizer whenever
feasible.} and that a minimizer can be selected.
The control input is $u_t:=u_t(c_t)$.\footnote{The variables \(c_t\) and \(u_t\) need to be \(\Fc_{t+1}\)-measurable.}
To quantify how the model mismatch affects the true barrier
condition, define the \emph{barrier loss}
\begin{equation}
    Y_t
    :=
    \left[
        h\bigl(\hat f(x_t,u_t)\bigr)-h(x_{t+1})
    \right]_+
    \label{eq:barrier-loss}
\end{equation}
where \([\chi]_+:=\max\{\chi,0\}\). Since \([\chi]_+\geq \chi\) for every \(\chi\in\mathbb R\), we know that the following inequality is always satisfied:
\begin{equation}
    h(x_{t+1})
    \geq h\bigl(\hat f(x_t,u_t)\bigr)-Y_t.
    \label{eq:barrier-loss-bound}
\end{equation}
Using equation
\eqref{eq:barrier-loss-bound}, we can now present a sufficient condition for our goal. Indeed, knowing that $u_t\in \Uc(x_t,c)$, we have
\begin{equation}
   Y_t \leq c_t
    \quad\Longrightarrow\quad
    h(x_{t+1})\geq(1-\gamma)h(x_t),
    \label{eq:true-cbf-condition}
\end{equation}
i.e., $Y_t\leq c_t$ is sufficient to ensure the ‘‘then" part of \eqref{eq:true-cbf-condition}, which is the true barrier condition \cite{agrawal2017discrete}. Hence, if $x_0\in\Cc$ and $Y_t\leq c_t$ for all $t\in\mathbb N_0$,
then $x_t\in\Cc$ for all $t\in\mathbb N_0$.

We note that the  loss \(Y_t\) becomes observable only after
\(u_t\) is applied and \(x_{t+1}\) is observed. It is hence
\(\Fc_{t+1}\)-measurable and unavailable when \(c_t\) and \(u_t\) are
selected. At  time \(t\), only the previous losses
\(Y_0,\ldots,Y_{t-1}\) are available.

\subsection{Problem Statement}
\label{subsec:problem-statement}
Let us select a CBF-violation probability
$\alpha_0\in(0,1)$. At every time step at which a margin $c_t$ is deployed
and an input $u_t$ satisfying \eqref{eq:cbf-safety-filter} is
applied, we seek to ensure
\begin{equation*}
    \Prob\!\left(
        h(x_{t+1})<(1-\gamma)h(x_t)
        \,\middle|\,\Fc_t
    \right)
    \leq\alpha_0.
\end{equation*}
In words, given all information available at time \(t\), we want that the
probability that the barrier condition $h(x_{t+1})\geq (1-\gamma)h(x_t)$ is violated does not
exceed \(\alpha_0\). By the sufficient condition in \eqref{eq:true-cbf-condition}, we can achieve this goal by ensuring
\begin{equation*}
    p_t(c_t)\leq\alpha_0
\end{equation*}
where $p_t(c)$ denotes the exceedance risk at time $t$:
\begin{equation}
    p_t(c):=\Prob(Y_t>c\mid\Fc_t)
    \label{eq:conditional-exceedance}
\end{equation}
 i.e., the probability that the loss $Y_t$ exceeds the margin $c$.

The difficulty here is circular: $c_t$ must be selected before
$Y_t$ is observed, while $Y_t$ depends on $c_t$. In particular,
$c_t$ influences the control input in \eqref{eq:cbf-safety-filter} and hence the resulting
barrier loss $Y_t$ in \eqref{eq:barrier-loss}. Additionally, the exceedance risk $p_t(c_t)$ is unknown.  Before sufficient data is available to certify
a margin $c_t$, we therefore require
a pre-validated margin that guarantees feasibility
and bounds the barrier loss.

\begin{assumption}[Safe Margin]
\label{as:conservative-margin}
There exists a known margin $c_{\max}>0$ such that
$\Uc(x_t,c_{\max})\neq\emptyset$ almost surely for all $x_t\in\Xc$.
If $c_t=c_{\max}$, then applying
$u_t=u_t(c_{\max})\in\Uc(x_t,c_{\max})$ ensures $ Y_t\leq c_{\max}$ almost surely  for all $x_t\in\Xc$.
\end{assumption}

Under this assumption, any margin $c<c_{\max}$ has the property that $\Uc(x_t,c_{\max})\subseteq\Uc(x_t,c)$ so that the
safety filter \eqref{eq:cbf-safety-filter} remains feasible.  We restrict the candidate margins to a finite ordered
grid of $M$ candidates
\begin{equation*}
    \Gamma
    :=
    \{c^{(1)}<\cdots<c^{(M)}\}
    \subseteq[0,c_{\max}).
\end{equation*}
The deployed margin $c_t$ will be selected from
$\Gamma\cup\{c_{\max}\}$.

\begin{problem}
\label{prob:margin-selection}
Consider the system in
\eqref{eq:true-stochastic-system} under the safety filter in
\eqref{eq:cbf-safety-filter} and let Assumption \ref{as:conservative-margin} hold. Design a causal rule that, at each time $t\in\mathbb N_0$, uses the data in $\Fc_t$
to select the smallest margin $c_t$ that satisfies $p_t(c_t)\leq\alpha_0$.
\end{problem}

\section{E-Values and Sequential Tests}
\label{sec:sequential-evidence}

\subsection{Composite Null Hypotheses and E-Values}
We first review the statistical background used throughout; further information
is provided in \cite{ramdas2025hypothesis,ramdas2023savi,grunwald2024safe,vovk2021evalues}. A null hypothesis $H_0$ is a set of probability measures
on $(\Omega,\Fc)$ satisfying a specific event.
It is called simple if it contains a single probability measure and
composite otherwise. Throughout this section, $\mathbb Q\in H_0$ denotes a generic probability measure satisfying the hypothesis $H_0$, and
$\E_{\mathbb Q}$ denotes expectation under $\mathbb Q$.
This notation is used only in this section, the closed-loop probability measure remains $\Prob$.

For example, let $Z_t\in\{0,1\}$ indicate whether a specific event (e.g., $Y_t\leq c_t$) occurs at time $t$. Fix $\alpha\in(0,1)$ and consider the composite null hypothesis
\begin{equation}
    H_0
    :=
    \left\{
        \mathbb Q:
        \mathbb Q(Z_t=1\mid\Fc_t)\geq\alpha
    \right\}, \quad t\in\mathbb N_0.
    \label{eq:binary-null-example}
\end{equation}
This is a composite null hypothesis because many probability measures
may satisfy the inequality $Q(Z_t=1\mid\Fc_t)\geq\alpha$. Under any such probability measure, the
event indicated by $Z_t=1$ occurs with probability at least
$\alpha$.

An \emph{e-variable} converts observed data into a nonnegative measure of
evidence against the null hypothesis. Formally, a nonnegative random variable
$E$ is an e-variable for $H_0$ if
\begin{equation}
    \E_{\mathbb Q}[E]\leq1
    ~~
    \text{for every $\mathbb Q$ satisfying $H_0$.}
\label{eq:evariable}
\end{equation}
The realized value of $E$ after the data are observed is called an
\emph{e-value}. Equation \eqref{eq:evariable} ensures that large e-values
cannot occur frequently when $H_0$ is true, so a larger e-value provides
stronger evidence against $H_0$.

To make this interpretation precise, fix an error confidence level
$\delta\in(0,1)$. Markov's inequality then gives
\begin{equation*}
\mathbb Q\!\left(
        E\geq\frac{1}{\delta}
    \right)
    \leq\delta
    ~~
    \text{for every $\mathbb Q$ satisfying $H_0$.}
\end{equation*}
Thus, observing an e-value of at least $1/\delta$ justifies rejecting $H_0$ at
level $\delta$, i.e., with a confidence of $1-\delta$.

\subsection{Test Supermartingales}
We seek to accumulate evidence as new data becomes available. One-step
e-variables can be multiplied over time to form a
\emph{test supermartingale}, provided their $\Fc_t$-conditional
expectation under the null hypothesis are at most one
\cite{grunwald2024safe}. This construction allows  evidence
to be assessed at every time, making it suitable for
online monitoring systems
\eqref{eq:true-stochastic-system}.

A process $(E_t)_{t\geq0}$ is
adapted to the filtration
$(\mathcal{F}_t)_{t\geq 0}$ if $E_t$ is $\Fc_t$-measurable, meaning that its
value uses only the information available at time $t$.
A nonnegative adapted process with $E_0=1$ is a
\emph{test supermartingale} for $H_0$ if, for every
$\mathbb Q\in H_0$, it is integrable and satisfies
\begin{equation}
    \E_{\mathbb Q}[E_{t+1}\mid\Fc_t]\leq E_t,
    \quad t\in\mathbb N_0.
    \label{eq:test-supermartingale}
\end{equation}
Thus, under the null hypothesis, the
conditional expected value of the evidence cannot increase,
although its observed value may rise or fall.
Taking expectations recursively gives
$\E_{\mathbb Q}[E_t]\leq1$, so each $E_t$ is an e-variable.

The key advantage is a validity guarantee over all times $t\in\mathbb N_0$.
For any $\delta\in(0,1)$, Ville's inequality
\cite{ville1939} gives
\begin{equation*}
    \mathbb Q\!\left(
        \sup_{t\geq0}E_t\geq\frac{1}{\delta}
    \right)
    \leq\delta,
    \qquad \mathbb Q\in H_0.
\end{equation*}
Under any $\mathbb Q\in H_0$, the probability that the
evidence ever reaches $1/\delta$ is therefore at most
$\delta$. We may check $E_t$ at every time step and reject
$H_0$ at the first time $E_t$ reaches the threshold. This property is called
\emph{anytime-validity}.

\begin{example}
Let $Z_t=1$ indicate if a robot ends up closer to an obstacle than predicted, and $Z_t=0$ otherwise. For the null hypothesis in \eqref{eq:binary-null-example}, we define
this event to have probability at least
$\alpha$ at every time $t$. Starting from $E_0=1$, update the evidence by
multiplying it by a factor determined by the latest
observation, e.g. $E_{t+1}
    =E_t\left[1+\frac12(\alpha-Z_t)\right].$ For every $\mathbb Q\in H_0$, the nonnegative
update factor satisfies $\E_{\mathbb Q}\!\left[
        1+\frac12(\alpha-Z_t)\mid\Fc_t
    \right] =  1+\frac12\left[
        \alpha-\mathbb Q(Z_t=1\mid\Fc_t)
    \right] \leq 1$.
Thus, each factor is an $\Fc_t$-conditioned e-variable, i.e., their product
satisfies \eqref{eq:test-supermartingale}
and forms a test supermartingale \cite{grunwald2024safe}. For $\alpha=0.1$, the evidence $E_t$ is multiplied by
$1.05$ without an exceedance and by $0.55$ otherwise.
With $\delta=0.05$, rejection occurs at $E_t\geq20$, e.g., $62$ consecutive non-exceedances give
$E_{62}=1.05^{62}>20$.
Under $H_0$, Ville's inequality bounds the probability
of ever reaching this threshold by $5\%$.
\end{example}

\section{Anytime-Valid Robustness Margin Certification}
\label{sec:certification}
Let us now fix $t\in\mathbb N_0$ to be the current time. Our goal is to construct an anytime-valid monitor that uses online data to find candidate robustness margins $c\in \Gamma$ which have satisfied the condition $Y_s \leq c$ for some time $s\leq t$ in the past.
\subsection{Simultaneous Margin Certification}
\label{subsec:false-certification}
At time  $t\geq1$, we use the  data
$(x_{t-1},u_{t-1},x_t)$ to compute the barrier loss $Y_{t-1}$.
For each candidate $c\in\Gamma$, we define
\begin{align}
     Z_{t-1}(c):=\mathbf{1}_{\{Y_{t-1}>c\}}. \label{Zt1}
\end{align}
While $Z_{t-1}(c)$
records whether the observed loss during the current transition exceeded $c$, we recall that the exceedance probability $p_t(c)$ in \eqref{eq:conditional-exceedance} was defined for the next transition.

Next,  choose the   margin insufficiency $\alpha\in(0,\alpha_0]$ where we recall that $\alpha_0$ was the CBF-violation probability from  Section~\ref{subsec:problem-statement}. The difference $\alpha_0-\alpha$ allows for an exceedance buffer to account for an increase
in exceedance risk from $p_s(c)$ in the past to $p_t(c)$ now. Additionally, choose the error confidence $\delta\in(0,1)$. The parameters $\alpha$ and $\delta$ have different roles:
$\alpha$ defines margin insufficiency through $p_t(c)\geq\alpha$,
whereas $\delta$ bounds the probability of ever falsely
certifying any candidate.
For each candidate, we test the null
hypothesis
\begin{equation}
    H_0(c):=
    \left\{
        \Prob:
        \Prob(Y_t>c\mid\Fc_t)\geq\alpha
    \right\}, \quad t\in\mathbb N_0.
\label{eq:persistent-insufficiency-null}
\end{equation}
Under $H_0(c)$, the candidate remains $\alpha$-insufficient throughout
monitoring. We aim to certify the opposite.

For this reason, we define the variable $E_t^\lambda(c)$ that will gather evidence against the null hypothesis $H_0(c)$. Initially, we set $E_0^\lambda(c)=1$ and update $E_t^\lambda(c)$ as
\begin{align}
    E_t^\lambda(c)
    =
    E_{t-1}^\lambda(c)
    \left[1+\lambda\bigl(\alpha-Z_{t-1}(c)\bigr)\right],
    \qquad t\geq1 \label{eq:wealth-recursion}
\end{align}
where $\lambda\in[0,(1-\alpha)^{-1}]$ is a gain that scales the contribution of each observation to the evidence, i.e., larger $\lambda$ increases the factor $1+\lambda\alpha$
when $Z_{t-1}(c)=0$ and decreases the factor
$1-\lambda(1-\alpha)$ when $Z_{t-1}(c)=1$.
The admissible range of $\lambda$ ensures nonnegativity of $E_t^\lambda(c)$.

Under the null hypothesis
$H_0(c)$, we notice that
\begin{equation}
\begin{aligned}
    \E\!\left[
        1+\lambda\bigl(\alpha-Z_t(c)\bigr)
        \,\middle|\,\Fc_t
    \right]
    =
    1+\lambda\bigl(\alpha-p_t(c)\bigr)
    \leq1,
\end{aligned}
\label{eq:one-step-evariable}
\end{equation}
i.e., $1+\lambda\bigl(\alpha-Z_t(c)\bigl)$ is an $\Fc_t$-conditioned e-variable. Their product again form a test supermartingale, i.e., $(E_t^\lambda(c))_{t\geq0}$ is a test
supermartingale under every $\Prob\in H_0(c)$. Therefore, we can apply
Ville's inequality  to $E_t^\lambda(c)$ by which we get
\begin{align*}
    \mathbb \Prob\!\left(
        \sup_{t\geq0}E_t^\lambda(c)\geq\frac{1}{\delta}
    \right)
    \leq\delta,
    \qquad \mathbb \Prob\in H_0.
\end{align*}

\textbf{Reducing sensitivity. }The rate at which evidence accumulates depends on the gain $\lambda$. To
reduce sensitivity to a particular choice, we select a finite set of gains
\[
    \Lambda
    :=
    \{\lambda^{(1)},\ldots,\lambda^{(L)}\}
    \subset (0,(1-\alpha)^{-1})
\]
that we use in parallel for every candidate margin $c\in\Gamma$. Indeed, we combine
the corresponding processes through
\begin{equation}
    E_t(c)
    :=
    \frac{1}{|\Lambda|}
    \sum_{\lambda\in\Lambda}E_t^\lambda(c),
    \qquad E_0(c)=1.
    \label{eq:mixture-eprocess}
\end{equation}
By linearity of conditional expectations,
$(E_t(c))_{t\geq0}$ is again a test supermartingale
under every $\Prob\in H_0(c)$. Thus, combining the gains
preserves the same anytime-valid guarantee, i.e., we can apply Ville's inequality to $E_t(c)$ and obtain \begin{equation}
    \mathbb \Prob\!\left(
        \sup_{t\geq0}E_t(c)\geq\frac{1}{\delta}
    \right)
    \leq\delta,
    \qquad \mathbb \Prob\in H_0.
    \label{eq:E_t_c}
\end{equation}

 For each candidate $c$, define
its first certification time as
\begin{equation}
    \tau(c)
    :=
    \inf\left\{
        t\geq1:
        E_t(c)\geq\frac{1}{\delta}
    \right\}.
    \label{eq:certification-time}
\end{equation}
If the threshold is never crossed, we set $\tau(c)=+\infty$. At decision time $t$, the evidence $E_t(c)$ depends only on
$Y_0,\ldots,Y_{t-1}$. The candidates certified at that time are
\begin{equation*}
    \mathcal S_t
    :=
    \{c\in\Gamma:\tau(c)\leq t\}.
\end{equation*}
If $\mathcal S_t=\emptyset$, no certificate is available and the controller continues to use $c_{\max}$ (see Section \ref{sec:main-results} for details).

\textbf{Simultaneous Guarantees.} So far, we tested all candidate margins $c\in \Gamma$ separately. However, we would ideally like to obtain simultaneous guarantees for all candidate margins. A naive union bounding argument  of \eqref{eq:E_t_c} would give
\begin{align*}
    \mathbb \Prob\!\left(
        \sup_{t\geq0}E_t(c)\geq\frac{1}{\delta}, \; \forall c\in\Gamma
    \right)
    \leq M \delta,
    \qquad \mathbb \Prob\in H_0.
\end{align*}

However, this is overly conservative, increasing the error confidence from $\delta$ to $M\delta$. We can avoid this by noticing that $ c'\geq c \implies E_t(c')\geq E_t(c)$. Thus, certification of $c$ also certifies every larger
candidate $c'$; see Lemma~\ref{prop:monotone-certification}
in the appendix. Since $\{Y_t>c'\}\subseteq\{Y_t>c\}$ for $c<c'$,
we have $p_t(c')\leq p_t(c)$. Hence, if
$p_t(c')\geq\alpha$ throughout monitoring, then we have
$p_t(c)\geq\alpha$ over the same period.

The next theorem uses these two properties to bound
the probability of any false certification for all candidates by $\delta$. Hence, define the event that no candidate is falsely certified:
\begin{equation}
\begin{aligned}
    \mathcal G:=\Bigl\{& \forall c\in\Gamma,\ \tau(c)<\infty \implies\\
        &\exists\,s\in\{0,\ldots,\tau(c)-1\}:
        p_s(c)<\alpha
    \Bigr\}.
\end{aligned}
\label{eq:no-false-certification-event}
\end{equation}
On $\mathcal G$, every certified candidate $c$ (indicated by $\tau(c)<\infty$) had an
exceedance risk $p_s(c)$ below $\alpha$ for some time $s\in\{0,\ldots,\tau(c)-1\}$ in the past.\footnote{The event of incorrectly certified robustness margins is $\mathcal G^c:=\Bigl\{
        \exists\,c\in\Gamma,\ \tau(c)<\infty
        \ \text{and }
        \forall s\in\{0,\ldots,\tau(c)-1\}:
        p_s(c)\ge\alpha$, indicating $c$ that are certified even though its exceedance risk $p_s(c)$ was never below $\alpha$.}

\begin{theorem}[Simultaneous Margin Certification]
\label{thm:false-certification}
Consider the system in
\eqref{eq:true-stochastic-system} under the safety filter in
\eqref{eq:cbf-safety-filter}.\footnote{\Cref{thm:false-certification} holds for any sequence of robustness margin $\{c_s\}_{t\geq s}$. Indeed, \Cref{thm:false-certification} holds more generally for any control sequence $\{u_s\}_{t\geq s}$.} Let $\alpha\in (0,1)$ and $\delta\in(0,1)$. Given the grid of candidate margins $\Gamma$, the set of gains
$\Lambda$, and the certification rule
\eqref{eq:certification-time}, then we have
\begin{equation*}
    \Prob(\mathcal G)\geq1-\delta.
\end{equation*}
\end{theorem}
    \begin{proof}
  The proof is given in Appendix~\ref{app:cd-proof}.
\end{proof}
Theorem~\ref{thm:false-certification} ensures that, with
probability at least $1-\delta$, every certified candidate
$c$ satisfies $p_s(c)<\alpha$ for some $s<\tau(c)$.
All candidates use the same threshold $1/\delta$, without
increasing the error confidence or requiring
independent or stationary observations.
On $\mathcal G$, this guarantee concerns past observations:
it neither identifies the exact certification time $s$ nor ensures
$p_{\tau(c)}(c)<\alpha$. For this reason, Section~\ref{sec:main-results}
will combine certificates $c\in \mathcal S_t$ with an exceedance risk buffer to compute safe control inputs in the future.

\begin{remark}
The one-step e-variables in \eqref{eq:wealth-recursion}
generate a test supermartingale. For finitely constrained
conditional hypotheses such as ours, Clerico~\cite{clerico2026}
shows that every e-process (a general anytime-valid evidence
process that does not necessarily satisfy \eqref{eq:test-supermartingale}) is pointwise bounded from above by a test supermartingale.
Thus, using test supermartingales does not limit the evidence
attainable in principle.
\end{remark}
\subsection{Evidence Growth and Certification Time}
\label{cert}
To analyze the time needed for certification, we assume that the null hypothesis \eqref{eq:persistent-insufficiency-null} is violated at all times, i.e., that $p_t(c)<\alpha$ for all times $t\ge 0$. Indeed, we impose the slightly
stronger condition that $p_t(c)$
remains uniformly below $\alpha$.
Fix $c\in\Gamma$ and assume that
there exists $\bar p(c)\in[0,\alpha)$ such that:
\begin{equation}
    p_t(c)\leq\bar p(c),
    \qquad \forall t\in\mathbb N_0,
    \quad\text{almost surely}.
    \label{eq:risk-bound-for-growth}
\end{equation}
 This assumption is used only for the following result, i.e., all other results in the paper remain valid without it.
\begin{proposition}[Evidence growth and certification time]
\label{thm:evidence-growth}
Assume that \eqref{eq:risk-bound-for-growth} holds. Define
$g_q(\lambda):=(1-q)\log(1+\lambda\alpha)
+q\log\!\left(1-\lambda(1-\alpha)\right)$.
Then, for any $\lambda\in\Lambda$ with
$g_{\bar p(c)}(\lambda)>0$:
\begin{equation}
    \liminf_{t\to\infty}\frac{\log E_t^\lambda(c)}{t}
    \geq g_{\bar p(c)}(\lambda)
    \quad\text{a.s.}
    \label{eq:e-power-limit}
\end{equation}and the expected certification time satisfies
\begin{equation}
    \E[\tau(c)]
    \leq
    \frac{\log(|\Lambda|/\delta)+\log(1+\lambda\alpha)}
         {g_{\bar p(c)}(\lambda)}.
    \label{eq:expected-certification-time}
\end{equation}
\end{proposition}
\begin{proof}
See Appendix~\ref{app:evidence-growth-proof}.
\end{proof}

Equation~\eqref{eq:e-power-limit} shows that the
evidence grows without bound almost surely when
$g_{\bar p(c)}(\lambda)>0$.
Since $E_t(c)$ averages the nonnegative evidence
across gains, one such gain is enough to ensure
certification in finite time almost surely.
Equation~\eqref{eq:expected-certification-time} bounds the expected certification time.

\begin{figure}[t]
    \centering
    \includegraphics[width=\columnwidth]{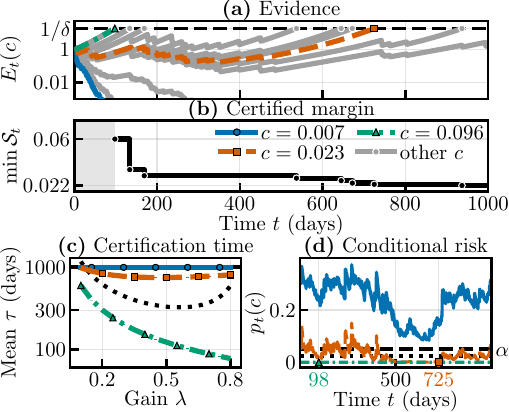}
    \caption{Sequential certification with time-varying  stock losses.\hspace{0.33em}One test per candidate, $\alpha=\delta=0.05$; $\alpha_0$ is not used.
(a) Mixture evidence, shown above $0.001$; markers: certification.
(b) Smallest certified margin.
(c) Mean capped certification times $\pm2$ standard errors over 5000
trials; dotted: bound for $c=0.096$.
(d) Conditional risks; dotted: $\bar p(0.096)$; vertical dashed:
certification times. The legend in (b) applies to all panels.}
    \label{fig:certification-illustration}
\end{figure}

\begin{example}
   We illustrate these results with  daily losses of synthetic stock data. Here, $Y_t$ is the relative drop of the stock price during day~$t$, or zero if the price rises. The daily log return, i.e., the logarithm of the ratio of consecutive prices, follows a time-varying generalized autoregressive conditional heteroskedasticity (TV-GARCH) model~\cite{chen2014tvgarch}. Hence, the losses are dependent and their conditional distribution varies over time. More details about the model and the candidate grid are provided in Appendix~\ref{app:certification-details}.

We set $\alpha=\delta=0.05$ and $\Lambda=\{0.2,0.5,0.8\}$. For
$c=0.096$, the assumption of
Proposition~\ref{thm:evidence-growth} is satisfied with $p_t(c)\leq\bar p(c)=0.0233<\alpha$. Panels~(a), (b), and (d) in Figure \ref{fig:certification-illustration}
show one realization over 1000 trading days. In it, $c=0.096$
is certified at day 98 without risk exceedance, while $c=0.023$ is certified at
day 725 after 21 risk exceedances. The exceedance risk of $c=0.007$ stays above
$\alpha$, and this margin is not certified. The smallest certified margin
 $\min\mathcal S_t$  takes eight values, from $0.06$
at day 98 to $0.022$ at day 937. Panel~(c) compares fixed-gain tests for  three  margins
over 5000 independent trials, showing averaged certification times. The black dotted curve is the bound
\eqref{eq:expected-certification-time} for $c=0.096$. Finally, note that the exceedance risk of
$c=0.023$ is below $\alpha$ when $c$ is certified, but later
exceeds $\alpha$ again. Hence, a certificate does not ensure that exceedance risk
remains low. The next section accounts for such exceedance risk increases by restarting tests.
\end{example}

\section{Robustness Margins for Safe Control}
\label{sec:main-results}

Section~\ref{sec:certification} provides statistical evidence that
a candidate's exceedance risk $p_s(c)$ was below
$\alpha$ for some time $s<t$ in the past. However, the exact time $s$ is unknown. To use a certified margin for control, we have to account for a possible increase in exceedance risk from time $s$ to the current time~$t$.

In Section  \ref{sec:1}, we develop a method to transfer a certified margin to the current time $t$ via a deployment rule using one certificate per candidate. To reduce conservatism in this transfer, we also present repeated certification via parallel tests in Section \ref{subsec:distribution-shift}. Throughout, we make the following assumption to  bound the increase in exceedance risk.

\begin{assumption}[Bounding  exceedance risk]
\label{as:bounded-shift}

Let $c\in\Gamma$ be a margin, $u_t:=u_t(c)$ be the control input at time $t$ where $u_t(c)$ follows
\eqref{eq:cbf-safety-filter}\footnote{Past control inputs $\{u_s\}_{t> s}$ need not be generated under the same $c$.}, and  $\sigma,\tau\ge 0$ be times with $\sigma<\tau\leq t$. Then, there exist known constants $L_x,L_u\geq0$ and a
 $\Fc_t$-measurable bound $\Delta_{t}\ge 0$ such that, for any $s\in\{\sigma,\hdots,\tau-1\}$,
 \begin{equation}
    p_t(c)\leq p_s(c)+d_t(c;\sigma,\tau)
    \label{eq:predictive-shift}
\end{equation}
holds almost surely with $d_t(c;\sigma,\tau)$ being defined as
\begin{equation}
\begin{aligned}
    d_t(c;\sigma,\tau):=
    \max_{\sigma\leq s<\tau}
    \bigl\{&L_x\|x_t-x_s\|\\
           &+L_u\|u_t-u_s\|+\Delta_{t}\bigr\}.
\end{aligned}
\label{eq:candidate-shift-bound}
\end{equation}
\end{assumption}

We note that $\sigma$ and $\tau$ denote the starting and certification times of a deployed test. The exceedance risk $p_t(c)$ is bounded via \eqref{eq:predictive-shift}, where the state and input terms in \eqref{eq:candidate-shift-bound} account for changes in operating conditions, while $\Delta_{t}$ accounts for additional risk increase. Appendix~\ref{app:conditional-risk-bound} provides
sufficient conditions for obtaining $L_x$ and $L_u$ from
derivative bounds on $f$, $\hat f$, and $h$ and a
bound on the density function of
$h(\hat f(x_t,u_t))-h(x_{t+1})$. We can obtain $\Delta_{t}$ via a bound on changes in the distribution of the disturbance $w_t$. This bound has to be known or estimated from data.

\subsection{Certificate Transfer and Deployment Guarantee}
\label{sec:1}

 In this section, we set $\sigma:=0$ and $\tau=\tau(c)$, resembling the setting in Section~\ref{sec:certification}. For each certified
candidate, define $d_t(c):=d_t(c;0,\tau(c))$.
If $p_s(c)<\alpha$ at some time $s<\tau(c)$,
deploying $c$ with $d_t(c)\leq\alpha_0-\alpha$ ensures $ p_t(c)<\alpha+d_t(c)\leq\alpha_0.$ We therefore define the deployable set of margins as
\begin{equation}
    \mathcal D_t:=
    \bigl\{c\in\mathcal S_t:
        d_t(c)\leq\alpha_0-\alpha\bigr\}.
    \label{eq:deployable-set}
\end{equation}
Based on this reasoning, the controller $u_t$ selects the smallest deployable margin,
using $c_{\max}$ if none qualifies:
\begin{equation}
    u_t := u_t(c_t), \qquad
    c_t :=
    \begin{cases}
        \min\mathcal D_t, & \mathcal D_t\neq\emptyset,\\
        c_{\max}, & \mathcal D_t=\emptyset.
    \end{cases}
    \label{eq:deployment-rule}
\end{equation}
Certificates are retained, but their eligibility
for deployment is checked at every decision time $t$
using the current bound $d_t(c)$. Finally, define the event that the deployed margin satisfies
the conditional risk limit at every decision time:
\begin{equation}
    \mathcal G^d:=
    \bigl\{p_t(c_t)\leq\alpha_0
    \text{ for all }t\in\mathbb N_0\bigr\}.
    \label{eq:first-deployment-event}
\end{equation}

\begin{theorem}[Deployment Guarantee]
\label{thm:certified-deployment}
Consider the system in
\eqref{eq:true-stochastic-system} under the safety filter in
\eqref{eq:deployment-rule}. Let $\alpha\in (0,1)$ and $\delta\in(0,1)$. If Assumptions~\ref{as:conservative-margin} and \ref{as:bounded-shift} hold, then we have that
\begin{equation}
    \Prob(\mathcal G^d)\geq1-\delta.
    \label{eq:certified-cbf-guarantee}
\end{equation}
\end{theorem}
\begin{proof}
See Appendix~\ref{app:main-proof}.
\end{proof}

On $\mathcal G^d$,    \eqref{eq:true-cbf-condition} gives $  \Prob\!\left(
        h(x_{t+1})<(1-\gamma)h(x_t)
        \,\middle|\,\Fc_t
    \right)
   \leq p_t(c_t)\leq\alpha_0$
simultaneously for all $t\in\mathbb N_0$.
Thus, with probability at least $1-\delta$,
the conditional exceedance risk of a next-step CBF violation
remains at most $\alpha_0$ along the entire
closed-loop trajectory.  Later, in Corollary \ref{cor:finite-horizon-safety}, we convert this guarantee into a joint safety bound for $\Prob\!\left(\exists t: x_t\notin\Cc\right)$.

A single certificate can be conservative as operating conditions change. We therefore introduce parallel tests next.
\subsection{Repeated Certification and Parallel Tests}
\label{subsec:distribution-shift}
A certified margin $c$ no longer supports deployment when
$d_t(c)>\alpha_0-\alpha$. Starting new tests in parallel allows the
controller to obtain certificates from more recent
observations. In particular, for each candidate
$c\in\Gamma$, let $r\in\mathbb N_{>0}$ index tests in
the order they start, and let $\sigma_r$ and $\tau_r(c)$
denote the starting and certification times of test $r$,
respectively.

We start new tests for all candidates every $K$ time steps,
which gives $\sigma_r=(r-1)K$ for $K\in\mathbb N_{>0}$.
Unfinished tests continue in parallel.
At time $t$, each unfinished test started before $t$
uses the same newly observed loss $Y_{t-1}$, but
accumulates evidence from its own starting time
$\sigma_r$.

To obtain a guarantee that holds jointly over all tests,
we assign each test index $r$ the error level $\delta_r:=\frac{\delta}{\kappa(r+1)\ln^2(r+1)}$ with $\kappa
    :=\sum_{n=2}^{\infty}\frac{1}{n\ln^2 n}
    \approx2.11$. The constant $\kappa$ normalizes the allocation so that
$\sum_{r\geq1}\delta_r=\delta$.
The same level $\delta_r$ is used for every candidate $c \in \Gamma$.
The common starting times $\sigma_r$ and gains $\Lambda$ allow the ordering argument of Theorem~\ref{thm:false-certification}
to give a joint error bound $\delta_r$ over $\Gamma$
for each test index $r$.

We proceed similarly as before. For each $\sigma_r$,
we initialize $E_{\sigma_r}^{\lambda,r}(c)=1$ for every
$\lambda\in\Lambda$ and define
\begin{equation*}
    E_t^{\lambda,r}(c)
    =E_{t-1}^{\lambda,r}(c)
      \bigl[1+\lambda(\alpha-Z_{t-1}(c))\bigr]
\end{equation*}
for $t>\sigma_r$.
Let $E_t^r(c)$ be the uniform mixture of these
processes, as in \eqref{eq:mixture-eprocess}, and define
\begin{equation}
    \tau_r(c):=
    \inf\left\{t>\sigma_r:
        E_t^r(c)\geq\frac{1}{\delta_r}\right\}
    \label{eq:restarted-certification-time}
\end{equation}
with $\inf\emptyset=\infty$.
If $\tau_r(c)<\infty$, the test stops and
certificate $(c,r)$ is stored. Overlapping tests may certify in a different order from the one in which they started.

For each certificate $(c,r)$ with $\tau_r(c)\leq t$,
define its current risk-increase bound as
\begin{equation}
    d_t^{(r)}(c):=
    d_t\bigl(c;\sigma_r,\tau_r(c)\bigr).
    \label{eq:repeated-shift-bound}
\end{equation}
The set $\mathcal S_t$ now contains every candidate
with at least one certificate issued by time $t$,
i.e., $\mathcal S_t
    :=
    \bigl\{c\in\Gamma:
        \exists r\in\mathbb N_{>0},\
        \tau_r(c)\leq t\bigr\}.$ Next, to obtain the set $\mathcal D_t$, we replace
$d_t(c)$ by
$\min\limits_{r:\,\tau_r(c)\leq t}d_t^{(r)}(c)$
in \eqref{eq:deployable-set}, that is:
\begin{equation}
    \mathcal D_t:=\bigl\{c\in\mathcal S_t:
        \min\limits_{r:\,\tau_r(c)\leq t}
        d_t^{(r)}(c)\leq\alpha_0-\alpha\bigr\}.
    \label{eq:D_updated}
\end{equation}
Thus, a margin qualifies if at least one
of its certificates satisfies the deployment condition. The controller selects $c_t$ and $u_t$ according
to \eqref{eq:deployment-rule}, with $\mathcal D_t$
defined by \eqref{eq:D_updated}.
We next extend the guarantee for
$\mathcal G^d$ in \eqref{eq:first-deployment-event}
to repeated testing.

\begin{theorem}[Deployment Guarantee for Repeated Tests]
\label{prop:repeated-deployment}
Consider the system in
\eqref{eq:true-stochastic-system} under the safety
filter in \eqref{eq:deployment-rule}, with
$\mathcal D_t$ as in \eqref{eq:D_updated} and
certification rule \eqref{eq:restarted-certification-time}.
Let $\alpha\in(0,1)$ and $\delta\in(0,1)$.
If Assumptions~\ref{as:conservative-margin} and
\ref{as:bounded-shift} hold, then
\begin{equation}
    \Prob(\mathcal G^d)\geq1-\delta.
    \label{eq:repeated-deployment-guarantee}
\end{equation}
\end{theorem}
\begin{proof}
See Appendix~\ref{app:repeated-deployment-proof}.
\end{proof}
Allocating the budget $\delta$ across test starts via $\delta_r$ preserves the joint deployment guarantee, i.e.,
with probability at least $1-\delta$,
$p_t(c_t)\leq\alpha_0$ holds simultaneously for all $t\in\mathbb N_0$.

Theorem \ref{prop:repeated-deployment} also yields a bound on the probability of the system being unsafe over bounded horizons.

\begin{corollary}[Finite-horizon safety]
\label{cor:finite-horizon-safety}
Under the conditions of
Theorem~\ref{prop:repeated-deployment},
let $T\in\mathbb N_{>0}$ and $x_0\in\Cc$. Further, define
\[
   n_0:=
    \left\lceil
        \frac{\log(1/\delta)}
             {\log(1+\max \limits _{\lambda\in\Lambda}\lambda\alpha)}
    \right\rceil
\]
Then, it holds that
\begin{equation}
        \begin{split}
            &\Prob\!\left(\exists t\in\{0,\ldots,T\}:
        x_t\notin\Cc\right)\\
    &\quad\leq
    \mathbf1_{\{T>n_0\}}
    \min\!\left\{1,\delta+1-(1-\alpha_0)^{T-n_0}\right\}.
        \end{split}
\label{eq:finite-horizon-safety}
\end{equation}
\end{corollary}
\begin{proof}
See Appendix~\ref{corpoof}.
\end{proof}
Equation~\eqref{eq:finite-horizon-safety} bounds the probability that the state $x_t$ leaves the set  $\Cc$ within $T$ steps, accounting for the
initial $n_0$ time steps in which the controller
necessarily uses the safe margin $c_{\max}$.
It therefore concerns safety over an entire finite horizon, whereas the coverage guarantees of online conformal prediction  \cite{gibbs2021adaptive,dixit2023adaptive}
are averaged over time.

\section{Numerical Examples}
\label{sec:empirical-evaluation}
We evaluate the safety controller
\eqref{eq:deployment-rule} with repeated certification
in two robot-navigation scenarios: a corridor
with cones (Corridor) and a maze with walls and
cones (Maze). We assess the theoretical guarantees
empirically and compare safety and navigation
performance with existing methods. For clarity of presentation we present
the Maze results in this section, and the
Corridor results in Appendix~\ref{app:sim-corridor}.
\subsection{Setup}
\label{sec:empirical-setup}
We model the robot as a second-order unicycle
navigating a maze with walls and cones. The goal of the robot is to reach a goal, using a nominal controller $u_t^{\mathrm{nom}}$, while avoiding the obstacles.
The barrier function accounts for obstacle clearance
at a fixed point ahead of the robot and its
braking distance. The robot model and environment are detailed in
Appendix~\ref{app:sim-scenarios}.
The true system differs from the model through an additive disturbance $w_t$, which models a bounded position
displacement with a time-varying magnitude distribution.
Its direction is uniformly distributed over
$[0,2\pi)$ conditional on its magnitude and $\Fc_t$.
To compute the risk-increase bounds
\eqref{eq:repeated-shift-bound}, we
use this directional model, a known magnitude
bound, and a bound on changes in the conditional
disturbance distribution.
The magnitude distribution itself is unknown to the controller.
We compute the risk-increase bounds
\eqref{eq:repeated-shift-bound} using the analysis
in Appendix~\ref{app:unicycle-risk-bound}.
The parameter values and simulation details
are given in Appendix~\ref{app:sim-scenarios}.
The trajectories are simulated in Python, and the trajectories are visualized in NVIDIA Isaac Sim. The simulation code, data, and supplementary ablation report are available at
\href{https://github.com/omidmirza/ecbf-ablation}{\nolinkurl{github.com/omidmirza/ecbf-ablation}}.

A setting $(\alpha_0,\alpha,\delta,K)$ consists of the CBF-violation
probability $\alpha_0$, the margin insufficiency $\alpha$, the error
confidence $\delta$, and the test period $K$ (measured in control steps $t$). The tests follow
Section~\ref{subsec:distribution-shift}, with the deployable set $\mathcal D_t$ in
\eqref{eq:D_updated}.

We compare five settings, S1-S5,
with four variants of our method at
each setting, as described below.
The settings and results are reported in
Table~\ref{tab:empirical-topfive}.
Additional parameter sweeps are provided in
the supplementary ablation report~\cite{trakas2026ablation}.

\begin{table}[t]
\centering
\caption{Maze results, $\alpha_0=0.30$. Means $\pm$ SD. Event frequencies in \%
($95\%$ upper confidence bounds). $\widehat T_{\rm mis}$ in s and $\widehat q^+_{\rm CBF}$ in \%.
Bold: best variant.}
\label{tab:empirical-topfive}
{\footnotesize\setlength{\tabcolsep}{1.5pt}\renewcommand{\arraystretch}{1.0}%
\begin{tabular}{@{}lr@{}lr@{}lr@{}lr@{}lr@{}lr@{}l@{}}
\guaranteeboxrow{T1}\toprule
Method & \multicolumn{2}{c}{$\widehat m_c/c_{\max}{\downarrow}$} & \multicolumn{2}{c}{$\widehat T_{\rm mis}{\downarrow}$} & \multicolumn{2}{c}{$\widehat q^{+}_{\rm CBF}{\downarrow}$} & \multicolumn{2}{c}{$\widehat m_{p^{\max}}{\downarrow}$} & \multicolumn{2}{c}{$\widehat q_{\rm dep}{\downarrow}$} & \multicolumn{2}{c}{$\widehat q_{\rm cert}{\downarrow}$}\\\midrule
Unfiltered & \multicolumn{2}{c}{---} & $42.2$&{\scriptsize${\pm}.1$} & $55.7$&{\scriptsize${\pm}1.6$} & \multicolumn{2}{c}{---} & \multicolumn{2}{c}{---} & \multicolumn{2}{c}{---}\\
$c=0$ & $0.00$&{\scriptsize${\pm}.00$} & $48.7$&{\scriptsize${\pm}.3$} & $24.9$&{\scriptsize${\pm}2.6$} & $0.67$&{\scriptsize${\pm}.13$} & 100&\,(100) & \multicolumn{2}{c}{---}\\
$c_{\max}$ & $1.00$&{\scriptsize${\pm}.00$} & $54.4$&{\scriptsize${\pm}.3$} & $0.0$&{\scriptsize${\pm}0.0$} & \multicolumn{2}{c}{---} & 0.0&\,(1.5) & \multicolumn{2}{c}{---}\\
ACI & $0.22$&{\scriptsize${\pm}.04$} & $49.7$&{\scriptsize${\pm}.3$} & $15.8$&{\scriptsize${\pm}1.9$} & $0.47$&{\scriptsize${\pm}.09$} & 99.5&\,(100) & \multicolumn{2}{c}{---}\\
QT & $0.50$&{\scriptsize${\pm}.03$} & $51.2$&{\scriptsize${\pm}.3$} & $8.0$&{\scriptsize${\pm}1.3$} & $0.34$&{\scriptsize${\pm}.08$} & 65.5&\,(71.1) & \multicolumn{2}{c}{---}\\
\midrule
\multicolumn{13}{@{}l}{\textbf{S1}: $(\alpha_0,\alpha,\delta,K)=$ (0.30, 0.25, 0.20, 50)}\\\cmidrule{1-13}
V-U & $0.97$&{\scriptsize${\pm}.01$} & $54.4$&{\scriptsize${\pm}.3$} & $\mathbf{0.1}$&{\scriptsize${\pm}0.3$} & $\mathbf{0.11}$&{\scriptsize${\pm}.05$} & 0.0&\,(1.5) & 0.0&\,(1.5)\\
V-$\Delta$ & $0.97$&{\scriptsize${\pm}.01$} & $54.3$&{\scriptsize${\pm}.3$} & $0.2$&{\scriptsize${\pm}0.3$} & $\mathbf{0.11}$&{\scriptsize${\pm}.05$} & 0.0&\,(1.5) & 0.0&\,(1.5)\\
V-L & $0.62$&{\scriptsize${\pm}.05$} & $51.9$&{\scriptsize${\pm}.4$} & $4.8$&{\scriptsize${\pm}1.4$} & $0.17$&{\scriptsize${\pm}.04$} & 0.0&\,(1.5) & 0.0&\,(1.5)\\
V-C & $\mathbf{0.52}$&{\scriptsize${\pm}.07$} & $\mathbf{51.3}$&{\scriptsize${\pm}.5$} & $7.4$&{\scriptsize${\pm}2.3$} & $0.20$&{\scriptsize${\pm}.04$} & 0.0&\,(1.5) & 0.5&\,(2.4)\\
\cmidrule{1-13}
\multicolumn{13}{@{}l}{\textbf{S2}: $(\alpha_0,\alpha,\delta,K)=$ (0.30, 0.25, 0.15, 50)}\\\cmidrule{1-13}
V-U & $0.98$&{\scriptsize${\pm}.01$} & $54.4$&{\scriptsize${\pm}.3$} & $\mathbf{0.1}$&{\scriptsize${\pm}0.2$} & $\mathbf{0.10}$&{\scriptsize${\pm}.05$} & 0.0&\,(1.5) & 0.0&\,(1.5)\\
V-$\Delta$ & $0.97$&{\scriptsize${\pm}.01$} & $54.4$&{\scriptsize${\pm}.3$} & $\mathbf{0.1}$&{\scriptsize${\pm}0.2$} & $\mathbf{0.10}$&{\scriptsize${\pm}.05$} & 0.0&\,(1.5) & 0.0&\,(1.5)\\
V-L & $0.64$&{\scriptsize${\pm}.05$} & $51.9$&{\scriptsize${\pm}.4$} & $4.6$&{\scriptsize${\pm}1.4$} & $0.17$&{\scriptsize${\pm}.04$} & 0.0&\,(1.5) & 0.0&\,(1.5)\\
V-C & $\mathbf{0.53}$&{\scriptsize${\pm}.08$} & $\mathbf{51.3}$&{\scriptsize${\pm}.5$} & $7.3$&{\scriptsize${\pm}2.4$} & $0.19$&{\scriptsize${\pm}.04$} & 0.0&\,(1.5) & 0.0&\,(1.5)\\
\cmidrule{1-13}
\multicolumn{13}{@{}l}{\textbf{S3}: $(\alpha_0,\alpha,\delta,K)=$ (0.30, 0.20, 0.20, 50)}\\\cmidrule{1-13}
V-U & $0.82$&{\scriptsize${\pm}.04$} & $53.2$&{\scriptsize${\pm}.4$} & $\mathbf{1.5}$&{\scriptsize${\pm}0.8$} & $\mathbf{0.12}$&{\scriptsize${\pm}.04$} & 0.0&\,(1.5) & 2.5&\,(5.2)\\
V-$\Delta$ & $0.82$&{\scriptsize${\pm}.04$} & $53.2$&{\scriptsize${\pm}.4$} & $\mathbf{1.5}$&{\scriptsize${\pm}0.8$} & $\mathbf{0.12}$&{\scriptsize${\pm}.04$} & 0.0&\,(1.5) & 2.5&\,(5.2)\\
V-L & $0.62$&{\scriptsize${\pm}.07$} & $51.9$&{\scriptsize${\pm}.5$} & $5.0$&{\scriptsize${\pm}1.9$} & $0.15$&{\scriptsize${\pm}.04$} & 0.0&\,(1.5) & 2.0&\,(4.6)\\
V-C & $\mathbf{0.61}$&{\scriptsize${\pm}.08$} & $\mathbf{51.8}$&{\scriptsize${\pm}.6$} & $5.4$&{\scriptsize${\pm}2.2$} & $0.15$&{\scriptsize${\pm}.04$} & 0.0&\,(1.5) & 2.0&\,(4.6)\\
\cmidrule{1-13}
\multicolumn{13}{@{}l}{\textbf{S4}: $(\alpha_0,\alpha,\delta,K)=$ (0.30, 0.25, 0.025, 100)}\\\cmidrule{1-13}
V-U & $1.00$&{\scriptsize${\pm}.00$} & $54.4$&{\scriptsize${\pm}.3$} & $\mathbf{0.0}$&{\scriptsize${\pm}0.0$} & $\mathbf{0.04}$&{\scriptsize${\pm}.03$} & 0.0&\,(1.5) & 0.0&\,(1.5)\\
V-$\Delta$ & $1.00$&{\scriptsize${\pm}.00$} & $54.4$&{\scriptsize${\pm}.3$} & $\mathbf{0.0}$&{\scriptsize${\pm}0.0$} & $\mathbf{0.04}$&{\scriptsize${\pm}.03$} & 0.0&\,(1.5) & 0.0&\,(1.5)\\
V-L & $0.81$&{\scriptsize${\pm}.04$} & $53.0$&{\scriptsize${\pm}.4$} & $2.1$&{\scriptsize${\pm}1.0$} & $0.13$&{\scriptsize${\pm}.03$} & 0.0&\,(1.5) & 0.0&\,(1.5)\\
V-C & $\mathbf{0.60}$&{\scriptsize${\pm}.06$} & $\mathbf{51.7}$&{\scriptsize${\pm}.5$} & $5.8$&{\scriptsize${\pm}1.9$} & $0.17$&{\scriptsize${\pm}.03$} & 0.0&\,(1.5) & 0.0&\,(1.5)\\
\cmidrule{1-13}
\multicolumn{13}{@{}l}{\textbf{S5}: $(\alpha_0,\alpha,\delta,K)=$ (0.30, 0.25, 0.10, 10)}\\\cmidrule{1-13}
V-U & $0.99$&{\scriptsize${\pm}.01$} & $54.4$&{\scriptsize${\pm}.3$} & $\mathbf{0.0}$&{\scriptsize${\pm}0.1$} & $\mathbf{0.07}$&{\scriptsize${\pm}.05$} & 0.0&\,(1.5) & 0.0&\,(1.5)\\
V-$\Delta$ & $0.99$&{\scriptsize${\pm}.01$} & $54.4$&{\scriptsize${\pm}.3$} & $\mathbf{0.0}$&{\scriptsize${\pm}0.1$} & $\mathbf{0.07}$&{\scriptsize${\pm}.05$} & 0.0&\,(1.5) & 0.0&\,(1.5)\\
V-L & $0.66$&{\scriptsize${\pm}.05$} & $52.1$&{\scriptsize${\pm}.4$} & $4.2$&{\scriptsize${\pm}1.2$} & $0.15$&{\scriptsize${\pm}.04$} & 0.0&\,(1.5) & 0.0&\,(1.5)\\
V-C & $\mathbf{0.55}$&{\scriptsize${\pm}.07$} & $\mathbf{51.4}$&{\scriptsize${\pm}.5$} & $6.8$&{\scriptsize${\pm}2.2$} & $0.18$&{\scriptsize${\pm}.03$} & 0.0&\,(1.5) & 0.0&\,(1.5)\\
\guaranteeboxend{T1}\bottomrule\end{tabular}}

\vspace{-6pt}
\end{table}

\subsection{Methods and Metrics}
\label{sec:empirical-metrics}
\paragraph{Method variations} We implement four variants that differ only in
the risk-increase bound $d_t^{(r)}(c)$.
We derive these bounds using the nominal dynamics,
obstacle geometry, and a bound on changes
in the disturbance distribution. Recall that smaller bounds $d_t^{(r)}(c)$ make the deployment condition
$d_t^{(r)}(c)\leq\alpha_0-\alpha$ easier to satisfy.
We next summarize these four variants,  but refer  to Appendix~\ref{app:unicycle-risk-bound} for a detailed description. Variant V-U uses uniform bounds for $L_x$, $L_u$, and $\Delta_t$ across all candidate margins $c\in\Gamma$. In the remaining three variants $L_x$, $L_u$, or $\Delta_t$ depend on the candidate $c$.
V-$\Delta$ tightens these bounds for each candidate
 depending on the shape of nearby obstacles, here walls and cones. V-L instead tightens the bound on risk changes
caused by changes in the distance between
the predicted nose point of the robot and nearby obstacles.
V-C combines both refinements and also bounds
the effect of disturbance-distribution changes
more tightly.
All variants account for changes in the disturbance
distribution: V-U, V-$\Delta$, and V-L add the
full drift bound $\Delta_t$, whereas V-C uses
the geometry and candidate margin to reduce
its contribution to the risk-increase bound.

\paragraph{Baselines} The baselines are unfiltered nominal control
$u_t^{\mathrm{nom}}$, CBF filters with fixed margins $0$
and $c_{\max}$; adaptive conformal inference
(ACI)~\cite{gibbs2021adaptive}, as adapted
in~\cite{huriot2026adaptive}; and quantile
tracking (QT)~\cite{angelopoulos2023conformal},
as adapted in~\cite{cherenson2026staggered}. ACI and QT target an asymptotic average loss-exceedance
frequency of $\alpha_0$, i.e., $\frac1T\sum_{t=0}^{T-1}Z_t(c_t)
    \approx\alpha_0$ for large $T$ with the exceedance indicator $Z_t$ defined in~\eqref{Zt1}. This target allows the exceedance risk $p_t(c_t)$ to
be larger than $\alpha_0$ at some time instants. In contrast, Theorem~\ref{prop:repeated-deployment}
gives $ \Prob\!\left(
        p_t(c_t)\leq\alpha_0,
        \forall t\in\mathbb N_0
    \right)\geq1-\delta.$
Thus, with probability at least $1-\delta$,
the exceedance risk limit holds along the
entire trajectory.
Since $\alpha_0>0$, this does not rule out
individual loss exceedances or CBF violations. Implementation details are provided in Appendix~\ref{app:sim-methods}.

\paragraph{Metrics} For each method, scenario, and setting, we simulate
$N_{\mathrm{MC}}=200$  trajectories for validation.

We report the mean mission time
$\widehat T_{\mathrm{mis}}$, assigning the full
simulation horizon to unsuccessful trajectories. We also report the normalized mean robustness margin
$\widehat m_c/c_{\max}$ and mean CBF-violation
fraction $\widehat q_{\mathrm{CBF}}^+$. For this, we compute the average margin
and the average fraction of time steps with a CBF violation along each trajectory, and
then average these values across all trajectories resulting in $\widehat m_c$ and $\widehat q_{\mathrm{CBF}}^+$.

We also report $\widehat m_{p^{\max}}$ as the average   of the largest
exceedance risk $ p_t(c_t)$ across all trajectories that use a margin below
$c_{\max}$. We approximate $ p_t(c_t)$ through an estimate $\widehat p_t(c_t)$, see Appendix~\ref{app:empirical-risk-classification} for details. The metric $\widehat m_{p^{\max}}$ focuses on certified margins;
the fallback has $p_t(c_{\max})=0$.

Lastly, we report $\widehat q_{\mathrm{dep}}$ as the
fraction
of trajectories in which the deployed margin
is such that $p_t(c_t)>\alpha_0$ at least once.\footnote{When the estimated risk $\widehat p_t(c)$ is
close to $\alpha_0$, finite sample errors may
prevent us from determining whether the true
risk $p_t(c)$ exceeds $\alpha_0$.
Table~\ref{tab:empirical-topfive} conservatively
counts these uncertain cases as failures.
Parentheses give $95\%$ upper confidence
bounds based on the possible-event counts.} This approximates the complement of  $\mathcal G^d$, which is upper bounded by $\delta$ following Theorem~\ref{prop:repeated-deployment}.
We also report
$\widehat q_{\mathrm{cert}}$ as the
fraction
of trajectories containing a false certification, i.e.,
a candidate that is certified although its exceedance risk remained
above $\alpha$ at all times. Theorem~\ref{prop:repeated-deployment} bounds this false-certification probability by $\delta$. Details and additional metrics are provided
in Appendix~\ref{app:empirical-validation}.

\begin{figure}[t]
\centering
\includegraphics{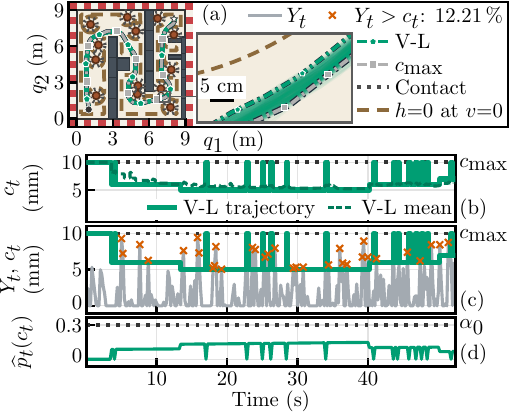}
\caption{Maze, S3 with V-L.
(a) Paths and clearance boundaries, with an
enlarged view near an obstacle.
(b) Selected margin for one trajectory and
the mean across trajectories.
(c) Loss and margin; crosses mark $Y_t>c_t$.
(d) Estimated exceedance risk.
Panels~(b)-(d) show the same example trajectory.}
\label{fig:example-maze-winner}
\end{figure}

\subsection{Results}
\label{sec:empirical-results}
\paragraph{Statistical guarantees}
Across all five settings in
Table~\ref{tab:empirical-topfive}, every certified
variant has $\widehat q_{\mathrm{dep}}=0\%$
over $200$ trajectories, with a $95\%$ upper confidence bound of $1.5\%$.
False-certification frequencies satisfy
$\widehat q_{\mathrm{cert}}\leq2.5\%$.
For both events, the upper confidence bounds are below the corresponding budget $\delta$,
consistent with
Theorem~\ref{prop:repeated-deployment}. In comparison, ACI and QT have
$\widehat q_{\mathrm{dep}}=99.5\%$ and $65.5\%$,
respectively.
Their objective concerns the exceedance frequency
averaged over time.
Our guarantee instead ensures, with probability
at least $1-\delta$, that
$p_t(c_t)\leq\alpha_0$ simultaneously at all
decision times.

\paragraph{Margin and mission time}
V-C achieves the smallest mean margin
$\widehat m_c/c_{\max}$ among the certified
variants at every setting in
Table~\ref{tab:empirical-topfive}.
At S1, it reduces the mean margin by
$48\%$ relative to $c_{\max}$ and shortens
$\widehat T_{\mathrm{mis}}$ from $54.4$ to
$51.3$\,s.
Its mean margin and mission time are close
to those of QT ($0.52$ versus $0.50$ and
$51.3$ versus $51.2$\,s), with a lower
CBF-violation fraction
$\widehat q_{\mathrm{CBF}}^+$
($7.4\%$ versus $8.0\%$).

At the same setting, V-U uses a mean margin
of $0.97c_{\max}$ and has
$\widehat q_{\mathrm{CBF}}^+=0.1\%$.
Thus, the smaller margins enabled by the
tighter bounds in V-C improve navigation performance
but allow more realized CBF violations.
These violations can occur even when the
exceedance risk remains within its limit.

\paragraph{Illustrative trajectory}
Figure~\ref{fig:example-maze-winner} shows V-L
at S3.
Panel~(a) compares its paths with those obtained
using $c_{\max}$.
Panel~(b) shows how the selected margin varies, including temporary
returns to $c_{\max}$.
Panel~(c) marks the observed loss exceedances
$Y_t>c_t$, which occur at $12.21\%$ of the steps.
Despite these exceedances, the
exceedance risk $ p_t(c_t)$, measured via $\widehat p_t(c_t)$, remains
below $\alpha_0=0.30$  in panel~(d).
This illustrates the distinction between
realized exceedances and the exceedance risk
bounded by our theory.

\paragraph{Effect of the parameters}
Reducing $\delta$ from $0.20$ in S1
to $0.15$ in S2 raises the certification
threshold and slightly increases the mean
margins in Table~\ref{tab:empirical-topfive}.

Changing $\alpha$ has two competing effects:
a smaller $\alpha$ makes certification harder
but increases the deployment buffer
$\alpha_0-\alpha$.
From S1 to S3, $\alpha$ decreases
from $0.25$ to $0.20$, with all other parameters
unchanged.
The mean margin increases from $0.52c_{\max}$
to $0.61c_{\max}$ for V-C, but decreases from
$0.97c_{\max}$ to $0.82c_{\max}$ for V-U and
V-$\Delta$.
The resulting trade-off therefore depends
on the risk-increase bound used.

Further comparisons of $\alpha$, $\delta$,
and the test period $K$, together
with additional ablations, are provided
in the supplementary ablation report~\cite{trakas2026ablation}.

\paragraph{Additional results and checks}

Appendix~\ref{app:arxiv-restored-stress}
examines sensitivity to reducing the supplied
risk-increase bounds $d_t^{(r)}(c)$.
These reductions improve the overall performance but may violate
Assumption~\ref{as:bounded-shift}, in which
case the guarantees no longer apply.
Further comparisons, including single and
parallel tests, are provided in
the supplementary ablation report~\cite{trakas2026ablation}.

\bibliographystyle{IEEEtran}
\bibliography{references}

@article{ramdas2023savi,
  author = {Ramdas, Aaditya and Gr{\"u}nwald, Peter and Vovk, Vladimir and Shafer, Glenn},
  title = {Game-theoretic statistics and safe anytime-valid inference},
  journal = {Stat. Sci.},
  volume = {38},
  number = {4},
  pages = {576--601},
  year = {2023}
}

@article{jagtap2020formal,
  title={Formal synthesis of stochastic systems via control barrier certificates},
  author={Jagtap, Pushpak and Soudjani, Sadegh and Zamani, Majid},
  journal={IEEE Transactions on Automatic Control},
  volume={66},
  number={7},
  pages={3097--3110},
  year={2020},
  publisher={IEEE}
}

@article{fochesato2025synthesis,
  title={Synthesis of safety certificates for discrete-time uncertain systems via convex optimization},
  author={Fochesato, Marta and Wang, Han and Papachristodoulou, Antonis and Goulart, Paul},
  journal={arXiv preprint arXiv:2505.08559},
  year={2025}
}

@article{wabersich2023data,
  author = {Wabersich, Kim P. and Taylor, Andrew J. and Choi, Jason J. and Sreenath, Koushil and Tomlin, Claire J. and Ames, Aaron D. and Zeilinger, Melanie N.},
  title = {Data-driven safety filters: {Hamilton--Jacobi} reachability, control barrier functions, and predictive methods for uncertain systems},
  journal = {IEEE Control Syst. Mag.},
  volume = {43},
  number = {5},
  pages = {137--177},
  year = {2023}
}

@inproceedings{wei2025conformal,
  author = {Wei, Sihang and Ornik, Melkior and Tsukamoto, Hiroyasu},
  title = {Conformal contraction for robust nonlinear control with distribution-free uncertainty quantification},
  booktitle = {Proc. IEEE Conf. Decis. Control},
  pages = {3502--3507},
  year = {2025}
}

@article{hsu2025statistical,
  author = {Hsu, Ting-Wei and Tsukamoto, Hiroyasu},
  title = {Statistical guarantees in data-driven nonlinear control: conformal robustness for stability and safety},
  journal = {IEEE Control Syst. Lett.},
  volume = {9},
  pages = {997--1002},
  year = {2025}
}

@inproceedings{lederer2019uniform,
  author = {Lederer, Armin and Umlauft, Jonas and Hirche, Sandra},
  title = {Uniform error bounds for {Gaussian} process regression with application to safe control},
  booktitle = {Adv. Neural Inf. Process. Syst.},
  volume = {32},
  year = {2019}
}

@article{grunwald2024safe,
  author = {Gr{\"u}nwald, Peter and de Heide, Rianne and Koolen, Wouter},
  title = {Safe testing},
  journal = {J. R. Stat. Soc. Ser. B Stat. Methodol.},
  volume = {86},
  number = {5},
  pages = {1091--1128},
  year = {2024}
}

@article{vovk2021evalues,
  author = {Vovk, Vladimir and Wang, Ruodu},
  title = {E-values: calibration, combination and applications},
  journal = {Ann. Statist.},
  volume = {49},
  number = {3},
  pages = {1736--1754},
  year = {2021}
}

@book{ville1939,
  author = {Ville, Jean},
  title = {{\'E}tude critique de la notion de collectif},
  publisher = {Gauthier-Villars},
  address = {Paris},
  year = {1939}
}

@article{shafer2021testing,
  author = {Shafer, Glenn},
  title = {Testing by betting: a strategy for statistical and scientific communication},
  journal = {J. R. Stat. Soc. Ser. A Stat. Soc.},
  volume = {184},
  number = {2},
  pages = {407--431},
  year = {2021}
}

@inproceedings{xu2024active,
  author = {Xu, Ziyu and Karampatziakis, Nikos and Mineiro, Paul},
  title = {Active, anytime-valid risk controlling prediction sets},
  booktitle = {Adv. Neural Inf. Process. Syst.},
  volume = {37},
  pages = {60110--60132},
  year = {2024}
}

@article{ames2017cbfqp,
  author = {Ames, Aaron D. and Xu, Xiangru and Grizzle, Jessy W. and Tabuada, Paulo},
  title = {Control barrier function based quadratic programs for safety critical systems},
  journal = {IEEE Trans. Autom. Control},
  volume = {62},
  number = {8},
  pages = {3861--3876},
  year = {2017}
}

@inproceedings{agrawal2017discrete,
  author = {Agrawal, Ayush and Sreenath, Koushil},
  title = {Discrete control barrier functions for safety-critical control of discrete systems with application to bipedal robot navigation},
  booktitle = {Proc. Robot.: Sci. Syst.},
  year = {2017}
}

@article{wabersich2021predictive,
  author = {Wabersich, Kim P. and Zeilinger, Melanie N.},
  title = {A predictive safety filter for learning-based control of constrained nonlinear dynamical systems},
  journal = {Automatica},
  volume = {129},
  year = {2021},
  note = {Art. no. 109597}
}

@article{santoyo2021barrier,
  author = {Santoyo, Cesar and Dutreix, Maxence and Coogan, Samuel},
  title = {A barrier function approach to finite-time stochastic system verification and control},
  journal = {Automatica},
  volume = {125},
  year = {2021},
  note = {Art. no. 109439}
}

@article{cosner2024bounding,
  title={Bounding stochastic safety: Leveraging freedman’s inequality with discrete-time control barrier functions},
  author={Cosner, Ryan K and Culbertson, Preston and Ames, Aaron D},
  journal={IEEE Control Systems Letters},
  volume={8},
  pages={1937--1942},
  year={2024},
  publisher={IEEE}
}

@inproceedings{cosner2023robust,
  author = {Ryan K. Cosner and Preston Culbertson and Andrew J. Taylor and Aaron D. Ames},
  title = {Robust safety under stochastic uncertainty with discrete-time control barrier functions},
  booktitle = {Proc. Robot.: Sci. Syst.},
  year = {2023}
}

@inproceedings{mestres2026probabilistic,
  author = {Mestres, Pol and Werner, Blake and Cosner, Ryan K. and Ames, Aaron D.},
  title = {Probabilistic control barrier functions: safety in probability for discrete-time stochastic systems},
  booktitle = {Proc. Am. Control Conf.},
    pages={1942-1949},
  year = {2026}
}

@article{das2026safe,
  author = {Da{\c{s}}, E. and Nanayakkara, R. and Tan, X. and Bena, R. M. and Burdick, J. W. and Tabuada, P. and Ames, A. D.},
  title = {Safe navigation under state uncertainty: online adaptation for robust control barrier functions},
  journal = {IEEE Robot. Autom. Lett.},
  volume = {11},
  number = {3},
  pages = {3398--3405},
  year = {2026}
}

@article{lindemann2025formal,
  author = {Lindemann, Lars and Zhao, Yiqi and Yu, Xinyi and Pappas, George J. and Deshmukh, Jyotirmoy V.},
  title = {Formal verification and control with conformal prediction: practical safety guarantees for autonomous systems},
  journal = {IEEE Control Syst. Mag.},
  volume = {45},
  number = {6},
  pages = {72--122},
  year = {2025}
}

@inproceedings{gibbs2021adaptive,
  author = {Gibbs, Isaac and Cand{\`e}s, Emmanuel},
  title = {Adaptive conformal inference under distribution shift},
  booktitle = {Adv. Neural Inf. Process. Syst.},
  volume = {34},
  pages = {1660--1672},
  year = {2021}
}

@misc{mirzaeedodangeh2026robust,
  author = {Mirzaeedodangeh, Omid and Shekhtman, Eliot and Matni, Nikolai and Lindemann, Lars},
  title = {Robust conformal {CBF} and {CLF} controllers via iterative policy updates},
  howpublished = {arXiv:2606.15366},
  year = {2026}
}

@misc{cherenson2026staggered,
  author = {Cherenson, Daniel M. and Panagou, Dimitra},
  title = {Staggered integral online conformal prediction for safe dynamics adaptation with multi-step coverage guarantees},
  howpublished = {arXiv:2604.06058},
  year = {2026}
}

@inproceedings{huriot2026adaptive,
  author = {Huriot, Sacha and Tabbara, Ihab and Sibai, Hussein},
  title = {Safe control using learned safety filters and adaptive conformal inference},
  booktitle = {Proc. Learn. Dyn. Control Conf.},
  series = {Proc. Mach. Learn. Res.},
  volume = {331},
  pages = {833--847},
  year = {2026}
}

@misc{gauthier2025evalues,
  author = {Gauthier, Etienne and Bach, Francis and Jordan, Michael I.},
  title = {E-values expand the scope of conformal prediction},
  howpublished = {arXiv:2503.13050},
  year = {2025}
}

@article{shin2023edetectors,
  author = {Shin, Jaehyeok and Ramdas, Aaditya and Rinaldo, Alessandro},
  title = {E-detectors: a nonparametric framework for sequential change detection},
  journal = {New Engl. J. Stat. Data Sci.},
  volume = {2},
  number = {2},
  pages = {229--260},
  year = {2024}
}

@inproceedings{mirzaeedodangeh2026safe,
  author = {Mirzaeedodangeh, Omid and Shekhtman, Eliot Seo and Matni, Nikolai and Lindemann, Lars},
  title = {Safe planning in interactive environments via iterative policy updates and adversarially robust conformal prediction},
  booktitle = {Proc. Learn. Dyn. Control Conf.},
  series = {Proc. Mach. Learn. Res.},
  volume = {331},
  pages = {678--704},
  year = {2026}
}

@article{ramdas2025hypothesis,
  author = {Ramdas, Aaditya and Wang, Ruodu},
  title = {Hypothesis testing with e-values},
  journal = {Found. Trends Stat.},
  volume = {1},
  number = {1--2},
  pages = {1--390},
  year = {2025}
}

@misc{clerico2026,
  author = {Eugenio Clerico},
  title = {Sequential testing of conditionally constrained hypotheses},
  howpublished = {arXiv:2606.06769},
  year = {2026}
}

@article{wang2025ebacktesting,
  author = {Wang, Qiuqi and Wang, Ruodu and Ziegel, Johanna F.},
  title = {E-backtesting},
  journal = {Manage. Sci.},
  volume = {72},
  number = {6},
  pages = {4952--4973},
  year = {2026}
}

@article{fan2025testing,
  author = {Fan, Yixuan and Jiao, Zhanyi and Wang, Ruodu},
  title = {Testing the mean and variance by e-processes},
  journal = {Biometrika},
  volume = {112},
  number = {1},
  year = {2025},
  note = {Art. no. asae049}
}

@article{jankovic2018robust,
  author = {Jankovic, Mrdjan},
  title = {Robust control barrier functions for constrained stabilization of nonlinear systems},
  journal = {Automatica},
  volume = {96},
  pages = {359--367},
  year = {2018}
}

@inproceedings{luo2024online,
  author = {Luo, Rachel and Sinha, Rohan and Sun, Yixiao and Hindy, Ali and Zhao, Shengjia and Savarese, Silvio and Schmerling, Edward and Pavone, Marco},
  title = {Online distribution shift detection via recency prediction},
  booktitle = {Proc. IEEE Int. Conf. Robot. Autom.},
  pages = {16251--16263},
  year = {2024}
}

@misc{hindy2024diagnostic,
  author = {Hindy, Ali and Luo, Rachel and Banerjee, Somrita and Kuck, Jonathan and Schmerling, Edward and Pavone, Marco},
  title = {Diagnostic runtime monitoring with martingales},
  howpublished = {arXiv:2407.21748},
  year = {2024}
}

@inproceedings{angelopoulos2023conformal,
  author = {Angelopoulos, Anastasios N. and Cand{\`e}s, Emmanuel J. and Tibshirani, Ryan J.},
  title = {Conformal {PID} control for time series prediction},
  booktitle = {Adv. Neural Inf. Process. Syst.},
  volume = {36},
  pages = {23047--23074},
  year = {2023}
}

@article{clopper1934use,
  author = {Clopper, C. J. and Pearson, E. S.},
  title = {The use of confidence or fiducial limits illustrated in the case of the binomial},
  journal = {Biometrika},
  volume = {26},
  number = {4},
  pages = {404--413},
  year = {1934}
}

@article{chen2014tvgarch,
  author = {Chen, Cathy W. S. and Gerlach, Richard and Lin, Edward M. H.},
  title = {{Bayesian} estimation of smoothly mixing time-varying parameter {GARCH} models},
  journal = {Comput. Stat. Data Anal.},
  volume = {76},
  pages = {194--209},
  year = {2014}
}

@article{wang2021pricelimits,
  author = {Wang, Donghua and Ding, Jin and Chu, Guoqing and Xu, Dinghai and Wirjanto, Tony S.},
  title = {Modelling asset returns in the presence of price limits with {Markov}-switching mixture of truncated normal {GARCH} distribution: evidence from {China}},
  journal = {Appl. Econ.},
  volume = {53},
  number = {7},
  pages = {781--804},
  year = {2021}
}

@inproceedings{dixit2023adaptive,
  author = {Dixit, Anushri and Lindemann, Lars and Wei, Skylar X. and Cleaveland, Matthew and Pappas, George J. and Burdick, Joel W.},
  title = {Adaptive conformal prediction for motion planning among dynamic agents},
  booktitle = {Proc. Learn. Dyn. Control Conf.},
  series = {Proc. Mach. Learn. Res.},
  volume = {211},
  pages = {300--314},
  year = {2023}
}

@inproceedings{snyder2026policy,
  author = {Snyder, David and Badithela, Apurva and Matni, Nikolai and Pappas, George J. and Majumdar, Anirudha and Itkina, Masha and Nishimura, Haruki},
  title = {Beyond binary success: sample-efficient and statistically rigorous robot policy comparison},
  booktitle = {Proc. Robot.: Sci. Syst.},
  year = {2026}
}

@article{LN,
  author  = {Azuma, Kazuoki},
  title   = {Weighted Sums of Certain Dependent Random Variables},
  journal = {Tohoku Math. J.},
  volume  = {19},
  number  = {3},
  pages   = {357--367},
  year    = {1967},
  doi     = {10.2748/tmj/1178243286}
}

@misc{trakas2026ablation,
  author = {Trakas, Panagiotis S. and Mirzaeedodangeh, Omid and Lindemann, Lars},
  title = {Anytime-valid certification of robustness margins for safe control via e-values: supplementary ablation report},
  year = {2026},
  url = {https://github.com/omidmirza/ecbf-ablation/blob/main/pdf/ablation_report.pdf}
}

\useRomanappendicesfalse
\appendices
\section{Monotone certification}
\begingroup
\begin{lemma}
\label{prop:monotone-certification}
For every $t\geq0$ and $c,c'\in\Gamma$:
\[
    c'\geq c
    \quad\Longrightarrow\quad
    E_t(c')\geq E_t(c).
\]
Consequently, if $c\in\mathcal S_t$ and $c'\geq c$,
then $c'\in\mathcal S_t$.
\end{lemma}

\begin{proof}
If $c'\geq c$, then $Z_s(c')\leq Z_s(c)$ for every $s$.
Since $\lambda\geq0$:
\[
    1+\lambda\bigl(\alpha-Z_s(c')\bigr)
    \geq
    1+\lambda\bigl(\alpha-Z_s(c)\bigr)
    \geq0.
\]
Both processes start at one, i.e., $E_0^\lambda(c)=E_0^\lambda(c')=1$, and use the same gain, so
multiplying these factors yields
$E_t^\lambda(c')\geq E_t^\lambda(c)$.
Averaging over the common set $\Lambda$ gives
$E_t(c')\geq E_t(c)$.
If $c\in\mathcal S_t$, then $E_s(c)\geq1/\delta$
for some $s\leq t$. Hence $E_s(c')\geq1/\delta$,
which implies $c'\in\mathcal S_t$.
\end{proof}
\endgroup
\section{Proofs of Main Theoretical Results}
\subsection{Proof of Theorem~\ref{thm:false-certification}}
\label{app:cd-proof}
\begin{proof}
Set $\mathcal I_{-1}:=\Gamma$ and, for $t\geq0$, define
\[
    \mathcal I_t
    :=
    \left\{
        c\in\Gamma:
        p_s(c)\geq\alpha,\
        s=0,\ldots,t
    \right\}.
\]
This set contains the candidates that have remained
$\alpha$-insufficient through time $t$.
Since $p_s(c)$ is nonincreasing in $c$, membership of
$c$ within $\mathcal I_t$ implies membership of every smaller candidate.
Hence, $\mathcal I_t$ is an initial segment of $\Gamma$.
Moreover, we have that $\mathcal I_t\subseteq\mathcal I_{t-1}$.

Whenever $\mathcal I_t\neq\emptyset$, define
$b_t:=\max\mathcal I_t$, with $b_{-1}=c^{(M)}$.
The variable $b_t$ is $\Fc_t$-measurable and satisfies
$p_t(b_t)\geq\alpha$ and $b_t\leq b_{t-1}$.
Define the auxiliary process $B_0:=1$ and
\[
    B_{t+1}
    :=
    \begin{cases}
        E_{t+1}(b_t), & \mathcal I_t\neq\emptyset,\\
        0, & \mathcal I_t=\emptyset.
    \end{cases}
\]
On $\{\mathcal I_t\neq\emptyset\}$, equation
\eqref{eq:one-step-evariable} and linearity of the conditional
expectation give
\[
\begin{aligned}
    \E[B_{t+1}\mid\Fc_t]
    \leq E_t(b_t)
    \leq E_t(b_{t-1})
    =B_t.
\end{aligned}
\]
The second inequality follows from
Lemma~\ref{prop:monotone-certification}.
On $\{\mathcal I_t=\emptyset\}$, $B_{t+1}=0$,
so the same inequality holds.
All update factors are bounded, ensuring integrability
at each finite time. Thus, $(B_t)_{t\geq0}$ is a
nonnegative supermartingale with $B_0=1$.

If $\mathcal G^c$ occurs, there exists $c\in\Gamma$ with
$\tau(c)<\infty$ such that $p_s(c)\geq\alpha$ for every
$s<\tau(c)$. Setting $t=\tau(c)-1$ gives
$c\in\mathcal I_t$ and hence $c\leq b_t$. By monotonicity, at $t=\tau(c)-1$:
\[
    B_{\tau(c)}
    =E_{\tau(c)}(b_{\tau(c)-1})
    \geq E_{\tau(c)}(c)
    \geq\frac{1}{\delta}.
\]
Thus, whenever $\mathcal G^c$ occurs, $B_t$ reaches
$1/\delta$ at some finite time. Consequently, $\mathcal G^c
    \subseteq
    \left\{
        \sup_{t\geq0}B_t\geq\frac{1}{\delta}
    \right\}$. Since $(B_t)_{t\geq0}$ is a nonnegative supermartingale
with $B_0=1$, Ville's inequality yields
$\Prob(\mathcal G^c)\leq\delta$, or equivalently,
$\Prob(\mathcal G)\geq1-\delta$.
\end{proof}

\subsection{Proof of Theorem~\ref{thm:certified-deployment}}
\label{app:main-proof}
\begin{proof}
Consider the event $\mathcal G$ defined in
\eqref{eq:no-false-certification-event} and fix
a  time $t\in\mathbb N_0$.
If $\mathcal D_t=\emptyset$, we have that $c_t=c_{\max}$. In this case, Assumption~\ref{as:conservative-margin}
gives $p_t(c_t)=0$ almost surely.
Otherwise, i.e., $\mathcal D_t\neq \emptyset$, the deployment rule selects
$c_t=\min\mathcal D_t$ and it holds that
\[
    \tau(c_t)\leq t,
    \qquad d_t(c_t)\leq\alpha_0-\alpha,
    \qquad u_t=u_t(c_t).
\]
On $\mathcal G$, there is an integer
$0\leq s<\tau(c_t)$ such that $p_s(c_t)<\alpha$.
Applying Assumption~\ref{as:bounded-shift} with
$\sigma=0$ and $\tau=\tau(c_t)$ gives
\[
\begin{aligned}
    p_t(c_t)
    \leq p_s(c_t)+d_t(c_t)
    <\alpha+(\alpha_0-\alpha)
    =\alpha_0.
\end{aligned}
\]
Since decision times are countable, these bounds
hold almost surely simultaneously for all
$t\in\mathbb N_0$ on $\mathcal G$.
Thus, by \eqref{eq:first-deployment-event},
\[
    \mathcal G\subseteq\mathcal G^d
    \quad\text{almost surely}.
\]
Theorem~\ref{thm:false-certification} then gives
\[
    \Prob(\mathcal G^d)
    \geq\Prob(\mathcal G)
    \geq1-\delta
\]
which proves \eqref{eq:certified-cbf-guarantee}.
\end{proof}
\subsection{Proof of Theorem~\ref{prop:repeated-deployment}}
\label{app:repeated-deployment-proof}
\begin{proof}
Fix $r\in\mathbb N_{>0}$ and recall that all
candidates $c\in\Gamma$ start a test $r$ at $\sigma_r=(r-1)K$.
Define the event
\begin{equation}
\begin{aligned}
    \mathcal G_r:=\bigl\{&
        \forall c\in\Gamma:\ \tau_r(c)<\infty
        \ \Longrightarrow\\
        &\exists s\in\mathbb N_0:
        \ \sigma_r\leq s<\tau_r(c),\
        p_s(c)<\alpha
    \bigr\}.
\end{aligned}
\stepcounter{equation}\notag\label{eq:restarted-good-event}
\end{equation}

Following equation \eqref{eq:wealth-recursion}, we first define the variable
\[
    E_t^{\lambda,r}(c):=
    \prod_{s=\sigma_r}^{t-1}
    \bigl[1+\lambda(\alpha-Z_s(c))\bigr],
    \qquad t\geq\sigma_r
\]
with $E_{\sigma_r}^{\lambda,r}(c)=1$. Further, following equation \eqref{eq:mixture-eprocess}, let
$E_t^r(c)$ denote their uniform mixture.
These product variables $E_t^{\lambda,r}(c)$ are defined even after
certification $\tau_r(c)$. They agree with the
evidence up to and including $\tau_r(c)$, so
their first crossing of $1/\delta_r$ occurs
at the same certification time.

Since $Z_s(c)$ is nonincreasing in $c$ and the
admissible gains keep every factor nonnegative,
$E_t^{\lambda,r}(c)$ and $E_t^r(c)$ are
nondecreasing in $c$ (similar to Lemma \ref{prop:monotone-certification}).
All candidates share the same starting time,
gains, and threshold.
Applying Theorem~\ref{thm:false-certification}
to the shifted observations
$(Y_{\sigma_r+s})_{s\geq0}$, with filtration
$(\Fc_{\sigma_r+s})_{s\geq0}$ and error level
$\delta_r$, therefore gives
\begin{equation}
    \Prob(\mathcal G_r)\geq1-\delta_r
    \label{eq:restarted-certification-guarantee}
\end{equation}
where no independence between tests is required.

Consider the event
$\bigcap_{r\geq1}\mathcal G_r$ and fix a time $t$. Following the proof of Theorem \ref{thm:certified-deployment}, if $\mathcal D_t=\emptyset$, the fallback use $c_t=c_{\max}$ satisfies
$p_t(c_t)\leq\alpha_0$ by
Assumption~\ref{as:conservative-margin}.

Otherwise, set $c=c_t$.
By \eqref{eq:deployment-rule},
$c_t\in\mathcal D_t$ and $u_t=u_t(c)$.
The definition \eqref{eq:D_updated} therefore
provides an index $r\in\mathbb N_{>0}$ such that
\[
    \tau_r(c)\leq t,
    \qquad d_t^{(r)}(c)\leq\alpha_0-\alpha.
\]

Since $\tau_r(c)<\infty$, the event $\mathcal G_r$
provides an integer $s$ satisfying
\[
    \sigma_r\leq s<\tau_r(c_t),
    \qquad p_s(c_t)<\alpha.
\]
Apply Assumption~\ref{as:bounded-shift} with
$\sigma=\sigma_r$ and $\tau=\tau_r(c_t)$.
Using \eqref{eq:repeated-shift-bound} gives
\[
\begin{aligned}
    p_t(c_t)
    \leq p_s(c_t)+d_t^{(r)}(c_t)
    <\alpha+d_t^{(r)}(c_t)
    \leq\alpha_0.
\end{aligned}
\]
Because the candidates, test indices, and decision
times form countable sets, these implications hold
almost surely simultaneously.
Hence, by \eqref{eq:first-deployment-event}:
\[
    \bigcap_{r\geq1}\mathcal G_r
    \subseteq\mathcal G^d
    \quad\text{almost surely}.
\]

Finally, a union bound and
\eqref{eq:restarted-certification-guarantee} yield
\[
\begin{aligned}
    \Prob\bigl((\mathcal G^d)^c\bigr)
    \leq
    \Prob\!\left(\bigcup_{r\geq1}\mathcal G_r^c\right)
    \leq
    \sum_{r\geq1}\Prob(\mathcal G_r^c)
    \leq
    \sum_{r\geq1}\delta_r
    =\delta.
\end{aligned}
\]
This proves \eqref{eq:repeated-deployment-guarantee}.
\end{proof}

\subsection{Proof of Corollary~\ref{cor:finite-horizon-safety}}
\label{corpoof}
\begin{proof}
Since $Z_t(c)\in\{0,1\}$, each evidence factor
$[1+\lambda(\alpha-Z_s(c))]$ in $E_t^{\lambda,r}(c)$
is at most $1+\lambda_{\max}\alpha$.
After $n$ observations, the mixture evidence
of any test is therefore at most
$(1+\lambda_{\max}\alpha)^n$.
Since $\delta_r\leq\delta$, certification requires
\[
    (1+\lambda_{\max}\alpha)^n
    \geq\frac1{\delta_r}\geq\frac1\delta,
\]
which implies $n\geq n_0$.
Thus, no certificate is available before time
$n_0$, and $c_t=c_{\max}$ for
$t=0,\ldots,n_0-1$.

The control input satisfies the condition
\eqref{eq:tightened-cbf-condition}. Consequently,
\eqref{eq:true-cbf-condition} gives
\[
    x_t\in\Cc,\quad Y_t\leq c_t
    \quad\Longrightarrow\quad x_{t+1}\in\Cc.
\]

For $t<n_0$, the controller uses $c_{\max}$,
so $Y_t\leq c_t$ almost surely.
Starting from $x_0\in\Cc$, the implication above
therefore gives $x_t\in\Cc$ for
$t=0,\ldots,n_0$ almost surely.

For $t\geq n_0$, an exceedance may occur.
Since the state remains safe through time $n_0$,
a first exit within $T$ steps must follow
a decision $t\in\{n_0,\ldots,T-1\}$.
Consequently, almost surely,
\[
    \{\exists t\in\{0,\ldots,T\}:x_t\notin\Cc\}
    \subseteq
    \bigcup_{t=n_0}^{T-1}\{Y_t>c_t\}.
\]
An exceedance need not cause an exit, so this
inclusion can be strict.
For $T\leq n_0$, the union is empty and the
exit probability is zero.
For $T>n_0$, we bound its probability below.

For each $t\in\{n_0,\ldots,T-1\}$,
define the event
\[
    \mathcal V_t:=\{Y_t>c_t,\ p_t(c_t)\leq\alpha_0\}.
\]
Since $c_t$ and $p_t(c_t)$ are $\Fc_t$-measurable,
\[
\begin{aligned}
    \Prob(\mathcal V_t\mid\Fc_t)
    &=\mathbf1_{\{p_t(c_t)\leq\alpha_0\}}
      \Prob(Y_t>c_t\mid\Fc_t)\\
    &=\mathbf1_{\{p_t(c_t)\leq\alpha_0\}}
      p_t(c_t)
    \leq\alpha_0.
\end{aligned}
\]
Moreover,
$\bigcap_{s=n_0}^{t-1}\mathcal V_s^c$
is $\Fc_t$-measurable. Therefore,
\begin{equation}
\begin{aligned}
    \Prob\!\left(
        \bigcap_{s=n_0}^{t}\mathcal V_s^c
    \right)
    &=
    \E\!\left[
        \mathbf1_{\bigcap_{s=n_0}^{t-1}
        \mathcal V_s^c}
        \Prob(\mathcal V_t^c\mid\Fc_t)
    \right]\\
    &\geq
    (1-\alpha_0)
    \Prob\!\left(
        \bigcap_{s=n_0}^{t-1}\mathcal V_s^c
    \right).
\end{aligned}
\label{C1}
\end{equation}
Applying \eqref{C1} successively for
$t=n_0,\ldots,T-1$,
starting from probability one
before step $n_0$, gives
\[
    \Prob\!\left(
        \bigcap_{t=n_0}^{T-1}\mathcal V_t^c
    \right)
    \geq(1-\alpha_0)^{T-n_0}.
\]
No independence between the events is required.

On $\mathcal G^d$, every exceedance
$Y_t>c_t$ belongs to $\mathcal V_t$.
Thus, almost surely,
\[
    \{\exists t\in\{0,\ldots,T\}:x_t\notin\Cc\}
    \subseteq
    (\mathcal G^d)^c
    \cup\bigcup_{t=n_0}^{T-1}\mathcal V_t.
\]
Using Theorem~\ref{prop:repeated-deployment},
we obtain
\[
\begin{aligned}
    &\Prob\!\left(\exists t\in\{0,\ldots,T\}:
        x_t\notin\Cc\right)\\
    &\quad\leq
    \Prob\bigl((\mathcal G^d)^c\bigr)
    +\Prob\!\left(
        \bigcup_{t=n_0}^{T-1}\mathcal V_t
    \right)\\
    &\quad\leq
    \delta+1-(1-\alpha_0)^{T-n_0}.
\end{aligned}
\]

Taking the minimum with one and combining this
bound with the zero exit probability for
$T\leq n_0$ yields
\[
\begin{aligned}
    &\Prob\!\left(\exists t\in\{0,\ldots,T\}:
        x_t\notin\Cc\right)\\
    &\quad\leq
    \mathbf1_{\{T>n_0\}}
    \min\{1,\delta+1-(1-\alpha_0)^{T-n_0}\},
\end{aligned}
\]
which proves \eqref{eq:finite-horizon-safety}.
\end{proof}
\section{Evidence Growth and Certification Time}
\subsection{Proof of Proposition~\ref{thm:evidence-growth}}
\label{app:evidence-growth-proof}
\begingroup

\begin{proof}
By \eqref{eq:wealth-recursion} we have
\[
\begin{aligned}
    &\E\!\left[
        \log\frac{E_{t+1}^{\lambda}(c)}{E_t^{\lambda}(c)}
        \,\middle|\,\Fc_t
    \right]\\
    &\quad=(1-p_t(c))\log(1+\lambda\alpha)\\
    &\qquad+p_t(c)\log\!\left(1-\lambda(1-\alpha)\right)\\
    &\quad\geq g_{\bar p(c)}(\lambda)>0
\end{aligned}
\]
where the inequality follows from
$p_t(c)\leq\bar p(c)$ and the fact that the conditional
expectation decreases with $p_t(c)$.

Since $0<\lambda<(1-\alpha)^{-1}$, note that the logarithmic
increments are bounded:
\[
    \log\!\left(1-\lambda(1-\alpha)\right)
    \leq
    \log\frac{E_{t+1}^\lambda(c)}{E_t^\lambda(c)}
    \leq
    \log(1+\lambda\alpha).
\]
The strong law for uniformly bounded martingale
differences \cite[Cor.~1]{LN},
together with $E_0^\lambda(c)=1$, gives
\[
    \frac{\log E_t^\lambda(c)}{t}
    -
    \frac{1}{t}\sum_{s=0}^{t-1}
    \E\!\left[
        \log\frac{E_{s+1}^\lambda(c)}{E_s^\lambda(c)}
        \,\middle|\,\Fc_s
    \right]
    \xrightarrow[t\to\infty]{\mathrm{a.s.}}0.
\]
Since each conditional expectation within the sum is at least
$g_{\bar p(c)}(\lambda)$, it follows that
\eqref{eq:e-power-limit}.

To bound $\E[\tau(c)]$, note that, before certification, we have
\[
    \frac{E_t^\lambda(c)}{|\Lambda|}
    \leq E_t(c)<\frac{1}{\delta},
    \qquad t<\tau(c).
\]
Hence, $\log E_t^\lambda(c)<\log(|\Lambda|/\delta)$
before $\tau(c)$, and its increase at $\tau(c)$ is at
most $\log(1+\lambda\alpha)$.

For an integer $n\geq1$, write
\[
    \log E_{\min\{\tau(c),n\}}^\lambda(c)
    =
    \sum_{t=0}^{n-1}
    \mathbf{1}_{\{\tau(c)>t\}}
    \log\frac{E_{t+1}^\lambda(c)}{E_t^\lambda(c)}.
\]
This sum has at most $n$ bounded terms. Since
$\{\tau(c)>t\}\in\Fc_t$, taking expectations and using
the conditional log-growth bound yields
\[
\begin{aligned}
    g_{\bar p(c)}(\lambda)\,
    \E[\min\{\tau(c),n\}]
    &\leq
    \E\!\left[
        \log E_{\min\{\tau(c),n\}}^\lambda(c)
    \right]\\
    &\leq
    \log(|\Lambda|/\delta)+\log(1+\lambda\alpha).
\end{aligned}
\]
The upper bound includes the largest possible increment
at $\tau(c)$. Since $\min\{\tau(c),n\}$ increases to
$\tau(c)$ as $n\to\infty$, monotone convergence yields
\[
\begin{aligned}
    \E[\tau(c)]
    &=\lim_{n\to\infty}\E[\min\{\tau(c),n\}]\\
    &\leq
    \frac{\log(|\Lambda|/\delta)+\log(1+\lambda\alpha)}
         {g_{\bar p(c)}(\lambda)}.
\end{aligned}
\]
This proves \eqref{eq:expected-certification-time} and
implies $\Prob(\tau(c)<\infty)=1$.
\end{proof}
\subsection{Gain Selection and Numerical Setup}
\label{app:certification-details}

\paragraph{Growth and gain selection}
The quantity $g_{\bar p(c)}(\lambda)$ lower bounds
the conditional expected increase in
$\log E_t^\lambda(c)$ per observation.
This expected log increase is also called the
conditional \emph{e-power} of the one-step
e-variable \cite{grunwald2024safe,shafer2021testing}.

For $q\in(0,\alpha)$, differentiating the growth rate gives
\[
    g_q'(\lambda)
    =
    \frac{\alpha-q-\lambda\alpha(1-\alpha)}
         {(1+\lambda\alpha)(1-\lambda(1-\alpha))}.
\]
The denominator is positive for
$0<\lambda<(1-\alpha)^{-1}$, and the numerator
changes sign from positive to negative at
\begin{equation}
    \lambda^\star(q)
    :=\underset{0<\lambda<(1-\alpha)^{-1}}{\arg\max}
       \,g_q(\lambda)
    =\frac{\alpha-q}{\alpha(1-\alpha)}.
    \label{eq:optimal-growth-gain}
\end{equation}
Thus, when $0<\bar p(c)<\alpha$, choosing
$q=\bar p(c)$ maximizes the growth lower bound
in Proposition~\ref{thm:evidence-growth}.
Maximizing this growth bound does not necessarily
give the shortest expected certification time.
In practice, the monitor combines the gains in
$\Lambda$, chosen in advance, without needing
to know $\bar p(c)$.

\paragraph{Adaptive gains}
The gain can alternatively be selected online using
GREE \cite{wang2025ebacktesting,fan2025testing}, which
uses past observations to maximize empirical log growth.
Any predictable gain in $(0,(1-\alpha)^{-1})$ preserves
validity for an individual candidate. However,
candidate-dependent gains may break the evidence
ordering used in Theorem~\ref{thm:false-certification}.
A common predictable gain for all candidates preserves
this ordering and the simultaneous guarantee.

\paragraph{Details of Example~2}
In Example~2 and Fig.~\ref{fig:certification-illustration}, one time
step is one trading day. Let $P_t>0$ be the stock price at the start of
day~$t$ and $\xi_t:=\log(P_{t+1}/P_t)$ its log return; $\Fc_t$
contains $\xi_0,\ldots,\xi_{t-1}$. The volatility $\sigma_t$ follows
the TV-GARCH$(1,1)$ model of~\cite{chen2014tvgarch} with an asymmetric
logistic transition:
\begin{equation}
\begin{aligned}
    \sigma_t^2={}&a_0+a_1\xi_{t-1}^2+b_1\sigma_{t-1}^2\\
    &+\left(a_0^\star+a_1^\star \xi_{t-1}^2
       +b_1^\star\sigma_{t-1}^2\right)G_t,
    \qquad t\geq1,
\end{aligned}
\stepcounter{equation}\notag\label{eq:finance-tvgarch-scale}
\end{equation}
where
\begin{equation}
G_t:=
\begin{cases}
\bigl[1+\exp\{\zeta_1(t/1000-m_1)\}\bigr]^{-1},
    &t\leq t_{\mathrm c},\\[1mm]
\bigl[1+\exp\{-\zeta_2(t/1000-m_2)\}\bigr]^{-1},
    &t>t_{\mathrm c},
\end{cases}
\stepcounter{equation}\notag\label{eq:finance-tvgarch-transition}
\end{equation}
with $t_{\mathrm c}:=1000(\zeta_1m_1+\zeta_2m_2)/(\zeta_1+\zeta_2)$,
which makes $G_t$ continuous in~$t$. A large return of either sign
raises the next variance, so large losses occur in clusters. The weight
$G_t\in(0,1)$ moves the variance from a high-volatility regime
($G_t\approx1$) to a low-volatility regime ($G_t\approx0$) near
$t=1000m_1$ and back near $t=1000m_2$, at the speeds $\zeta_1$
and $\zeta_2$. Let $\varepsilon_t$ be independent of $\Fc_t$ and
Student-$t_\eta$ distributed with unit variance, where $\eta$ is the
number of degrees of freedom. Given $\Fc_t$, $\xi_t$ has the distribution of
$\sigma_t\varepsilon_t$ conditioned on $[\xi_{\min},\xi_{\max}]$ with
$\xi_{\min}:=\log0.9$ and $\xi_{\max}:=\log1.1$. This truncation models
a daily price limit of $\pm10\%$, as in GARCH models of price-limited
stocks~\cite{wang2021pricelimits}.

We use the posterior means of the Student-$t$ model that
\cite{chen2014tvgarch} fits to daily S\&P~500 returns:
\[
\begin{aligned}
 a_0&=0.0297\times10^{-4}, & a_1&=0.0480, & b_1&=0.8726,\\
 a_0^\star&=0.0111\times10^{-4}, & a_1^\star&=0.0382,
 & b_1^\star&=0.0289,\\
 m_1&=0.4047, & m_2&=0.7280, & \eta&=10.5233,\\
 \zeta_1&=22.6482, & \zeta_2&=46.2823.&&
\end{aligned}
\]
We rename the GARCH parameters to avoid conflicts with the notation of
the main text and scale the intercepts by $10^{-4}$, since
\cite{chen2014tvgarch} uses returns in percent. The recursion starts at
\[
    \sigma_0^2
    =\frac{a_0+a_0^\star G_0}
    {1-a_1-b_1-(a_1^\star+b_1^\star)G_0},
\]
the stationary variance of the model with $G_t\equiv G_0$ and without a
price limit. The model thus combines published parameters with a price
limit; it is not fitted to a particular stock.

The loss is the relative price drop,
\begin{equation}
    Y_t:=[1-\exp(\xi_t)]_+=\bigl[(P_t-P_{t+1})/P_t\bigr]_+,
    \stepcounter{equation}\notag\label{eq:finance-loss}
\end{equation}
so $0\leq Y_t\leq c_{\max}:=0.10$. The candidate grid is
\[
\begin{aligned}
\Gamma=\{&0.007,0.010,0.015,0.020,0.022,0.023,\\
&0.024,0.026,0.028,0.030,0.035,0.040,\\
&0.050,0.060,0.070,0.080,0.090,0.096\}.
\end{aligned}
\]
Let $F_\eta$ be the distribution function of $\varepsilon_t$. Since
$Y_t>c$ if and only if $\xi_t<\log(1-c)$, for $0<c<c_{\max}$,
\begin{equation}
    p_t(c)=
    \frac{
    F_\eta\!\left(\frac{\log(1-c)}{\sigma_t}\right)
    -F_\eta\!\left(\frac{\xi_{\min}}{\sigma_t}\right)}{
    F_\eta\!\left(\frac{\xi_{\max}}{\sigma_t}\right)
    -F_\eta\!\left(\frac{\xi_{\min}}{\sigma_t}\right)}.
    \label{eq:finance-conditional-risk}
\end{equation}
The tests use only $Z_t(c)=\mathbf 1_{\{Y_t>c\}}$;
\eqref{eq:finance-conditional-risk} serves only to evaluate the
exceedance risks in Fig.~\ref{fig:certification-illustration}.

Let $\phi_t$ be the density of $\sigma_t\varepsilon_t$; it is
symmetric and nonincreasing on $[0,\infty)$. If
$\log(1-c)\leq-\xi_{\max}$, which holds in $\Gamma$ only for $c=0.096$,
then $\phi_t\leq\phi_t(\xi_{\max})$ on $[\xi_{\min},\log(1-c)]$ and
$\phi_t\geq\phi_t(\xi_{\max})$ on $[-\xi_{\max},\xi_{\max}]$. Hence,
\[
    p_t(c)
    =\frac{\int_{\xi_{\min}}^{\log(1-c)} \phi_t(z)\,dz}
          {\int_{\xi_{\min}}^{\xi_{\max}} \phi_t(z)\,dz}
    \leq\frac{\log(1-c)-\xi_{\min}}{2\xi_{\max}}=:\bar p(c).
\]
Thus, $p_t(0.096)\leq\bar p(0.096)=0.0233<\alpha$ for every $t$
and every history, and \eqref{eq:risk-bound-for-growth} holds for
$c=0.096$ without a bound on $\sigma_t$.

Panel~(c) of Fig.~\ref{fig:certification-illustration} tests
$c\in\{0.007,0.023,0.096\}$ with each fixed gain
$\lambda\in\{0.10,0.15,\ldots,0.80\}$, i.e., $\Lambda=\{\lambda\}$. The
three tests of a trial share the losses of 1000 days, and each point
averages $\min\{\tau(c),1000\}$ over the trials. The dotted bound
\eqref{eq:expected-certification-time} therefore uses $|\Lambda|=1$.
\section{A Sufficient Construction of the Conditional Risk Bound}
\label{app:conditional-risk-bound}

The following model conditions give sufficient choices
of $L_x,L_u$ and $\Delta_{t}$ for the bound
$d_t(c;\sigma,\tau)$ in Assumption~\ref{as:bounded-shift}.
Define the signed barrier shortfall
\begin{subequations}
\begin{equation}
    \ell(x,u,w)
    :=h(\hat f(x,u))-h(f(x,u,w)).
    \label{eq:signed-barrier-shortfall}
\end{equation}
Set $\widetilde Y_t:=\ell(x_t,u_t,w_t)$, so
$Y_t=[\widetilde Y_t]_+$.
For $c\geq0$,
$\{Y_t>c\}=\{\ell(x_t,u_t,w_t)>c\}$.
Let $\mu_t$ denote the conditional distribution of $w_t$
given $\Fc_t$, and define its total-variation distance
from $\mu_s$ by
\begin{equation}
    d_{\mathrm{TV}}(\mu_t,\mu_s)
    :=\sup_A|\mu_t(A)-\mu_s(A)|,
\label{eq:total-variation-distance}
\end{equation}
\end{subequations}
where $A$ ranges over measurable disturbance sets.

Suppose known constants $K_x,K_u,K_p\geq0$ satisfy
\begin{equation}
\begin{aligned}
    |\ell(x,u,w)-\ell(x',u',w)|
    \leq{}&K_x\|x-x'\|\\
          &+K_u\|u-u'\|
\end{aligned}
\label{eq:shortfall-regularity}
\end{equation}
for all operating points $(x,u),(x',u')$ and relevant
disturbances $w$.
Also suppose that, almost surely,
\begin{equation}
    \mu_s\!\left(\{w:a<\ell(x,u,w)\leq b\}\right)
    \leq K_p(b-a)
    \label{eq:shortfall-interval-bound}
\end{equation}
for every $s$, operating point $(x,u)$, and $a<b$.
Condition~\eqref{eq:shortfall-interval-bound} holds if
$\ell(x,u,w_s)$ has a conditional density bounded by $K_p$,
uniformly in $s$, $(x,u)$, and the observed history.
This condition is imposed on the signed shortfall,
since $Y_t=[\widetilde Y_t]_+$ can equal zero
with positive probability even when the signed
shortfall has a continuous density.

\begin{proposition}[Model-based conditional risk bound]
\label{prop:model-risk-bound}
Suppose \eqref{eq:shortfall-regularity} and
\eqref{eq:shortfall-interval-bound} hold.
Let $\Delta_{t}\geq0$ be known, $\Fc_t$-measurable
bounds such that, whenever $c_t=c\in\Gamma$ is deployed
using a certificate with monitoring period
$[\sigma,\tau)$, where $0\leq \sigma<\tau\leq t$,
\[
    d_{\mathrm{TV}}(\mu_t,\mu_s)\leq\Delta_{t}
    \quad\text{almost surely}
\]
for every integer $\sigma\leq s<\tau$.
Then $d_t(c;\sigma,\tau)$ in
\eqref{eq:candidate-shift-bound}, with
\begin{equation}
    L_x=K_pK_x,\qquad L_u=K_pK_u,
\label{eq:model-risk-constants}
\end{equation}
satisfies Assumption~\ref{as:bounded-shift}.
\end{proposition}

\begin{proof}
Fix a decision time $t$ at which $c_t=c\in\Gamma$
is deployed using a certificate with monitoring
period $[\sigma,\tau)$, so $u_t=u_t(c)$.
For any integer $\sigma\leq s<\tau$, set
\[
    \varepsilon_{t,s}
    :=K_x\|x_t-x_s\|+K_u\|u_t(c)-u_s\|.
\]
Condition~\eqref{eq:shortfall-regularity} gives
\[
    \{w:\ell(x_t,u_t(c),w)>c\}
    \subseteq
    \{w:\ell(x_s,u_s,w)>c-\varepsilon_{t,s}\}.
\]
Since
$p_t(c)=\mu_t\{w:\ell(x_t,u_t(c),w)>c\}$,
the total-variation and interval bounds give
\[
\begin{aligned}
    p_t(c)
    &\leq
    \mu_s\{w:\ell(x_s,u_s,w)>c-\varepsilon_{t,s}\}
    +\Delta_{t}\\
    &=p_s(c)
      +\mu_s\{w:c-\varepsilon_{t,s}
                 <\ell(x_s,u_s,w)\leq c\}
      +\Delta_{t}\\
    &\leq p_s(c)+K_p\varepsilon_{t,s}+\Delta_{t}.
\end{aligned}
\]
Substituting $\varepsilon_{t,s}$ and using
$L_x=K_pK_x$, $L_u=K_pK_u$ yields
\[
    \max_{\sigma\leq s<\tau}
    \{p_t(c)-p_s(c)\}
    \leq d_t(c;\sigma,\tau).
\]
This proves \eqref{eq:predictive-shift} for every
$s$ in the certificate's monitoring period.
\end{proof}

\subsection{Deterministic bounds for general dynamics}

Consider the general dynamics $f(x,u,w)$ and signed
barrier shortfall
\[
    \ell(x,u,w)
    =h(\hat f(x,u))-h(f(x,u,w)).
\]
All bounds below are required uniformly over the
relevant operating region and disturbance support.
Bounds involving conditional distributions must
also hold uniformly in time and almost surely
with respect to the observed history.

Suppose $h$ is $H$-Lipschitz and $f,\hat f$ satisfy
\[
\begin{aligned}
    \|f(x,u,w)-f(x',u',w)\|
    &\leq a_x\|x-x'\|+a_u\|u-u'\|,\\
    \|\hat f(x,u)-\hat f(x',u')\|
    &\leq
    \hat a_x\|x-x'\|+\hat a_u\|u-u'\|.
\end{aligned}
\]
Then \eqref{eq:shortfall-regularity} holds with
\[
    K_x=H(a_x+\hat a_x),
    \qquad
    K_u=H(a_u+\hat a_u).
\]
For differentiable models, sufficient values follow
from uniform Jacobian bounds on suitable convex
regions containing the operating points and their
predictions.
A known disturbance-amplitude bound specifies the
disturbance set over which these bounds are required.

The interval-probability bound requires additional
regularity. One sufficient construction uses a scalar
disturbance component.
Write $w=(v,\zeta)$, where $v$ takes values in a known
interval $I$.
Suppose the conditional density of $v_s$, given
$\zeta_s$ and $\Fc_s$, is bounded by a known constant
$M_v$, without specifying that density.

Assume also that, for every relevant $(x,u,\zeta)$,
the map $v\mapsto\ell(x,u,(v,\zeta))$ is continuously
differentiable on $I$ and satisfies
\[
    \left|
        \frac{\partial\ell(x,u,(v,\zeta))}
             {\partial v}
    \right|
    \geq m_\ell>0,
\]
where $m_\ell$ is known.
This map is strictly monotone on $I$.
Consequently, the set of $v$ satisfying
$a<\ell(x,u,(v,\zeta))\leq b$ has length at most
$(b-a)/m_\ell$.
Conditioning on $\zeta_s$ and then integrating gives
\[
    \mu_s\{w:a<\ell(x,u,w)\leq b\}
    \leq\frac{M_v}{m_\ell}(b-a).
\]
Thus, a sufficient choice in
\eqref{eq:shortfall-interval-bound} is
\[
    K_p=\frac{M_v}{m_\ell}.
\]
This construction permits nonlinear dependence on
the disturbance. It requires a disturbance component
whose effect on the barrier does not vanish over
the specified operating region.

Combining these bounds with
Proposition~\ref{prop:model-risk-bound} yields
\begin{equation*}
\begin{aligned}
    L_x&=
    \frac{M_vH}{m_\ell}(a_x+\hat a_x),\\
    L_u&=
    \frac{M_vH}{m_\ell}(a_u+\hat a_u).
\end{aligned}
\end{equation*}

To allow changing conditional disturbance
distributions, suppose a known constant $L_\mu\geq0$
satisfies
\begin{subequations}
\begin{equation}
    d_{\mathrm{TV}}(\mu_{k+1},\mu_k)
    \leq L_\mu
\label{eq:distribution-drift-rate}
\end{equation}
almost surely for every $k$.
The triangle inequality then gives
\begin{equation}
    d_{\mathrm{TV}}(\mu_t,\mu_s)
    \leq\min\{1,L_\mu(t-s)\}.
\label{eq:distribution-drift-distance}
\end{equation}
Hence, for each $\sigma\leq s<\tau$, one may take the uniform bound
\begin{equation}
\Delta_{t}
    =\min\{1,L_\mu(t-\sigma)\}.
\label{eq:distribution-drift-bound}
\end{equation}
\end{subequations}
This bound is available before observing $Y_t$
and allows a nonzero increase in conditional risk.
Since \eqref{eq:predictive-shift} is required for each $s$ separately, one may also take the smaller value $\Delta_t=\min\{1,L_\mu(t-s)\}$ at each $\sigma\leq s<\tau$. Appendix~\ref{app:unicycle-risk-bound} uses this choice.

For example, a sufficient value of $L_\mu$ can be
obtained when the unknown conditional distributions
have densities $\varphi_k$ supported on a known set
$\mathcal W$ of finite volume and satisfy
\[
    |\varphi_{k+1}(w)-\varphi_k(w)|
    \leq B_\mu,
    \qquad w\in\mathcal W,
\]
with a known constant $B_\mu$.
Indeed,
\[
\begin{aligned}
    d_{\mathrm{TV}}(\mu_{k+1},\mu_k)
    &=\frac12\int_{\mathcal W}
        |\varphi_{k+1}(w)-\varphi_k(w)|\,dw\\
    &\leq
    \frac12\operatorname{vol}(\mathcal W)B_\mu.
\end{aligned}
\]
Thus,
\[
    L_\mu
    =\frac12\operatorname{vol}(\mathcal W)B_\mu
\]
is sufficient.

\label{app:general-dynamics-bounds}
\subsection{Risk-Increase Terms for the Unicycle Examples}
\label{app:unicycle-risk-bound}

In this subsection we construct the risk-increase bounds for the Corridor
and Maze in Appendix~\ref{app:sim-scenarios}. Each certified variant $\mathrm V\in\{\mathrm U,\Delta,\mathrm L,\mathrm C\}$
of Section~\ref{sec:empirical-metrics} replaces the bound $d_t$ defined in \eqref{eq:candidate-shift-bound} by the
largest risk-increase term over the monitoring period,
\begin{equation}
 d_t(c;\sigma,\tau)=d_t^{\mathrm V}(c;\sigma,\tau),
\label{eq:unicycle-pairwise-bound}
\end{equation}
the maximum over $\sigma\leq s<\tau$ in \eqref{eq:unicycle-vu-term},
\eqref{eq:unicycle-vdelta-term}, \eqref{eq:unicycle-vl-term} or
\eqref{eq:unicycle-candidate-transfer},
and each risk-increase term is $\Fc_t$-measurable. The proofs of
Theorems~\ref{thm:certified-deployment} and~\ref{prop:repeated-deployment}
use Assumption~\ref{as:bounded-shift} only through
\eqref{eq:predictive-shift}.

\paragraph{Drift term}
The disturbance distribution changes with $t$ only through the weight
$\beta_t$ of the magnitude distribution $\mu_t^A$ defined in
\eqref{eq:empirical-disturbance} of Appendix~\ref{app:sim-scenarios}, and the
supports of its components $\mathrm U_1$ and $\mathrm U_2$ overlap only at
$c_{\max}/2$. Hence, with $d_{\mathrm{TV}}$ in \eqref{eq:total-variation-distance},
\begin{equation}
 d_{\mathrm{TV}}(\mu_t,\mu_s)=d_{\mathrm{TV}}(\mu_t^A,\mu_s^A)
 =|\beta_t-\beta_s|.
\stepcounter{equation}\notag\label{eq:unicycle-drift-term}
\end{equation}
Since $\sin$ is $1$-Lipschitz, the parameter $L_\mu$ of $\beta_t$ satisfies
\eqref{eq:distribution-drift-rate}, so \eqref{eq:distribution-drift-distance} gives
$|\beta_t-\beta_s|\leq\min\{1,L_\mu(t-s)\}$, the drift term in
\eqref{eq:distribution-drift-bound} with the choice for each $s$ stated after it.

\paragraph{Clearance and shortfall}
Vector norms are Euclidean. In the Corridor, cones have centers $o_j$ and
radii $\rho_j$, and walls lie at $z_2=0$ and
$z_2=2$. The clearance of a disk of radius
$\rho_{\mathrm{body}}$ centered at $z$ is
\begin{equation}
\begin{aligned}
    \psi(z):=\min\bigl\{
        &\min_j(\|z-o_j\|-\rho_j),\\
        &z_2,\ 2-z_2
    \bigr\}-\rho_{\mathrm{body}}.
\end{aligned}
\label{eq:corridor-example-clearance}
\end{equation}
In the Maze, interior walls are closed rectangles
$\mathcal O_k$. Boundary walls have points
$z_m^\partial$ and inward unit normals
$n_m^\partial$, giving
\begin{equation}
\begin{aligned}
    \psi(z):=\min\bigl\{
        &\min_j(\|z-o_j\|-\rho_j),\\
        &\min_k\operatorname{sdist}(z,\mathcal O_k),\\
        &\min_m(n_m^\partial)^\top(z-z_m^\partial)
    \bigr\}-\rho_{\mathrm{body}}.
\end{aligned}
\label{eq:maze-example-clearance}
\end{equation}
Here, $\operatorname{sdist}$ is the distance to
the obstacle boundary, positive outside and
negative inside.
Distances and affine terms with unit normals are $1$-Lipschitz. For
$z\notin\mathcal O_k$ and $z'\in\mathcal O_k$, a point $\tilde z\in\partial\mathcal O_k$
on the segment $[z,z']$ gives
$0\leq\operatorname{sdist}(z,\mathcal O_k)-\operatorname{sdist}(z',\mathcal O_k)
\leq\|z-\tilde z\|+\|\tilde z-z'\|=\|z-z'\|$.
Each clearance term is $1$-Lipschitz, so
\begin{equation}
    |\psi(z)-\psi(z')|\leq\|z-z'\|.
    \label{eq:unicycle-clearance-lipschitz}
\end{equation}
With the nose point $q$ and the nose distance $l$ defined in
Appendix~\ref{app:sim-scenarios}, let
$z(x,u):=q(\hat f(x,u))$ be the predicted nose point.
The disturbance changes position but not heading
or speed, so $q(f(x,u,w))=z(x,u)+w$ with the transition $f$ defined in \eqref{eq:empirical-transition}. Thus, the braking term in
\eqref{eq:unicycle-barriers} cancels, and
the shortfall in \eqref{eq:signed-barrier-shortfall} is
\begin{equation}
    \ell(x,u,w)
    =\psi(z(x,u))-\psi(z(x,u)+w).
    \label{eq:unicycle-shortfall}
\end{equation}
Consequently,
$Y_t=[\ell(x_t,u_t,w_t)]_+\leq\|w_t\|\leq c_{\max}$, the support condition of Assumption~\ref{as:conservative-margin}.

For a feasible candidate $c$ with the filtered input $u_t(c)$ defined in \eqref{eq:cbf-safety-filter}, write
$z_t(c):=z(x_t,u_t(c))$.
At a past time $s$, use $z_s:=z(x_s,u_s)$,
where $u_s$ is the input actually applied.

\paragraph{Exceedance fraction of one feature}
Each term in the minima of $\psi$ belongs to one \emph{feature}: a cone, a
wall or a rectangle $\mathcal O_k$. Let $\psi_i$ be the term of feature $i$
minus $\rho_{\mathrm{body}}$, so that $\psi=\min_i\psi_i$. Near $z$, a feature
is either \emph{circular}, with $\psi_i(z)=\|z-o\|-R$ for a center $o$ and an
\emph{effective radius} $R$, or \emph{affine} (a wall or a rectangle face),
with $R:=\infty$. A cone has $R=\rho_j+\rho_{\mathrm{body}}$, and a rectangle
whose nearest point $o$ to $z$ is a corner has $R=\rho_{\mathrm{body}}$. For a
circular feature with clearance $g=\psi_i(z)$ and $\|w\|=A$, the law of
cosines gives $\psi_i(z)-\psi_i(z+w)>c$ if and only if the angle between $w$
and $o-z$ has cosine larger than $c/A+(A^2-c^2)/(2(g+R)A)$. Since the
disturbance has a uniform direction given its magnitude, the fraction of directions with this property is the
\emph{directional exceedance fraction}
\begin{equation}
 P_R(g,c,A):=\frac1\pi\arccos\Bigl(\frac cA+\frac{A^2-c^2}{2(g+R)A}\Bigr)
\label{eq:unicycle-radial-risk}
\end{equation}
for $c<A\leq2(g+R)-c$, where the argument lies in $[0,1]$, and
$P_R(g,c,A):=0$ otherwise. An affine feature gives the limit $R\to\infty$,
which does not depend on $g$. Its value at $A=c_{\max}$ is the
\emph{fraction bound} $\bar P(c)$,
\begin{equation}
 P_\infty(c,A):=\frac1\pi\arccos\frac cA,\qquad
 \bar P(c):=P_\infty(c,c_{\max}),
\label{eq:unicycle-affine-risk}
\end{equation}
for $A>c$, and $P_\infty(c,A):=0$ otherwise. For
$c\in\Gamma\subseteq[0,c_{\max})$ (Section~\ref{subsec:problem-statement}),
$\bar P(c)\in(0,1/2]$. The argument in \eqref{eq:unicycle-radial-risk}
decreases in $g+R$ and is at least $c/A$, and $P_\infty(c,A)$ increases with
$A$. Hence $P_R$ increases with $g$ and $R$, and
$0\leq P_R(g,c,A)\leq\bar P(c)$ for $A\leq c_{\max}$.

\paragraph{Sensitivity bound}
If $h(x)\geq0$ and $u\in\Uc(x,c)$, the tightened
CBF condition \eqref{eq:tightened-cbf-condition} and \eqref{eq:unicycle-barriers}
give $\psi(z(x,u))\geq l$.
Set $g_{\min}:=l$ and, with $\mathcal R$ the set
of finite effective radii, define
\begin{equation}
    \varrho_{\min}:=g_{\min}+\min\mathcal R.
    \label{eq:unicycle-minimum-distance}
\end{equation}
Both scenarios satisfy $c_{\max}<\varrho_{\min}$,
with $\varrho_{\min}=0.675$\,m in the Corridor
and $0.37$\,m in the Maze.

For $g\geq g_{\min}$, $R\in\mathcal R$ and $c<A\leq c_{\max}$,
\begin{equation}
    \partial_gP_R(g,c,A)
    =
    \frac{\sqrt{A^2-c^2}}
    {\pi(g+R)\sqrt{(2(g+R)-c)^2-A^2}}.
    \label{eq:unicycle-radial-derivative}
\end{equation}
The denominator is positive since $A+c<2c_{\max}<2\varrho_{\min}\leq2(g+R)$.
This derivative decreases with $g+R$ and increases
with $A$. Therefore,
\begin{equation}
    L_{\mathrm{geom}}(c):=
    \frac{\sqrt{c_{\max}^2-c^2}}
    {\pi\varrho_{\min}
    \sqrt{(2\varrho_{\min}-c)^2-c_{\max}^2}}
    \label{eq:unicycle-clearance-sensitivity}
\end{equation}
bounds the sensitivity to clearance, uniformly
over $g\geq g_{\min}$, $A\leq c_{\max}$, and
$R\in\mathcal R\cup\{\infty\}$. Hence $L_{\mathrm{geom}}(c)$ is a Lipschitz constant of $P_R(\cdot,c,A)$ on $[g_{\min},\infty)$.
Let $L_{\mathrm{geom}}:=\max_{c\in\Gamma}L_{\mathrm{geom}}(c)$.

\paragraph{Geometric check and comparison}
A geometric check ensures that one feature determines $\psi$ near the
predicted nose point $z_t(c)$. The \emph{nearest feature} $i_z$ at $z$ is the smallest
index $i$ that minimizes $\psi_i(z)$. If
\begin{equation}
 \min_{i\neq i_z}\psi_i(z)\geq\psi(z)+2c_{\max},
\label{eq:unicycle-feature-check}
\end{equation}
then $\psi(z+w)=\psi_{i_z}(z+w)$ for every $\|w\|\leq c_{\max}$, since each
$\psi_i$ is $1$-Lipschitz. If $i_z$ is a rectangle
$[\underline z_1,\bar z_1]\times[\underline z_2,\bar z_2]$, the
\emph{rectangle condition} requires the disk $\{z+w:\|w\|\leq c_{\max}\}$ to
lie in one exterior face or corner region, given in coordinates in Section~1.1 of~\cite{trakas2026ablation}, so that the feature
is affine or circular on the disk. The \emph{geometric check} for $c$ passes
at $(t,s)$ if $\Uc(x_t,c)\neq\emptyset$, \eqref{eq:unicycle-feature-check} and
the rectangle condition hold at $z=z_t(c)$, and
$\psi(z_t(c)),\psi(z_s)\geq g_{\min}$. The \emph{uniform geometric check}
passes at $(t,s)$ if the geometric check for every $c'\in\Gamma$ passes there.

If the geometric check for $c$ passes at $(t,s)$, let $g_t(c):=\psi(z_t(c))$
and $g_s:=\psi(z_s)$, and let $R_t(c)$ and $R_s$ be the effective radii of
the nearest features at $z_t(c)$ and $z_s$, where a rectangle at $z_s$ is
treated as circular with $R_s=\rho_{\mathrm{body}}$ (justified after
\eqref{eq:unicycle-one-sided-comparison}). Since $P_R$ depends on $(g,R)$
only through $g+R$ and increases with it, $P_{R_t(c)}(g,c,A)-P_{R_s}(g,c,A)\leq0$ if
$R_t(c)\leq R_s$. Otherwise this difference is the integral of
\eqref{eq:unicycle-radial-derivative} over $[g+R_s,g+R_t(c))$, an improper
integral if $R_t(c)=\infty$, which decreases in $g$ and increases in $A$. For
$g\geq g_{\min}$ and $A\leq c_{\max}$, it is therefore at most the
\emph{feature-change term}
\begin{equation}
\begin{aligned}
 \eta_{t,s}(c):=\bigl[&P_{R_t(c)}(g_{\min},c,c_{\max})\\
                      &-P_{R_s}(g_{\min},c,c_{\max})\bigr]_+.
\end{aligned}
\label{eq:unicycle-curvature-correction}
\end{equation}
With $P_t(A):=P_{R_t(c)}(g_t(c),c,A)$ and $P_s(A):=P_{R_s}(g_s,\allowbreak c,\allowbreak A)$,
\eqref{eq:unicycle-shortfall} gives
\begin{equation}
 p_t(c)=\int P_t\,d\mu_t^A,\qquad p_s(c)\geq\int P_s\,d\mu_s^A.
\label{eq:unicycle-one-sided-comparison}
\end{equation}
The equality holds because the nearest feature at $z_t(c)$ determines $\psi$
on the disk. The inequality holds because $\psi\leq\psi_i$ with equality at
$z_s$ for the nearest feature $i$ at $z_s$. For a rectangle $\mathcal O_k$
with nearest point $o$ to $z_s$, moreover,
$\psi_i(z_s)=\|z_s-o\|-\rho_{\mathrm{body}}$ since
$\psi(z_s)\geq g_{\min}>0$ places $z_s$ outside $\mathcal O_k$, and
$\psi_i(z_s+w)\leq\|z_s+w-o\|-\rho_{\mathrm{body}}$ since $o\in\partial\mathcal O_k$
gives $\operatorname{sdist}(z_s+w,\mathcal O_k)\leq\|z_s+w-o\|$. The \emph{clearance
term} is
\begin{equation}
 G_{t,s}(c):=L_{\mathrm{geom}}(c)[g_t(c)-g_s]_+.
\stepcounter{equation}\notag\label{eq:unicycle-clearance-term}
\end{equation}

\paragraph{Lipschitz constants of the clearance term}
On a convex operating region containing the
current and past state-input pairs, uniform
Jacobian bounds $K_x^z$ on $\|\partial_xz\|$ and $K_u^z$ on
$\|\partial_uz\|$ (induced norms) give
\begin{equation}
\begin{aligned}
    \|z(x,u)-z(x',u')\|
    \leq{}&K_x^z\|x-x'\|\\
          &+K_u^z\|u-u'\|.
\end{aligned}
\label{eq:unicycle-tip-regularity}
\end{equation}
Let $\bar v$ bound absolute speed throughout
a sampling interval of length $T_s$.
For the Corridor, sufficient constants are
\begin{equation}
\begin{aligned}
    K_x^z&=\sqrt{1+(l+\bar vT_s)^2},\\
    K_u^z&=\sqrt{T_s^2+(lT_s+\bar vT_s^2/2)^2},
\end{aligned}
\label{eq:unicycle-first-order-constants}
\end{equation}
and, for the Maze,
\begin{equation}
\begin{aligned}
    K_x^z&=
    \sqrt{1+(l+\bar vT_s)^2+T_s^2
          +(lT_s+\bar vT_s^2/2)^2},\\
    K_u^z&=
    \sqrt{T_s^4/4+(lT_s^2/2+\bar vT_s^3/6)^2}.
\end{aligned}
\label{eq:unicycle-second-order-constants}
\end{equation}
In the Corridor, $\bar v$ follows from the input
limits.
In the Maze, speed varies linearly between its
current and predicted next values, both constrained
by the limits in Table~\ref{tab:sim-bindings}.
The Jacobian bounds behind \eqref{eq:unicycle-first-order-constants} and
\eqref{eq:unicycle-second-order-constants} are derived in Section~1.3 of the supplementary ablation report~\cite{trakas2026ablation}.
The bounds also hold for the numerical predictor:
its Gauss--Legendre quadrature has positive
weights and integrates the relevant polynomial
bounds exactly.
\paragraph{Geometric construction}
We construct the risk-increase bound directly
from the obstacle geometry and the conditional
uniformity of the disturbance direction.
This provides an alternative to
Proposition~\ref{prop:model-risk-bound} and does
not require its density bound $K_p$.

The geometric comparison separates changes in
clearance from changes in the effective radius:
\[
    P_t(A)-P_s(A)
    \leq G_{t,s}(c)+\eta_{t,s}(c).
\]
The first term bounds the effect of changing
clearance while holding the effective radius
fixed. The second accounts for the remaining
change in geometry.
Define
\begin{equation}
    B_{t,s}(c):=
    \min\{\bar P(c),G_{t,s}(c)+\eta_{t,s}(c)\}.
    \stepcounter{equation}\notag\label{eq:unicycle-common-mass-bound}
\end{equation}
Splitting $P_t(A)-P_s(A)$ at
$P_{R_t(c)}(g_s,c,A)$ and using the sensitivity
bound and the feature-change term \eqref{eq:unicycle-clearance-sensitivity} and \eqref{eq:unicycle-curvature-correction}, and $P_t(A)-P_s(A)\leq P_t(A)\leq\bar P(c)$, gives
\begin{equation}
    P_t(A)-P_s(A)\leq B_{t,s}(c),
    \qquad 0\leq A\leq c_{\max}.
    \label{eq:unicycle-geometric-comparison}
\end{equation}

Combining the clearance and predicted-position
bounds gives
\[
\begin{aligned}
    G_{t,s}(c)
    \leq{}
    L_x^{\mathrm{geom}}\|x_t-x_s\|
    +L_u^{\mathrm{geom}}\|u_t(c)-u_s\|
\end{aligned}
\]
where
\begin{equation}
    L_x^{\mathrm{geom}}:=L_{\mathrm{geom}}K_x^z,
    \qquad
    L_u^{\mathrm{geom}}:=L_{\mathrm{geom}}K_u^z.
    \label{eq:unicycle-risk-constants}
\end{equation}
These constants bound only the clearance
contribution. The feature-change term
$\eta_{t,s}(c)$ is included separately.

For each past time $s$, distribution changes
are bounded by
\[
    d_{\mathrm{TV}}(\mu_t^A,\mu_s^A)
    \leq\min\{1,L_\mu(t-s)\}.
\]
We retain this bound inside the maximum
over the certificate's monitoring period.

\paragraph{Certificate bounds}
When the uniform geometric check passes, define
the \emph{uniform feature-change term}
\begin{equation}
    \bar\eta_{t,s}:=
    \max_{c'\in\Gamma}\eta_{t,s}(c').
    \stepcounter{equation}\notag\label{eq:unicycle-curvature-terms}
\end{equation}
For a certificate with monitoring period
$[\sigma,\tau)$, V-U uses
\begin{equation}
\begin{aligned}
&d_t^{\mathrm U}(c;\sigma,\tau)
:=\min\Bigl\{1,\max_{\sigma\leq s<\tau}\bigl[
L_x^{\mathrm{geom}}\|x_t-x_s\|\\
&\quad+L_u^{\mathrm{geom}}\|u_t(c)-u_s\|
+\bar\eta_{t,s}+\Delta_t\bigr]\Bigr\}.
\end{aligned}
\label{eq:unicycle-vu-term}
\end{equation}
V-$\Delta$ replaces the uniform feature-change term
by the candidate-specific feature-change term:
\begin{equation}
\begin{aligned}
&d_t^\Delta(c;\sigma,\tau)
:=\min\Bigl\{1,\max_{\sigma\leq s<\tau}\bigl[
L_x^{\mathrm{geom}}\|x_t-x_s\|\\
&\quad+L_u^{\mathrm{geom}}\|u_t(c)-u_s\|
+\eta_{t,s}(c)+\Delta_t\bigr]\Bigr\}.
\end{aligned}
\label{eq:unicycle-vdelta-term}
\end{equation}
V-L uses the candidate-specific clearance
term directly:
\begin{equation}
\begin{aligned}
&d_t^{\mathrm L}(c;\sigma,\tau)
:=\min\Bigl\{1,\max_{\sigma\leq s<\tau}\bigl[
\min\{\bar P(c),G_{t,s}(c)\}\\
&\quad+\bar\eta_{t,s}+\Delta_t\bigr]\Bigr\}.
\end{aligned}
\label{eq:unicycle-vl-term}
\end{equation}
V-C combines both refinements and bounds the
effect of distribution changes using
$\bar P(c)$:
\begin{equation}
d_t^{\mathrm C}(c;\sigma,\tau)
:=\max_{\sigma\leq s<\tau}\bigl[
(1-\Delta_t)B_{t,s}(c)
+\Delta_t\bar P(c)\bigr].
\label{eq:unicycle-candidate-transfer}
\end{equation}

V-U and V-L require the uniform geometric
check throughout the monitoring period.
V-$\Delta$ and V-C require the check for $c$.
If a required check fails, the corresponding
certificate bound is set to $1$.

To verify the V-C bound, fix $\sigma\leq s<\tau$
and let
$\varepsilon:=d_{\mathrm{TV}}(\mu_t^A,\mu_s^A)
\leq\Delta_t$.
The common part of the magnitude distributions
has mass $1-\varepsilon$.
On this part, the increase is at most $B_{t,s}(c)$ by \eqref{eq:unicycle-geometric-comparison}. The
remaining current mass contributes at most
$\varepsilon\bar P(c)$. The remaining past mass enters with a nonpositive sign, since $P_s\geq0$.
Thus, when $u_t=u_t(c)$,
\eqref{eq:unicycle-one-sided-comparison} gives
\[
\begin{aligned}
    p_t(c)-p_s(c)
    &\leq(1-\varepsilon)B_{t,s}(c)
         +\varepsilon\bar P(c)\\
    &\leq(1-\Delta_t)B_{t,s}(c)
         +\Delta_t\bar P(c)\\
    &\leq d_t^{\mathrm C}(c;\sigma,\tau),
\end{aligned}
\]
where the second inequality uses
$B_{t,s}(c)\leq\bar P(c)$.
This holds for every $s$ in the monitoring period.
The distributions are used only in the proof.
Computing the bound does not require them.

\paragraph{Ordering and validity}
For the same history and monitoring period
$[\sigma,\tau)$, the certificate bounds satisfy
\begin{equation}
\begin{aligned}
    d_t^{\mathrm C}(c;\sigma,\tau)
    &\leq d_t^{\mathrm L}(c;\sigma,\tau)
    \leq d_t^{\mathrm U}(c;\sigma,\tau),\\
    d_t^{\mathrm C}(c;\sigma,\tau)
    &\leq d_t^\Delta(c;\sigma,\tau)
    \leq d_t^{\mathrm U}(c;\sigma,\tau).
\end{aligned}
\label{eq:unicycle-bound-ordering}
\end{equation}
These inequalities follow from
$\eta_{t,s}(c)\leq\bar\eta_{t,s}$,
the state-input bound on $G_{t,s}(c)$, $\Delta_t\leq1$
and $B_{t,s}(c)\leq\bar P(c)\leq1$.
Maximizing over the same monitoring period
preserves the ordering.
The ordering also holds when checks fail:
failure of the geometric check for $c$ sets
all four bounds to $1$, while failure of only
the uniform check sets the V-U and V-L bounds
to $1$.

The geometric comparison and distribution-drift
bound establish, whenever $u_t=u_t(c)$,
\[
    p_t(c)\leq p_s(c)+d_t^{\mathrm V}(c;\sigma,\tau),
    \qquad \sigma\leq s<\tau,
\]
for every variant
$\mathrm V\in\{\mathrm U,\Delta,\mathrm L,\mathrm C\}$.
Thus, these constructions satisfy
\eqref{eq:predictive-shift}.
A bound set to $1$ prevents deployment using
that certificate at the current decision time.

The ordering compares bounds on the same history, i.e.,
it does not imply an ordering of margins selected
along different closed-loop trajectories.
\endgroup

\section{Empirical Metrics and Checks}
\label{app:empirical-validation}

We describe how the recorded trajectories from
Section~\ref{sec:empirical-evaluation} are used
to assess the statistical guarantees and measure
performance over the simulated horizon.
Exceedance-risk evaluation uses the full simulated disturbance
distribution, which is not supplied to the controller.
Trajectory $k$ executes $N_k\leq T$ transitions, with the decision horizon $T$ of
Corollary~\ref{cor:finite-horizon-safety}, and the index $k$ marks its recorded
quantities, e.g., $p_{k,t}(c)$, $c_{k,t}$ and $\tau_{k,r}(c)$.

\subsection{False-Certification and Deployment Events}
\label{app:empirical-risk-classification}
\paragraph{Exceedance-risk evaluation}
At each recorded decision time $t$, we hold the state, applied input $u_t$, and observed history fixed.
We propagate disturbances $w$ through the true system $f(x_t,u_t,w)$ and
integrate the exceedance indicator of Section~\ref{subsec:false-certification}
numerically over the conditional distribution $\mu_t$ of $w_t$ in
Appendix~\ref{app:conditional-risk-bound}. Since $Y_t>c$ holds exactly when the signed barrier
shortfall $\ell$ in \eqref{eq:signed-barrier-shortfall} satisfies $\ell(x_t,u_t,w_t)>c$, this
approximates the exceedance risk $p_t(c)$ in \eqref{eq:conditional-exceedance} by the
\emph{estimated exceedance risk}
\begin{equation}
 \widehat p_t(c)\approx\int\mathbf 1_{\{\ell(x_t,u_t,w)>c\}}\,\mu_t(dw)=p_t(c)
\label{eq:empirical-risk-estimate}
\end{equation}
with the \emph{numerical risk bounds}
\begin{equation}
 p^-_t(c)\leq p_t(c)\leq p^+_t(c).
\stepcounter{equation}\notag\label{eq:empirical-risk-brackets}
\end{equation}
The computation of $\widehat p_t(c)$ and of its numerical risk bounds is given in
the supplementary ablation report~\cite{trakas2026ablation}.

\paragraph{Trajectory events}
With the starting times $\sigma_r$ of the tests in
Section~\ref{subsec:distribution-shift}, the trajectory events are
\begin{equation}
\begin{aligned}
 \mathcal A_{\mathrm{cert},k}:=\{\exists c\in\Gamma,\ r\geq1:\ &\tau_{k,r}(c)\leq N_k,\\
 &\min_{\sigma_r\leq s<\tau_{k,r}(c)}p_{k,s}(c)\geq\alpha\},
\end{aligned}
\label{eq:empirical-joint-certificate-event}
\end{equation}
and
\begin{equation}
 \mathcal A_{\mathrm{dep},k}:=
 \{\exists\,0\leq t<N_k:p_{k,t}(c_{k,t})>\alpha_0\}.
\label{eq:empirical-joint-deployment-event}
\end{equation}
The \emph{false-certification event} $\mathcal A_{\mathrm{cert},k}$ occurs if a test of
trajectory $k$ certifies a candidate $c$ whose exceedance risk stays at least $\alpha$ over the
monitoring period, and the \emph{deployment event} $\mathcal A_{\mathrm{dep},k}$ occurs if $p_{k,t}(c_{k,t})>\alpha_0$ at some
decision. For finite $K$, the proof of
Theorem~\ref{prop:repeated-deployment} bounds the probability of
$\mathcal A_{\mathrm{cert},k}$ by $\delta$, and the theorem that of
$\mathcal A_{\mathrm{dep},k}$. For $K=\infty$,
Theorems~\ref{thm:false-certification} and~\ref{thm:certified-deployment} bound
them by the error level $\delta_1$ of Section~\ref{subsec:distribution-shift}. The
bounds on $\mathcal A_{\mathrm{dep},k}$ require
Assumptions~\ref{as:conservative-margin} and~\ref{as:bounded-shift}.

\paragraph{Event frequencies}
An event is \emph{confirmed} if it occurs with every exceedance risk replaced by
its lower bound $p^-_{k,t}$, and \emph{possible} if it occurs with the upper
bounds $p^+_{k,t}$. Since both events are increasing in the exceedance risks, a confirmed
event occurs for all exceedance risks within $[p^-_{k,t},p^+_{k,t}]$, and an event that is not
possible occurs for none. Over the $N_{\mathrm{MC}}$ trajectories of a
method, scenario and setting, the \emph{trajectory frequency} of a trajectory event
$\mathcal A$ is
\begin{equation}
 \widehat q(\mathcal A):=\frac1{N_{\mathrm{MC}}}
 \sum_{k=1}^{N_{\mathrm{MC}}}\mathbf1_{\mathcal A_k}.
\label{eq:empirical-joint-frequencies}
\end{equation}
For \eqref{eq:empirical-joint-certificate-event} and
\eqref{eq:empirical-joint-deployment-event}, it is the
\emph{false-certification frequency} and the \emph{deployment-event frequency}.
They are denoted $\widehat q_{\mathrm{cert}}$ and
$\widehat q_{\mathrm{dep}}$.

\paragraph{Upper confidence bounds}
A trajectory frequency estimates the probability $q(\mathcal A):=\Prob(\mathcal A_k)$ from
$N_{\mathrm{MC}}$ independent trajectories. Its $95\%$ \emph{upper confidence bound}
$\overline q(\mathcal A)$ is the Clopper--Pearson bound~\cite{clopper1934use} computed from the
number $n_{\mathcal A}$ of trajectories with a possible event, so that
\begin{equation}
 \Prob\bigl(q(\mathcal A)\leq\overline q(\mathcal A)\bigr)\geq0.95.
\label{eq:empirical-binomial-upper}
\end{equation}
For example, zero possible events in $200$ trajectories gives an upper confidence bound of $1.5\%$.
An event frequency \emph{meets its level} if its upper confidence bound is at most
the level of the guarantee it checks, $\delta$ for finite $K$ and $\delta_1$ for
$K=\infty$.

\subsection{Performance and Safety Metrics}
\label{app:empirical-metrics}
Let $I_{\mathrm{goal},k}:=1$ if trajectory $k$ reaches the goal and $I_{\mathrm{goal},k}:=0$ otherwise, and
let $\mathcal K_+:=\{k:N_k\geq1\}$. For a decision indicator
$I_{k,t}\in\{0,1\}$ and the sampling period $T_s$ in Table~\ref{tab:sim-bindings},
define
\begin{equation}
\begin{aligned}
 T_{\mathrm{mis},k}&:=T_s[I_{\mathrm{goal},k}N_k+(1-I_{\mathrm{goal},k})T],\\
 m_{c,k}&:=\frac1{N_k}\sum_{t<N_k}c_{k,t},\qquad
 q_k^{+}:=\frac1{N_k}\sum_{t<N_k}I_{k,t},\\
 \widehat T_{\mathrm{mis}}&:=\frac1{N_{\mathrm{MC}}}
                         \sum_{k=1}^{N_{\mathrm{MC}}}T_{\mathrm{mis},k},\\
 \widehat m_c&:=\frac1{|\mathcal K_+|}\sum_{k\in\mathcal K_+}m_{c,k},\\
 \widehat q^{+}&:=\frac1{|\mathcal K_+|}
                         \sum_{k\in\mathcal K_+}q_k^{+}.
\end{aligned}
\label{eq:empirical-performance-definitions}
\end{equation}
They give the mean \emph{mission time} $\widehat T_{\mathrm{mis}}$, in which a trajectory that
does not reach the goal counts with the horizon $T$, the
\emph{normalized mean robustness margin} $\widehat m_c/c_{\max}$ and the
\emph{decision fraction} $\widehat q^+$ of $I$, the mean fraction of decisions with
$I_{k,t}=1$. The trajectory frequency \eqref{eq:empirical-joint-frequencies} of
$\{\exists t<N_k:I_{k,t}=1\}$ is denoted $\widehat q$. The \emph{CBF-violation fraction}
$\widehat q^+_{\mathrm{CBF}}$ and the \emph{safe-set-violation fraction}
$\widehat q^+_{\mathrm{unsafe}}$ are the decision fractions of a violation of the true barrier
condition in \eqref{eq:true-cbf-condition} and of leaving the safe set $\Cc$,
\begin{equation}
\begin{aligned}
 I_{\mathrm{CBF},k,t}&:=\mathbf1_{\{h(x_{k,t+1})<(1-\gamma)h(x_{k,t})\}},\\
 I_{\mathrm{unsafe},k,t}&:=\mathbf1_{\{h(x_{k,t+1})<0\}}.
\end{aligned}
\label{eq:empirical-performance-events}
\end{equation}
The estimated peak exceedance risk and the minimum barrier value are
\begin{equation}
\begin{aligned}
 \widehat p_k^{\max}
 &:=\max_{0\leq t<N_k,\ c_{k,t}<c_{\max}}\widehat p_{k,t}(c_{k,t}),\\
 h_k^{\min}&:=\min_{0\leq t\leq N_k}h(x_{k,t}),
\end{aligned}
\label{eq:empirical-peak-barrier}
\end{equation}
with $\widehat p_k^{\max}$ undefined if $c_{k,t}=c_{\max}$ for all $t$, and $h_k^{\min}<0$ exactly
when $x_{k,t}\notin\Cc$ for some $t$. Their means over the trajectories where they are defined
are the \emph{mean estimated peak exceedance risk} $\widehat m_{p^{\max}}$ and the \emph{mean
minimum barrier value} $\widehat m_{h^{\min}}$. The secondary metrics are defined in the supplementary ablation report~\cite{trakas2026ablation}.

\subsection{Empirical Checks of the Assumptions and the Tightened Condition}
\label{app:empirical-assumption-checks}
Assumption~\ref{as:conservative-margin} requires that $\Uc(x_t,c_{\max})$ is nonempty and
that the input $u_t(c_{\max})$ ensures the support condition $Y_t\leq c_{\max}$. At every
recorded decision, we check both conditions and the tightened condition
\eqref{eq:tightened-cbf-condition} for the applied input, that is, $u_t\in\Uc(x_t,c_t)$.

Assumption~\ref{as:bounded-shift} requires, through \eqref{eq:predictive-shift}, that the
exceedance risk $p_t(c)$ exceeds $p_s(c)$ by at most $d_t(c;\sigma,\tau)$ at every time $s$ of
the monitoring period $[\sigma,\tau)$. A decision is \emph{checked} if $c_t<c_{\max}$. At a
checked decision, $c_t$ is certified by the test with the supporting test index $r_t$ in
\eqref{eq:certificate-selection}, with monitoring period $[\sigma_{r_t},\tau_{r_t}(c_t))$. We check
\eqref{eq:predictive-shift} for this test with the estimated exceedance risk $\widehat p$ in place
of $p$, through the \emph{bound excesses}
\begin{equation}
\begin{aligned}
 e_{\mathrm{pw},t}&:=\max_{\sigma_{r_t}\leq s<\tau_{r_t}(c_t)}
        [\widehat p_t(c_t)-\widehat p_s(c_t)-d_t^{\mathrm V}(c_t;s,s+1)],\\
 e_{\mathrm{A2},t}&:=\widehat p_t(c_t)
         -\min_{\sigma_{r_t}\leq s<\tau_{r_t}(c_t)}\widehat p_s(c_t)-d_t^{(r_t)}(c_t).
\end{aligned}
\label{eq:empirical-bound-excesses}
\end{equation}
The excess $e_{\mathrm{A2},t}$ is the largest increase $\widehat p_t(c_t)-\widehat p_s(c_t)$ over the
monitoring period minus the current risk-increase bound $d_t^{(r_t)}(c_t)$ in
\eqref{eq:repeated-shift-bound}, and $e_{\mathrm{pw},t}$ compares each increase with the risk-increase term
of the deployed variant at the same $s$. A positive excess means that the bound is exceeded.
For a bound factor $\phi<1$ in Appendix~\ref{app:sim-assessments}, both bounds are scaled by $\phi$.
The frequencies $\widehat q_{\mathrm{pw}}$ and $\widehat q_{\mathrm{A2}}$ are the
trajectory frequencies \eqref{eq:empirical-joint-frequencies} of a positive $e_{\mathrm{pw},t}$ and
a positive $e_{\mathrm{A2},t}$ at some checked decision.

\subsection{Scenario-Specific Quantities}
\label{app:empirical-scenario}
In both scenarios, the barrier uses the clearance $\psi$ of
Appendix~\ref{app:unicycle-risk-bound}, and $y_{k,t}$ denotes the body center.
\emph{Contact} is the event $\psi(y_{k,t})\leq0$ for some $1\leq t\leq N_k$, and it
ends the trajectory. Since $\psi$
is $1$-Lipschitz \eqref{eq:unicycle-clearance-lipschitz} and $\|w_t\|\leq c_{\max}$,
the support condition $Y_t\leq c_{\max}$ holds almost surely.

\renewcommand{\dbltopfraction}{0.8}
\section{Simulation Details}
\label{app:sim-study}
\begin{table*}[t]
\centering
\caption{Terms of the certificate bounds $d_t^{\mathrm V}(c;\sigma,\tau)$ defined in \eqref{eq:unicycle-pairwise-bound} of Appendix~\ref{app:unicycle-risk-bound}. The drift term $\Delta_t$ is \eqref{eq:distribution-drift-bound} taken at each $s$, as stated after it.}
\label{tab:sim-bounds}
\normalsize\setlength{\tabcolsep}{4pt}
\begin{tabular}{@{}llllc@{}}\toprule
Variant & Sensitivity bound & Feature-change term & Drift term & Definition\\\midrule
V-U & $L_x^{\mathrm{geom}},L_u^{\mathrm{geom}}$ (uniform) & $\bar\eta_{t,s}$ (uniform) & $\Delta_t$ & \eqref{eq:unicycle-vu-term}\\
V-$\Delta$ & $L_x^{\mathrm{geom}},L_u^{\mathrm{geom}}$ (uniform) & $\eta_{t,s}(c)$ & $\Delta_t$ & \eqref{eq:unicycle-vdelta-term}\\
V-L & $L_{\mathrm{geom}}(c)$ & $\bar\eta_{t,s}$ (uniform) & $\Delta_t$ & \eqref{eq:unicycle-vl-term}\\
V-C & $L_{\mathrm{geom}}(c)$ & $\eta_{t,s}(c)$ & $\Delta_t(\bar P(c)-B_{t,s}(c))$ & \eqref{eq:unicycle-candidate-transfer}\\\bottomrule
\end{tabular}
\end{table*}

\begin{table*}[t]
\centering
\caption{Parameters of both scenarios, the main grid, and the selected settings.}
\label{tab:sim-bindings}
\normalsize\setlength{\tabcolsep}{3pt}\renewcommand{\arraystretch}{1.03}
\begin{tabular}{@{}>{\raggedright\arraybackslash\leavevmode}p{.30\textwidth}>{\raggedright\arraybackslash\leavevmode}p{.32\textwidth}>{\raggedright\arraybackslash\leavevmode}p{.32\textwidth}@{}}\toprule
Quantity & Corridor & Maze\\\midrule
Map; obstacles & $15\times2$\,m; $8$ cones & $9\times9$\,m; $6$ interior walls, $13$ cones\\
Start; goal; goal tolerance & $[0.3,1.1]\times[0.5,1.5]$\,m, $|\theta|\leq15^\circ$; $(14.4,1.0)$\,m; $0.25$\,m & $(1.0,0.75\pm0.08)$\,m, $\theta=90^\circ$; $(8.06,8.25)$\,m; $0.15$\,m\\
Sampling period $T_s$; decision horizon $T$ & $0.1$\,s; $700$ & $0.2$\,s; $600$\\
Input bounds & $|v|\leq0.6$\,m/s, $|\omega|\leq2$\,rad/s & $|a|\leq1.5$\,m/s$^2$, $|\nu|\leq8$\,rad/s$^2$\\
State bounds; braking $a_{\mathrm{br}}$ & -- & $v\in[-0.3,0.8]$\,m/s, $|\omega|\leq1.5$\,rad/s; $1.2$\,m/s$^2$\\
Cone radius $\rho_j$; body radius $\rho_{\mathrm{body}}$ & $0.305$\,m; $0.22$\,m & same\\
Nose distance $l$ & $0.15$\,m & same\\
Nominal input & $k_\theta=2$, $v_{\mathrm c}=0.5$\,m/s & look-ahead $0.5$\,m, $k_\theta=2.5$, $k_\omega=4$, $k_v=2$, $v_{\mathrm c}=0.6$\,m/s\\
Filter weight & $w_\omega=0.0225$ & $w_\nu=0.0225$\\
Decay $\gamma$; margin $c_{\max}$; candidates $\Gamma$ ($M=7$) & $0.25$; $0.02$\,m; $c_{\max}\{0.3,\ldots,0.9\}$ & $0.25$; $0.01$\,m; same\\
Gains $\Lambda$ & $\{0.2,0.4,0.6,0.8,1.05\}$ & same\\
$\varrho_{\min}$ in \eqref{eq:unicycle-minimum-distance}; $K_x^z$; $K_u^z$ in \eqref{eq:unicycle-first-order-constants}, \eqref{eq:unicycle-second-order-constants} & $0.675$\,m; $1.022$; $0.102$ & $0.37$\,m; $1.067$; $0.0204$\\
$L_{\mathrm{geom}}(c)$ in \eqref{eq:unicycle-clearance-sensitivity} over $\Gamma$ & $[3.09,6.69]\times10^{-3}$\,m$^{-1}$ & $[5.13,11.14]\times10^{-3}$\,m$^{-1}$\\
$L_x^{\mathrm{geom}}$; $L_u^{\mathrm{geom}}$ in \eqref{eq:unicycle-risk-constants} & $6.84\times10^{-3}$; $6.80\times10^{-4}$ & $1.19\times10^{-2}$; $2.27\times10^{-4}$\\
ACI $k_{\mathrm{ACI}}$; $N^{\mathrm{win}}$; initial margin & $0.02$; $500$; $c_{\max}$ & $0.005$; $500$; $c_{\max}$\\
QT $k_{\mathrm{QT}}$; initial margin & $0.3\,c_{\max}$; $c_{\max}$ & $0.02\,c_{\max}$; $c_{\max}$\\
\midrule
 Main grid: $\alpha_0$; $\alpha$ & \multicolumn{2}{>{\raggedright\arraybackslash\leavevmode}p{\dimexpr.64\textwidth+2\tabcolsep\relax}@{}}{$\{0.10,0.15,0.20,0.25,0.30\}$; $\{0.05,0.10,\ldots,\alpha_0-0.05\}$ ($15$ pairs)}\\
Main grid: $\delta$; $K$ & \multicolumn{2}{>{\raggedright\arraybackslash\leavevmode}p{\dimexpr.64\textwidth+2\tabcolsep\relax}@{}}{$\{0.025,0.05,0.10,0.15,0.20\}$; $\{10,50,100,200,\infty\}$ ($375$ settings)}\\
Default bound factor $\phi$ and budget normalizer $b$ (Appendix~\ref{app:sim-assessments}); $L_\mu$ in \eqref{eq:distribution-drift-rate} & \multicolumn{2}{>{\raggedright\arraybackslash\leavevmode}p{\dimexpr.64\textwidth+2\tabcolsep\relax}@{}}{$1$; $\kappa$, rounded up to $2.11$; $0.0005$}\\
Selected settings $(\alpha_0,\alpha,\delta,K)$ & \multicolumn{2}{>{\raggedright\arraybackslash\leavevmode}p{\dimexpr.64\textwidth+2\tabcolsep\relax}@{}}{S1~$(0.30,0.25,0.20,50)$, S2~$(0.30,0.25,0.15,50)$, S3~$(0.30,0.20,0.20,50)$, S4~$(0.30,0.25,0.025,100)$, S5~$(0.30,0.25,0.10,10)$}\\\bottomrule
\end{tabular}
\end{table*}

This appendix specifies the simulations of Section~\ref{sec:empirical-evaluation}.
Appendix~\ref{app:sim-methods} defines the compared methods and the simulation models, and
Appendix~\ref{app:sim-assessments} the error allocation and the bound factor.
Appendix~\ref{app:sim-corridor} and Appendix~\ref{app:sim-checks} report the Corridor results and the checks of the
guarantees and the assumptions. All methods share the initial
states and, for each $L_\mu$, the disturbance realizations. The supplementary ablation report~\cite{trakas2026ablation} gives
further comparisons and the implementation details, and the code is in the repository linked in
Section~\ref{sec:empirical-setup}.

\subsection{Compared Methods and Simulation Models}
\label{app:sim-methods}
\label{app:sim-scenarios}
All filtered methods use the same nominal input, input limits and weighted input deviation as the
objective of \eqref{eq:cbf-safety-filter}, and differ only in the robustness margin $c_t$. The certified
variants use the certificate bounds of Table~\ref{tab:sim-bounds}, derived in
Appendix~\ref{app:unicycle-risk-bound}.
Table~\ref{tab:sim-bindings} gives the parameters, and~\cite{trakas2026ablation} the obstacle coordinates.
Lengths are in meters, angles in radians unless marked in degrees, and times
in seconds.

\paragraph{Baselines}
Unfiltered nominal control applies $u_t^{\mathrm{nom}}$, and the CBF filters with fixed margins use $c_t=0$ and
$c_t=c_{\max}$ in \eqref{eq:cbf-safety-filter}. ACI and QT keep an online estimate $\hat c_t^{\mathrm M}$,
$\mathrm M\in\{\mathrm{ACI},\mathrm{QT}\}$, of the
$(1-\alpha_0)$-quantile of the barrier losses $Y_t$ and deploy $c_t=\min\{c_{\max},[\hat c_t^{\mathrm M}]_+\}$.
ACI takes the empirical $(1-\alpha_t^{\mathrm{ACI}})$-quantile of the last $N^{\mathrm{win}}$ losses and adapts
its level, starting from $\alpha_0^{\mathrm{ACI}}=\alpha_0$, while QT updates the estimate directly:
\begin{align}
 \alpha_{t+1}^{\mathrm{ACI}}
 &=\alpha_t^{\mathrm{ACI}}+k_{\mathrm{ACI}}
       (\alpha_0-\mathbf1_{\{Y_t>\hat c_t^{\mathrm{ACI}}\}}),
 \label{eq:empirical-baselines-aci}\\
 \hat c_{t+1}^{\mathrm{QT}}
 &=\hat c_t^{\mathrm{QT}}+k_{\mathrm{QT}}
       (\mathbf1_{\{Y_t>\hat c_t^{\mathrm{QT}}\}}-\alpha_0).
 \label{eq:empirical-baselines}
\end{align}
Both methods raise their estimate after an exceedance $Y_t>\hat c_t^{\mathrm M}$ and lower it
otherwise, which drives the long-run fraction of such exceedances toward $\alpha_0$.
The window $N^{\mathrm{win}}$, the step sizes $k_{\mathrm{ACI}},k_{\mathrm{QT}}>0$, and the initial
robustness margins are in Table~\ref{tab:sim-bindings}.

\paragraph{Supporting test index}
The \emph{supporting test index} of the deployed margin $c_t\in\mathcal D_t$ is the index of the most recently started test whose certificate of $c_t$ has the smallest current risk-increase bound,
\begin{equation}
 r_t:=\max\operatorname*{arg\,min}_{r:\tau_r(c_t)\leq t}
                      d_t^{(r)}(c_t),
\label{eq:certificate-selection}
\end{equation}
with $d_t^{(r)}$ in \eqref{eq:repeated-shift-bound}, and $r_t:=0$ if $\mathcal D_t=\emptyset$. Its certificate meets the deployment condition in
\eqref{eq:D_updated}, and Appendix~\ref{app:empirical-assumption-checks} checks
Assumption~\ref{as:bounded-shift} over its monitoring period.

\paragraph{Dynamics and barrier functions}
The robot has body center $y\in\R^2$, heading $\theta$, and nose point
$q(x):=y+le(\theta)$ with $e(\theta):=(\cos\theta,\sin\theta)$ and nose
distance $l>0$, and $\psi$ is the clearance
\eqref{eq:corridor-example-clearance} in the Corridor and
\eqref{eq:maze-example-clearance} in the Maze. The nominal continuous-time dynamics are
\begin{alignat}{2}
 &\text{Corridor:}\quad &\dot y&=ve(\theta),\ \ \dot\theta=\omega,
 \label{eq:corridor-example-dynamics}\\
 &\text{Maze:} &\dot y&=ve(\theta),\ \ \dot\theta=\omega,\ \ \dot v=a,\ \
 \dot\omega=\nu,
 \label{eq:maze-example-dynamics}
\end{alignat}
with $x=(y,\theta)$, $u=(v,\omega)$ in the Corridor and
$x=(y,\theta,v,\omega)$, $u=(a,\nu)$ in the Maze. With the braking
deceleration $a_{\mathrm{br}}>0$, the barrier function is
\begin{equation}
 h(x):=\psi(q(x))-l-v^2/(2a_{\mathrm{br}})
\label{eq:unicycle-barriers}
\end{equation}
in the Maze, and the same without the braking term in the Corridor. The
nominal one-step predictor $\hat f(x,u)$ in \eqref{eq:residual-decomposition}
is the state reached from $x$ after one sampling period $T_s$ (Table~\ref{tab:sim-bindings}) with $u$ held
constant.

\paragraph{Disturbance}
The true system adds the planar displacement $w_t=A_te(\Phi_t)$ to the
nominal prediction,
\begin{equation}
 f(x,u,w):=\hat f(x,u)+(w,0_{n-2}),
\label{eq:empirical-transition}
\end{equation}
with $n=3$ in the Corridor and $n=5$ in the Maze. Given $\Fc_t$, the direction $\Phi_t$ is uniform on
$[0,2\pi)$ and independent of the magnitude $A_t$, which has the distribution $\mu_t^A$,
\begin{equation}
\begin{aligned}
 A_t\mid\Fc_t&\sim(1-\beta_t)\,\mathrm U_1+\beta_t\,\mathrm U_2,\\
 \beta_t&:=\tfrac12[1+\sin(2L_\mu t)],
\end{aligned}
\label{eq:empirical-disturbance}
\end{equation}
where $\mathrm U_1$ and $\mathrm U_2$ are the uniform distributions on $[0,c_{\max}/2]$ and
$[c_{\max}/2,c_{\max}]$. The distribution $\mu_t^A$ shifts between small and large magnitudes, with
$|\beta_{t+1}-\beta_t|\leq L_\mu$. The controller knows neither $\mu_t^A$ nor $\beta_t$, and the simulations contain no
other model error.

\subsection{Error Allocation and Bound Factor}
\label{app:sim-assessments}
\paragraph{Error allocation}
The budget normalizer $b\geq\kappa$ replaces $\kappa$ in the error levels of
Section~\ref{subsec:distribution-shift}, which become
\begin{equation*}
 \delta_r=\frac{\delta}{b(r+1)\ln^2(r+1)},\qquad
 \sum_{r\geq1}\delta_r=\frac{\delta\kappa}{b}\leq\delta.
\end{equation*}
Theorem~\ref{prop:repeated-deployment} then holds with $\delta$ replaced by $\delta\kappa/b$. By default, $b=\kappa$, rounded up to $2.11$. For $K=\infty$,
only test $1$ starts, at $\sigma_1=0$, with the error level $\delta_1$.

\paragraph{Bound factor}
The bound factor $\phi\in[0,1]$ multiplies $d_t^{\mathrm V}(c;\sigma,\tau)$ in the current risk-increase bound
\eqref{eq:repeated-shift-bound}. The default $\phi=1$ gives the certificate bounds of
Appendix~\ref{app:unicycle-risk-bound}. Values $\phi<1$ shrink these bounds, which then need not
satisfy Assumption~\ref{as:bounded-shift}, and Appendix~\ref{app:arxiv-restored-stress}
and~\cite{trakas2026ablation} compare them.

\subsection{Corridor Results}
\label{app:sim-corridor}
In all certified rows of Table~\ref{tab:empirical-corridor}, both event frequencies meet their levels, with
deployment-event frequency
$0\%$, false-certification frequency at most $3.5\%$, and $\widehat m_{p^{\max}}\leq0.21$.
ACI and QT have deployment-event frequency $100\%$, with $\widehat m_{p^{\max}}$ at $1.91$ and $2.00$ times $\alpha_0$.
Relative to $c_{\max}$, S3 with V-L (Fig.~\ref{fig:sim-map-corridor-T3}) and S1 with V-C
reduce $\widehat m_c/c_{\max}$ by $38\%$ and $50\%$, respectively. V-C minimizes this mean at every setting, as in the Maze.
ACI, QT, and S1 with V-C reach the goal with mean mission times $33.1$, $33.4$, and $33.9\,$s.
The CBF-violation fraction $\widehat q^{+}_{\mathrm{CBF}}$ is $15.2$--$17.7\%$ for ACI and QT, versus $9.7\%$ for S1 with V-C.
\begin{table}[t]
\centering
\caption{Corridor results. Conventions as in Table~\ref{tab:empirical-topfive}. Unfiltered control has contact in
every trajectory and does not reach the goal, so its $\widehat T_{\mathrm{mis}}$ is the horizon $TT_s=70\,$s.}
\label{tab:empirical-corridor}
{\footnotesize\setlength{\tabcolsep}{1.5pt}\renewcommand{\arraystretch}{1.0}%
\begin{tabular}{@{}lr@{}lr@{}lr@{}lr@{}lr@{}lr@{}l@{}}
\guaranteeboxrow{T4}\toprule
Method & \multicolumn{2}{c}{$\widehat m_c/c_{\max}{\downarrow}$} & \multicolumn{2}{c}{$\widehat T_{\rm mis}{\downarrow}$} & \multicolumn{2}{c}{$\widehat q^{+}_{\rm CBF}{\downarrow}$} & \multicolumn{2}{c}{$\widehat m_{p^{\max}}{\downarrow}$} & \multicolumn{2}{c}{$\widehat q_{\rm dep}{\downarrow}$} & \multicolumn{2}{c}{$\widehat q_{\rm cert}{\downarrow}$}\\\midrule
Unfiltered & \multicolumn{2}{c}{---} & $70.0$&{\scriptsize${\pm}.0$} & $30.7$&{\scriptsize${\pm}13.5$} & \multicolumn{2}{c}{---} & \multicolumn{2}{c}{---} & \multicolumn{2}{c}{---}\\
$c=0$ & $0.00$&{\scriptsize${\pm}.00$} & $32.4$&{\scriptsize${\pm}.8$} & $28.0$&{\scriptsize${\pm}3.0$} & $0.79$&{\scriptsize${\pm}.09$} & 100&\,(100) & \multicolumn{2}{c}{---}\\
$c_{\max}$ & $1.00$&{\scriptsize${\pm}.00$} & $35.9$&{\scriptsize${\pm}.9$} & $0.0$&{\scriptsize${\pm}0.0$} & \multicolumn{2}{c}{---} & 0.0&\,(1.5) & \multicolumn{2}{c}{---}\\
ACI & $0.22$&{\scriptsize${\pm}.04$} & $33.1$&{\scriptsize${\pm}.8$} & $17.7$&{\scriptsize${\pm}1.6$} & $0.57$&{\scriptsize${\pm}.10$} & 100&\,(100) & \multicolumn{2}{c}{---}\\
QT & $0.30$&{\scriptsize${\pm}.03$} & $33.4$&{\scriptsize${\pm}.8$} & $15.2$&{\scriptsize${\pm}1.3$} & $0.60$&{\scriptsize${\pm}.11$} & 100&\,(100) & \multicolumn{2}{c}{---}\\
\midrule
\multicolumn{13}{@{}l}{\textbf{S1}: $(\alpha_0,\alpha,\delta,K)=$ (0.30, 0.25, 0.20, 50)}\\\cmidrule{1-13}
V-U & $0.79$&{\scriptsize${\pm}.03$} & $35.1$&{\scriptsize${\pm}.8$} & $\mathbf{2.6}$&{\scriptsize${\pm}0.9$} & $\mathbf{0.16}$&{\scriptsize${\pm}.04$} & 0.0&\,(1.5) & 0.5&\,(2.4)\\
V-$\Delta$ & $0.79$&{\scriptsize${\pm}.03$} & $35.1$&{\scriptsize${\pm}.8$} & $\mathbf{2.6}$&{\scriptsize${\pm}0.9$} & $\mathbf{0.16}$&{\scriptsize${\pm}.04$} & 0.0&\,(1.5) & 0.5&\,(2.4)\\
V-L & $0.63$&{\scriptsize${\pm}.05$} & $34.4$&{\scriptsize${\pm}.9$} & $5.9$&{\scriptsize${\pm}1.2$} & $0.18$&{\scriptsize${\pm}.04$} & 0.0&\,(1.5) & 0.0&\,(1.5)\\
V-C & $\mathbf{0.50}$&{\scriptsize${\pm}.06$} & $\mathbf{33.9}$&{\scriptsize${\pm}.9$} & $9.7$&{\scriptsize${\pm}2.1$} & $0.21$&{\scriptsize${\pm}.03$} & 0.0&\,(1.5) & 1.5&\,(3.9)\\
\cmidrule{1-13}
\multicolumn{13}{@{}l}{\textbf{S2}: $(\alpha_0,\alpha,\delta,K)=$ (0.30, 0.25, 0.15, 50)}\\\cmidrule{1-13}
V-U & $0.81$&{\scriptsize${\pm}.03$} & $35.1$&{\scriptsize${\pm}.8$} & $\mathbf{2.3}$&{\scriptsize${\pm}0.8$} & $\mathbf{0.15}$&{\scriptsize${\pm}.04$} & 0.0&\,(1.5) & 0.5&\,(2.4)\\
V-$\Delta$ & $0.81$&{\scriptsize${\pm}.03$} & $35.1$&{\scriptsize${\pm}.8$} & $\mathbf{2.3}$&{\scriptsize${\pm}0.8$} & $\mathbf{0.15}$&{\scriptsize${\pm}.04$} & 0.0&\,(1.5) & 0.5&\,(2.4)\\
V-L & $0.64$&{\scriptsize${\pm}.05$} & $34.4$&{\scriptsize${\pm}.9$} & $5.5$&{\scriptsize${\pm}1.2$} & $0.17$&{\scriptsize${\pm}.04$} & 0.0&\,(1.5) & 0.5&\,(2.4)\\
V-C & $\mathbf{0.52}$&{\scriptsize${\pm}.06$} & $\mathbf{33.9}$&{\scriptsize${\pm}.9$} & $9.4$&{\scriptsize${\pm}2.0$} & $0.20$&{\scriptsize${\pm}.03$} & 0.0&\,(1.5) & 0.5&\,(2.4)\\
\cmidrule{1-13}
\multicolumn{13}{@{}l}{\textbf{S3}: $(\alpha_0,\alpha,\delta,K)=$ (0.30, 0.20, 0.20, 50)}\\\cmidrule{1-13}
V-U & $0.67$&{\scriptsize${\pm}.05$} & $34.5$&{\scriptsize${\pm}.9$} & $\mathbf{5.0}$&{\scriptsize${\pm}1.1$} & $\mathbf{0.14}$&{\scriptsize${\pm}.04$} & 0.0&\,(1.5) & 3.5&\,(6.5)\\
V-$\Delta$ & $0.67$&{\scriptsize${\pm}.05$} & $34.5$&{\scriptsize${\pm}.9$} & $\mathbf{5.0}$&{\scriptsize${\pm}1.2$} & $\mathbf{0.14}$&{\scriptsize${\pm}.04$} & 0.0&\,(1.5) & 3.5&\,(6.5)\\
V-L & $0.62$&{\scriptsize${\pm}.06$} & $34.3$&{\scriptsize${\pm}.9$} & $6.4$&{\scriptsize${\pm}1.6$} & $0.15$&{\scriptsize${\pm}.04$} & 0.0&\,(1.5) & 3.5&\,(6.5)\\
V-C & $\mathbf{0.58}$&{\scriptsize${\pm}.07$} & $\mathbf{34.2}$&{\scriptsize${\pm}.9$} & $7.4$&{\scriptsize${\pm}2.2$} & $0.17$&{\scriptsize${\pm}.04$} & 0.0&\,(1.5) & 3.5&\,(6.5)\\
\cmidrule{1-13}
\multicolumn{13}{@{}l}{\textbf{S4}: $(\alpha_0,\alpha,\delta,K)=$ (0.30, 0.25, 0.025, 100)}\\\cmidrule{1-13}
V-U & $0.92$&{\scriptsize${\pm}.02$} & $35.6$&{\scriptsize${\pm}.9$} & $\mathbf{0.7}$&{\scriptsize${\pm}0.5$} & $\mathbf{0.11}$&{\scriptsize${\pm}.03$} & 0.0&\,(1.5) & 0.0&\,(1.5)\\
V-$\Delta$ & $0.92$&{\scriptsize${\pm}.02$} & $35.6$&{\scriptsize${\pm}.9$} & $\mathbf{0.7}$&{\scriptsize${\pm}0.5$} & $\mathbf{0.11}$&{\scriptsize${\pm}.03$} & 0.0&\,(1.5) & 0.0&\,(1.5)\\
V-L & $0.82$&{\scriptsize${\pm}.03$} & $35.1$&{\scriptsize${\pm}.9$} & $2.5$&{\scriptsize${\pm}0.9$} & $0.14$&{\scriptsize${\pm}.03$} & 0.0&\,(1.5) & 0.0&\,(1.5)\\
V-C & $\mathbf{0.58}$&{\scriptsize${\pm}.05$} & $\mathbf{34.1}$&{\scriptsize${\pm}.9$} & $7.7$&{\scriptsize${\pm}1.7$} & $0.18$&{\scriptsize${\pm}.03$} & 0.0&\,(1.5) & 0.0&\,(1.5)\\
\cmidrule{1-13}
\multicolumn{13}{@{}l}{\textbf{S5}: $(\alpha_0,\alpha,\delta,K)=$ (0.30, 0.25, 0.10, 10)}\\\cmidrule{1-13}
V-U & $0.77$&{\scriptsize${\pm}.04$} & $34.9$&{\scriptsize${\pm}.9$} & $\mathbf{2.9}$&{\scriptsize${\pm}0.8$} & $\mathbf{0.14}$&{\scriptsize${\pm}.04$} & 0.0&\,(1.5) & 0.5&\,(2.4)\\
V-$\Delta$ & $0.77$&{\scriptsize${\pm}.04$} & $34.9$&{\scriptsize${\pm}.9$} & $\mathbf{2.9}$&{\scriptsize${\pm}0.8$} & $\mathbf{0.14}$&{\scriptsize${\pm}.04$} & 0.0&\,(1.5) & 0.5&\,(2.4)\\
V-L & $0.66$&{\scriptsize${\pm}.05$} & $34.5$&{\scriptsize${\pm}.9$} & $5.1$&{\scriptsize${\pm}1.1$} & $0.16$&{\scriptsize${\pm}.04$} & 0.0&\,(1.5) & 0.0&\,(1.5)\\
V-C & $\mathbf{0.53}$&{\scriptsize${\pm}.06$} & $\mathbf{34.0}$&{\scriptsize${\pm}.9$} & $8.9$&{\scriptsize${\pm}1.9$} & $0.20$&{\scriptsize${\pm}.03$} & 0.0&\,(1.5) & 0.0&\,(1.5)\\
\guaranteeboxend{T4}\bottomrule\end{tabular}}

\vspace{-6pt}
\end{table}

\subsection{Guarantees and Assumptions}
\label{app:sim-checks}
\paragraph{Events}
In the $3000$ main-grid cells (Table~\ref{tab:sim-bindings}), deployment events occur in $5.13\%$,
only at $\alpha_0\leq0.25$. At finite $K$, $0.375\%$ of $2400$ cells miss a level: false-certification frequency
$1.0\%$ (upper confidence bound $3.2\%$) in the Corridor at $(\alpha,\delta,K)=(0.15,0.025,10)$.
At $K=\infty$, $26.67\%$ of $600$ cells miss a level: $20\%$ at $\delta=0.025$, where $\delta_1=1.23\%$ lies below
the $1.5\%$ upper confidence bound at zero possible events, and $6.67\%$ in the Corridor at $\alpha=0.05$,
$\delta\in\{0.05,0.10\}$. Section~2.8 of~\cite{trakas2026ablation}
details the events and single tests ($K=\infty$).

\paragraph{Assumptions}
In all simulations with $\phi=1$, here and in~\cite{trakas2026ablation},
Assumption~\ref{as:conservative-margin} and the tightened condition
\eqref{eq:tightened-cbf-condition} hold at every recorded decision, no bound excess
\eqref{eq:empirical-bound-excesses} is positive at a checked decision, and every trajectory
reaches the goal without contact (Appendix~\ref{app:empirical-scenario}).
In some main-grid cells, V-U, V-$\Delta$ and V-L never deploy a margin below $c_{\max}$, so no
decision is checked there.

\clearpage\onecolumn
\section{Simulation Figures and Tables}
\label{app:arxiv-restored-figures}
\noindent\textbf{Figures and tables in this appendix}
\par\smallskip
{\renewcommand{\arraystretch}{0.85}\setlength{\tabcolsep}{3pt}
\noindent\begin{tabular}{@{}p{0.49\linewidth}p{0.49\linewidth}@{}}
Table~\ref{tab:arxiv-stress-representative}: bound factor and distribution-drift bound. &
Figs.~\ref{fig:arxiv-large-corridor}--\ref{fig:arxiv-large-maze}: trajectories.\\
Fig.~\ref{fig:arxiv-distributions}: per-trajectory distributions. &
Fig.~\ref{fig:arxiv-margin-heatmap}: robustness-margin heatmap.\\
Fig.~\ref{fig:arxiv-parameter-sweeps}: parameter sweeps. &
\end{tabular}}
\subsection{Bound Factor and Distribution-Drift Bound}
\label{app:arxiv-restored-stress}
Table~\ref{tab:arxiv-stress-representative} varies the bound factor $\phi$
defined in Appendix~\ref{app:sim-assessments} and the distribution-drift bound $L_\mu$
defined in \eqref{eq:distribution-drift-rate}, for S3 with V-L in both scenarios.
These are the setting and variant used in the trajectory figures.
Values $\phi<1$ need not satisfy Assumption~\ref{as:bounded-shift} and carry no guarantee.
Section~2.7 of the supplementary ablation report~\cite{trakas2026ablation}
gives the full comparison. Its Tables~17 and~18
report events and bound-excess checks for S1--S5, and its
Section~2.2.3 gives sweeps for all four variants.
\begin{inlinetable}
\caption{Bound factor $\phi$ and distribution-drift bound $L_\mu$ at S3, V-L. Means $\pm$ SD.
Decision fractions and event frequencies in \%, with $95\%$ upper confidence bounds in parentheses for events.
$N_{\mathrm{MC}}=200$ per row.}
\label{tab:arxiv-stress-representative}
\normalsize
\renewcommand{\arraystretch}{1.0}
{\setlength{\tabcolsep}{30.42723pt}\noindent\begin{tabular}{!{\vrule width 0.8pt}w{c}{17.77783pt}|w{c}{45.15997pt}|w{c}{45.15997pt}|w{c}{45.15997pt}|w{c}{45.15997pt}!{\vrule width 0.8pt}}\Xhline{0.8pt}
\multicolumn{5}{!{\vrule width 0.8pt}@{\hskip 4pt}p{\dimexpr \linewidth-20.8pt-2\tabcolsep + 2\tabcolsep - 2\dimexpr 4pt\relax\relax}@{\hskip 4pt}!{\vrule width 0.8pt}}{\raggedright \textbf{S3}: \textcolor{vl}{V-L} $(\alpha_0,\alpha,\delta,K)$ = (0.30, 0.20, 0.20, 50)\quad $N_{\rm MC}=200$}\\\hline
$\phi$ & $\widehat m_c/c_{\max}\!\downarrow$ & $\widehat q_{\rm CBF}^+\!\downarrow$\,(\%) & $\widehat q_{\rm dep}\!\downarrow$\,(\%) & $\widehat q_{\rm cert}\!\downarrow$\,(\%)\rule[-5.10004pt]{0pt}{5.10004pt}\\
\Xhline{0.6pt}
\multicolumn{5}{!{\vrule width 0.8pt}@{\hskip 4pt}l!{\vrule width 0.8pt}}{\textbf{Corridor}: $L_\mu=0.0005$}\\\Xhline{0.4pt}
$1$ & $0.62{\pm}0.06$ & $6.38{\pm}1.57$ & 0.0 (1.5) & 3.5 (6.5)\\
$0.75$ & $0.59{\pm}0.07$ & $7.15{\pm}1.96$ & 0.0 (1.5) & 3.5 (6.5)\\
$0.5$ & $0.58{\pm}0.07$ & $7.44{\pm}2.16$ & 0.0 (1.5) & 3.5 (6.5)\\
$0.25$ & $0.58{\pm}0.07$ & $7.44{\pm}2.16$ & 0.0 (1.5) & 3.5 (6.5)\\
$0$ & $0.57{\pm}0.08$ & $7.44{\pm}2.15$ & 16.0 (20.9) & 3.5 (6.5)\\
\Xhline{1.0pt}
\multicolumn{5}{!{\vrule width 0.8pt}@{\hskip 4pt}l!{\vrule width 0.8pt}}{\textbf{Corridor}: $L_\mu=0.005$}\\\Xhline{0.4pt}
$1$ & $1.00{\pm}0.00$ & $0.02{\pm}0.07$ & 0.0 (1.5) & 1.5 (3.9)\\
$0.75$ & $1.00{\pm}0.00$ & $0.08{\pm}0.18$ & 0.0 (1.5) & 1.5 (3.9)\\
$0.5$ & $0.97{\pm}0.01$ & $0.36{\pm}0.43$ & 0.0 (1.5) & 1.5 (3.9)\\
$0.25$ & $0.83{\pm}0.03$ & $2.61{\pm}0.88$ & 1.5 (3.9) & 1.5 (3.9)\\
$0$ & $0.64{\pm}0.09$ & $7.79{\pm}3.15$ & 20.0 (25.3) & 1.5 (3.9)\\
\Xhline{1.0pt}
\multicolumn{5}{!{\vrule width 0.8pt}@{\hskip 4pt}l!{\vrule width 0.8pt}}{\textbf{Maze}: $L_\mu=0.0005$}\\\Xhline{0.4pt}
$1$ & $0.62{\pm}0.07$ & $5.04{\pm}1.93$ & 0.0 (1.5) & 2.0 (4.6)\\
$0.75$ & $0.61{\pm}0.08$ & $5.36{\pm}2.18$ & 0.0 (1.5) & 2.0 (4.6)\\
$0.5$ & $0.61{\pm}0.08$ & $5.36{\pm}2.18$ & 0.0 (1.5) & 2.0 (4.6)\\
$0.25$ & $0.61{\pm}0.08$ & $5.36{\pm}2.18$ & 0.0 (1.5) & 2.0 (4.6)\\
$0$ & $0.59{\pm}0.08$ & $5.42{\pm}2.13$ & 5.5 (9.0) & 2.0 (4.6)\\
\Xhline{1.0pt}
\multicolumn{5}{!{\vrule width 0.8pt}@{\hskip 4pt}l!{\vrule width 0.8pt}}{\textbf{Maze}: $L_\mu=0.005$}\\\Xhline{0.4pt}
$1$ & $1.00{\pm}0.00$ & $0.00{\pm}0.00$ & 0.0 (1.5) & 1.0 (3.2)\\
$0.75$ & $0.99{\pm}0.01$ & $0.00{\pm}0.03$ & 0.0 (1.5) & 1.0 (3.2)\\
$0.5$ & $0.97{\pm}0.01$ & $0.26{\pm}0.31$ & 0.0 (1.5) & 1.5 (3.9)\\
$0.25$ & $0.83{\pm}0.04$ & $1.84{\pm}1.05$ & 1.5 (3.9) & 1.5 (3.9)\\
$0$ & $0.66{\pm}0.09$ & $6.07{\pm}3.41$ & 12.0 (16.5) & 1.5 (3.9)\\
\Xhline{0.8pt}\end{tabular}}\par

\end{inlinetable}
\clearpage
\subsection{Trajectories and Estimated Exceedance Risk}
\label{app:arxiv-restored-trajectories}
\begin{inlinefigure}
\setlength{\abovecaptionskip}{2pt}\setlength{\belowcaptionskip}{0pt}
\includegraphics{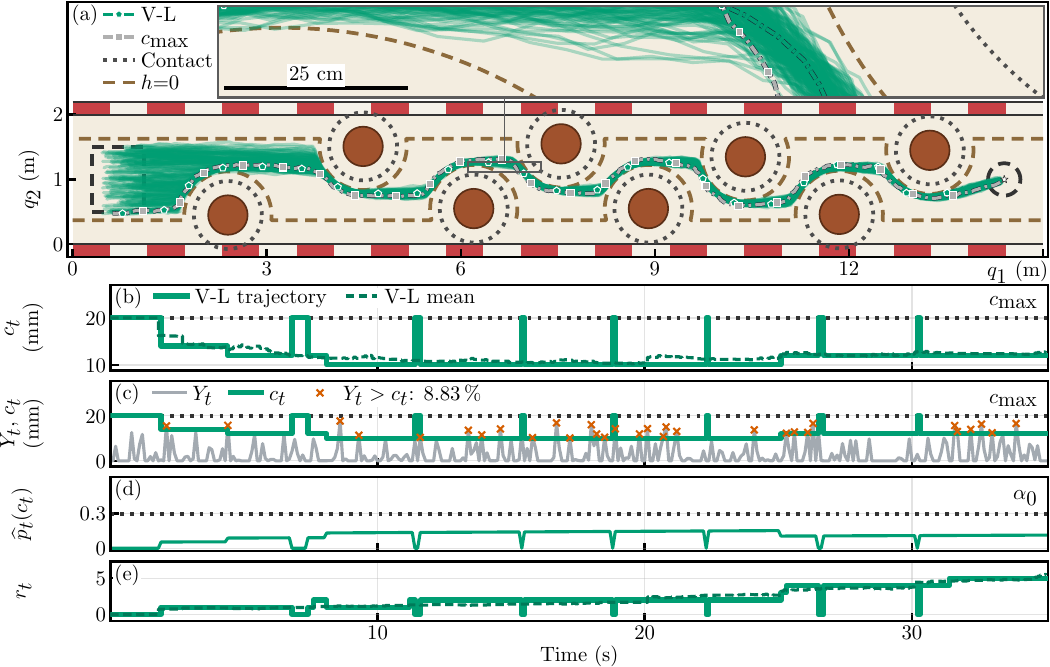}
\caption{Corridor, S3 with V-L. Panels (a)--(d) as in Fig.~\ref{fig:example-maze-winner},
and (e) the supporting test index $r_t$.}
\label{fig:arxiv-large-corridor}\label{fig:sim-map-corridor-T3}
\par\vspace{4pt}
\includegraphics{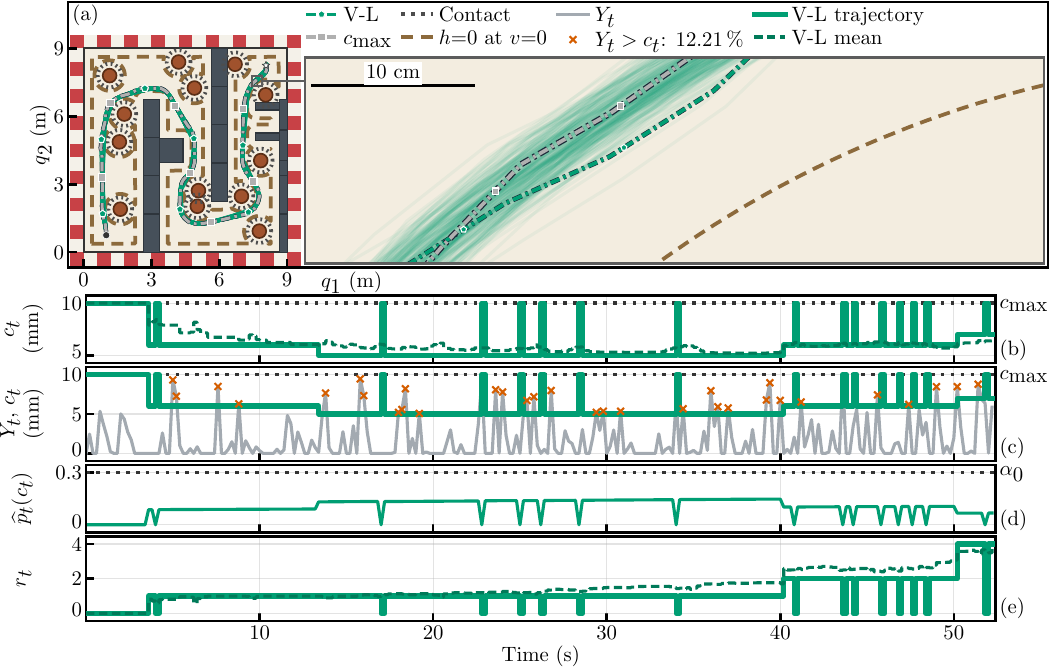}
\caption{Maze, S3 with V-L. Panels as in Fig.~\ref{fig:arxiv-large-corridor}.
The legend in (a) serves all panels.}
\label{fig:arxiv-large-maze}
\end{inlinefigure}

\clearpage
\subsection{Distributions and Robustness-Margin Heatmap}
\label{app:arxiv-restored-distributions}
In Fig.~\ref{fig:arxiv-distributions}, points are trajectories, boxes span the quartiles with the median,
and whiskers reach the most extreme values within $1.5$ interquartile ranges of the quartiles.
Unfiltered control is omitted for scale. Fig.~\ref{fig:arxiv-margin-heatmap} compares the certified variants over $\alpha$ and $K$.
\begin{inlinefigure}
\includegraphics{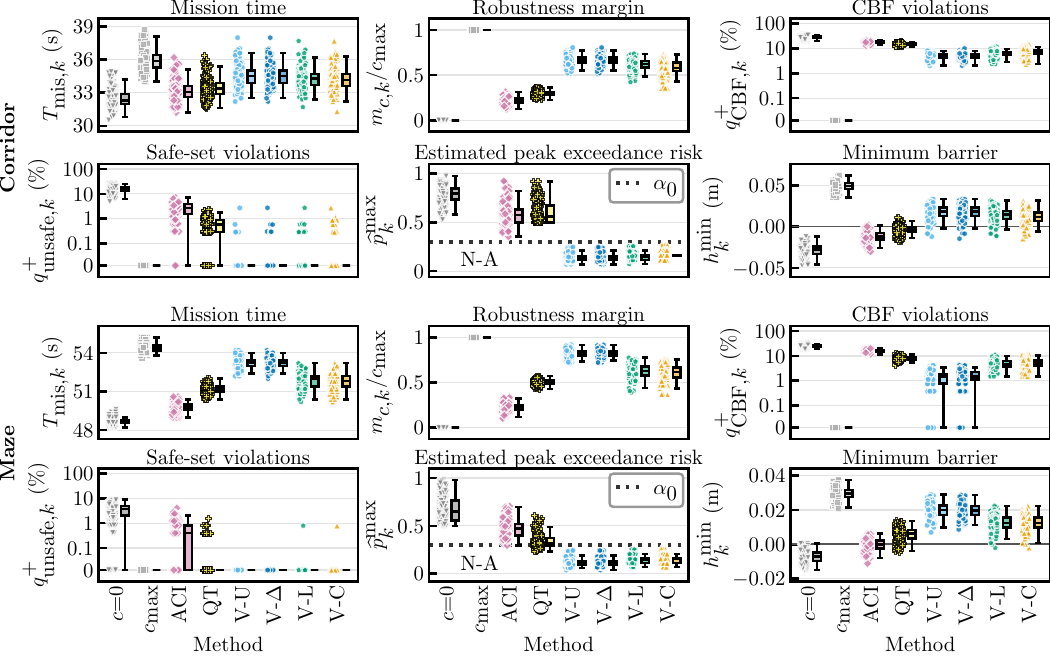}
\caption{Distributions at S3, Corridor (top) and Maze (bottom). N-A: no smaller-margin deployment.
Decision fractions are percentages.}
\label{fig:arxiv-distributions}
\end{inlinefigure}
\begin{inlinefigure}
\includegraphics{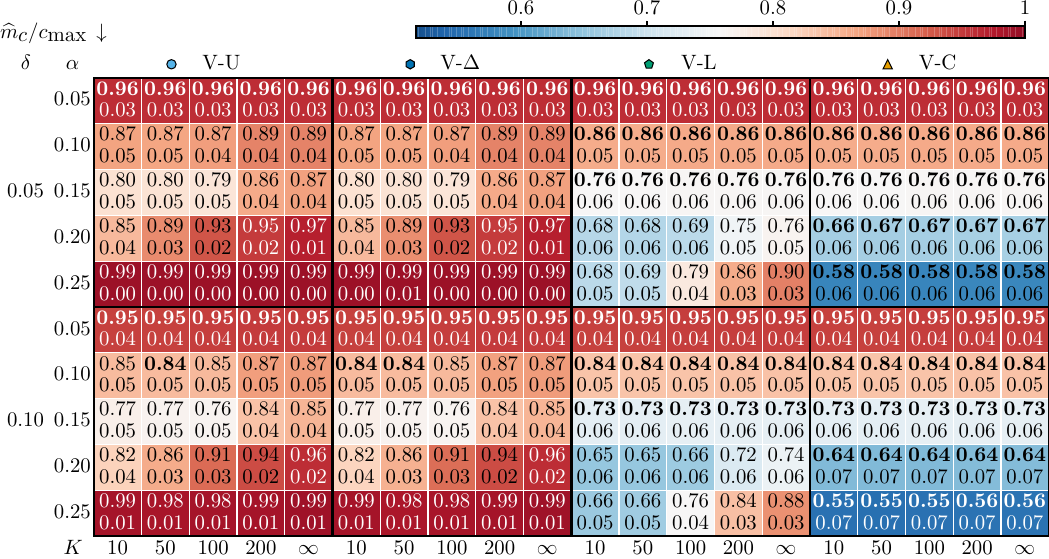}
\caption{Maze, $\alpha_0=0.30$: $\widehat m_c/c_{\max}$ over $\alpha$ and $K$ at
$\delta\in\{0.05,0.10\}$. Cells give mean (top) and SD (bottom). Bold: smallest displayed mean among variants,
including ties.}
\label{fig:arxiv-margin-heatmap}
\end{inlinefigure}

\clearpage
\subsection{Parameter Sweeps}
Fig.~\ref{fig:arxiv-parameter-sweeps} varies one parameter at a time around S3. Boxes span the
quartiles with the median, whiskers mark the 5th and 95th percentiles, and markers show the means.
\begin{inlinefigure}
\includegraphics{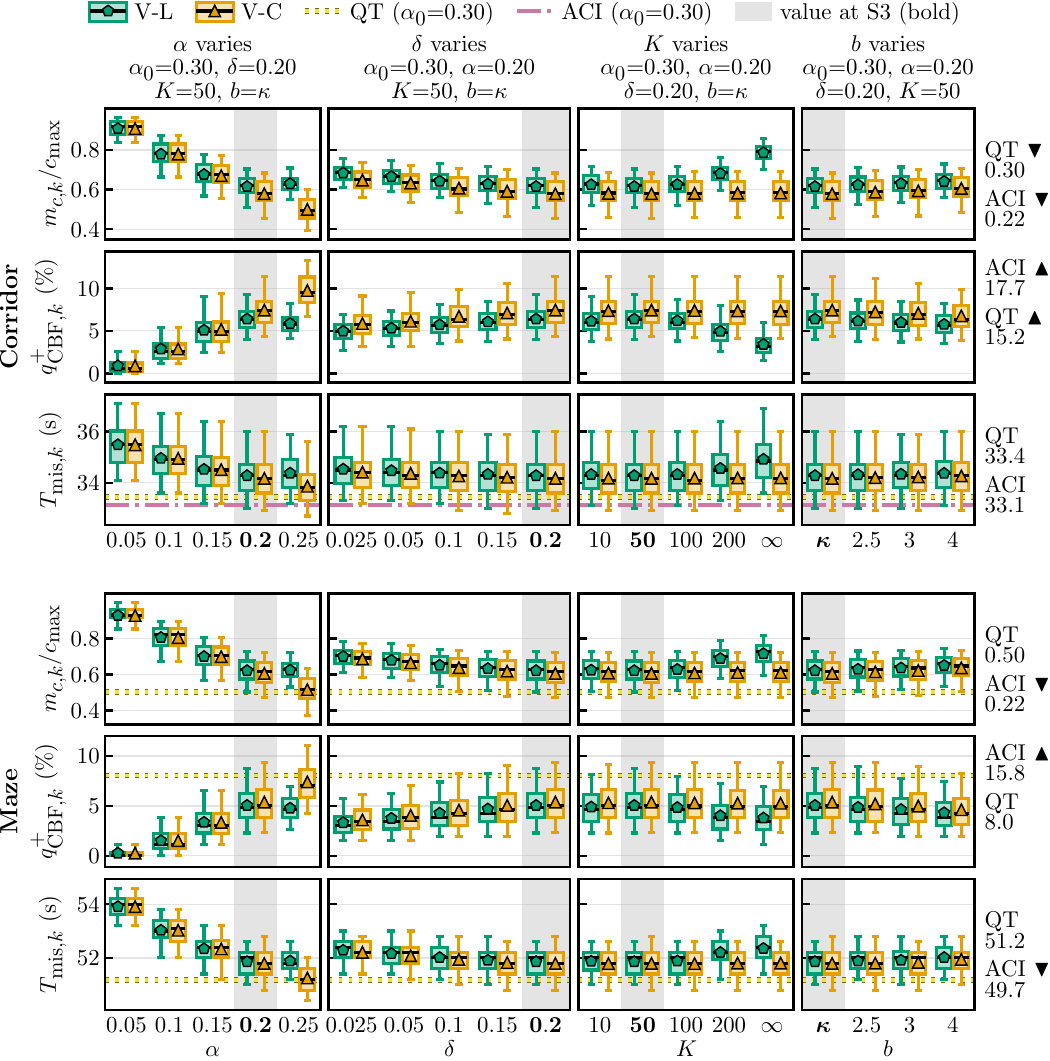}
\caption{One-parameter sweeps around S3, Corridor (top) and Maze (bottom), for V-L and V-C.
Right-hand labels: means of QT and ACI, with $\blacktriangle$/$\blacktriangledown$ for values above/below the axis range.}
\label{fig:arxiv-parameter-sweeps}
\end{inlinefigure}

\end{document}